%% file: main.tex
\documentclass[%
 reprint,
 amsmath,amssymb,
 aps,
prb,
]{revtex4-2}

\usepackage[utf8]{inputenc}
\usepackage{graphicx}
\usepackage{overpic}
 \usepackage{rotating}
 \usepackage{mwe, tikz}
 \usepackage{amsmath,amssymb}
\usepackage{pgfplots}
\usetikzlibrary{calc,arrows,decorations.markings,shapes.geometric} 
\usepackage{amssymb}

\usepackage[finalnew]{trackchanges}

\usepackage{lipsum}
\usepackage{romannum}
\usepackage{xfp}

\soulregister\ref7
\soulregister\cite7
\soulregister\citeA7
\soulregister\romannum7

\usepackage{graphicx}
\usepackage{amsmath}
\usepackage{mathtools} 
\usepackage{color}
\usepackage{overpic}
\usepackage[export]{adjustbox}
\usepackage{mwe, tikz}
\usepackage{rotating}
\usepackage{latexsym}
\usepackage{csquotes}
\usepackage{tikz} 
\usetikzlibrary{calc,arrows,decorations.markings} 
\usepackage{pict2e}
\usepackage{pgfplots}
\usepackage{float}
\usepackage{hyperref}
\usepackage{xr}
\usepackage{rotating}
\usepackage{amsmath,amssymb}
\usepackage{fp}
\usetikzlibrary{calc} 
\usepackage{booktabs, tabularx}
\usepackage{romannum}
\usepackage{bm}
\usepackage{float}
\usepackage{adjustbox}
\usepackage{booktabs, tabularx}
\usepackage{longtable}

\input{macros}
\newcommand*{\Labelxy}[4]{\put(#1,#2) {\setlength{\fboxsep}{0pt}{\strut\textcolor{black}{\begin{turn}{#3}{#4}\end{turn}}}}}
\newcommand*{\LabelFig}[3]{\put(#1,#2) {\setlength{\fboxsep}{0pt}\colorbox{white}{\textcolor{black}{#3}}} }
\definecolor{darkolivegreen}{rgb}{0.33, 0.42, 0.18}
\definecolor{darkspringgreen}{rgb}{0.09, 0.45, 0.27}
\definecolor{darkslategray}{rgb}{0.18, 0.31, 0.31}
\definecolor{darkred}{rgb}{0.55, 0.0, 0.0}
\addeditor{KK}

\newcommand{\glseven}{NiCoCrFeMn~}
\newcommand{\glsix}{NiCoCr~}

\newcommand{\ie}{{\it i.e.}~}
\newcommand{\cf}{{\it cf.}~}

\newcommand*{\romnum}[1]{\textbf{\uppercase\expandafter{\romannumeral#1\relax}}}
\usepackage{xr}
\makeatletter
\newcommand*{\addFileDependency}[1]{
  \typeout{(#1)}
  \@addtofilelist{#1}
  \IfFileExists{#1}{}{\typeout{No file #1.}}
}
\makeatother
\newcommand*{\myexternaldocument}[1]{
    \externaldocument{#1}
    \addFileDependency{#1.tex}
    \addFileDependency{#1.aux}
}
\myexternaldocument{./sm}

\begin{document}

\title{Multiscale modeling of kinetic sluggishness in equiatomic NiCoCr and NiCoCrFeMn single-phase solid solutions}

\author{Kamran Karimi$^1$}
\email{kamran.karimi@ncbj.gov.pl}
\author{Stefanos Papanikolaou$^1$}
\email{stefanos.papanikolaou@ncbj.gov.pl}
\affiliation{%
 $^1$ NOMATEN Centre of Excellence, National Center for Nuclear Research, ul. A. Sołtana 7, 05-400 Swierk/Otwock, Poland\\
}%

\begin{abstract}
Complex, concentrated, multi-component alloys have been shown to display outstanding thermo-mechanical properties, that have been typically attributed to sluggish diffusion, entropic, and lattice distortion effects. Here, we investigate two metal alloys with such exemplary properties, the equiatomic, single-phase, face-centered-cubic (FCC) alloys \glsix and \glseven\!\!\!, and we compare their microstructural kinetics to the  behaviors in a pure-Ni FCC metal. We perform long-time, kinetic Monte Carlo (kMC) simulations, and we analyze in detail the kinetics of atomic vacancies. We find that vacancies in both concentrated alloys exhibit subdiffusive thermally driven dynamics, in direct contrast to the diffusive dynamics of pure Ni. Subdiffusive dynamics shall be attributed to dynamical sluggishness, that is modeled by a  fractional Brownian random walk. Furthermore, we analyze the statistics of waiting times, and we interpret long power-law-distributed rest periods as a direct consequence of barriers' energy-scales and lattice distortions.
\end{abstract}

\maketitle

\begin{figure}[tbh]
    \raggedright
    \begin{overpic}[width=0.22\textwidth]{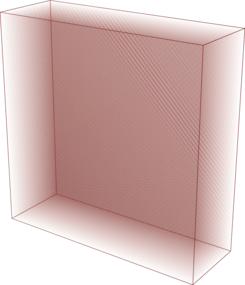}
        \put(24,43){\includegraphics[width=0.05\textwidth]{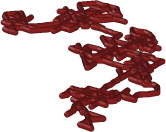}}
        \put(50,36){\includegraphics[width=0.05\textwidth]{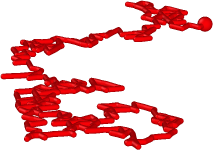}}
        \put(38,67){\includegraphics[width=0.05\textwidth]{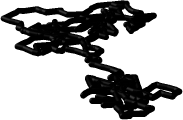}}
        
        \LabelFig{0}{99}{$a)$} 
        \LabelFig{84}{99}{$b)$} 

        \put(90,40){\includegraphics[width=0.49\columnwidth]{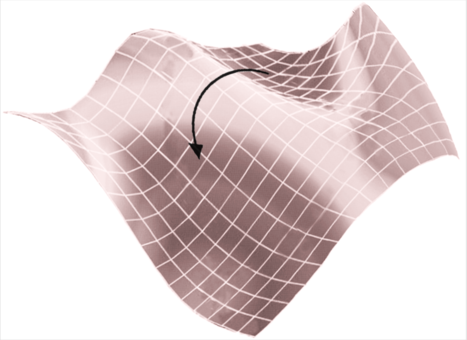}}

        \Labelxy{96}{92}{102}{\scriptsize Energy} 
        \Labelxy{107}{51}{-40}{\scriptsize reaction} 
        \Labelxy{107}{45}{-40}{\scriptsize coordinate}

        \LabelFig{149}{93}{\scriptsize \romnum{1}} 
        \LabelFig{120}{93}{\scriptsize \romnum{2}}
        \LabelFig{127}{65}{\scriptsize \romnum{3}}
        
        \put(90,-7){\includegraphics[width=0.1\textwidth]{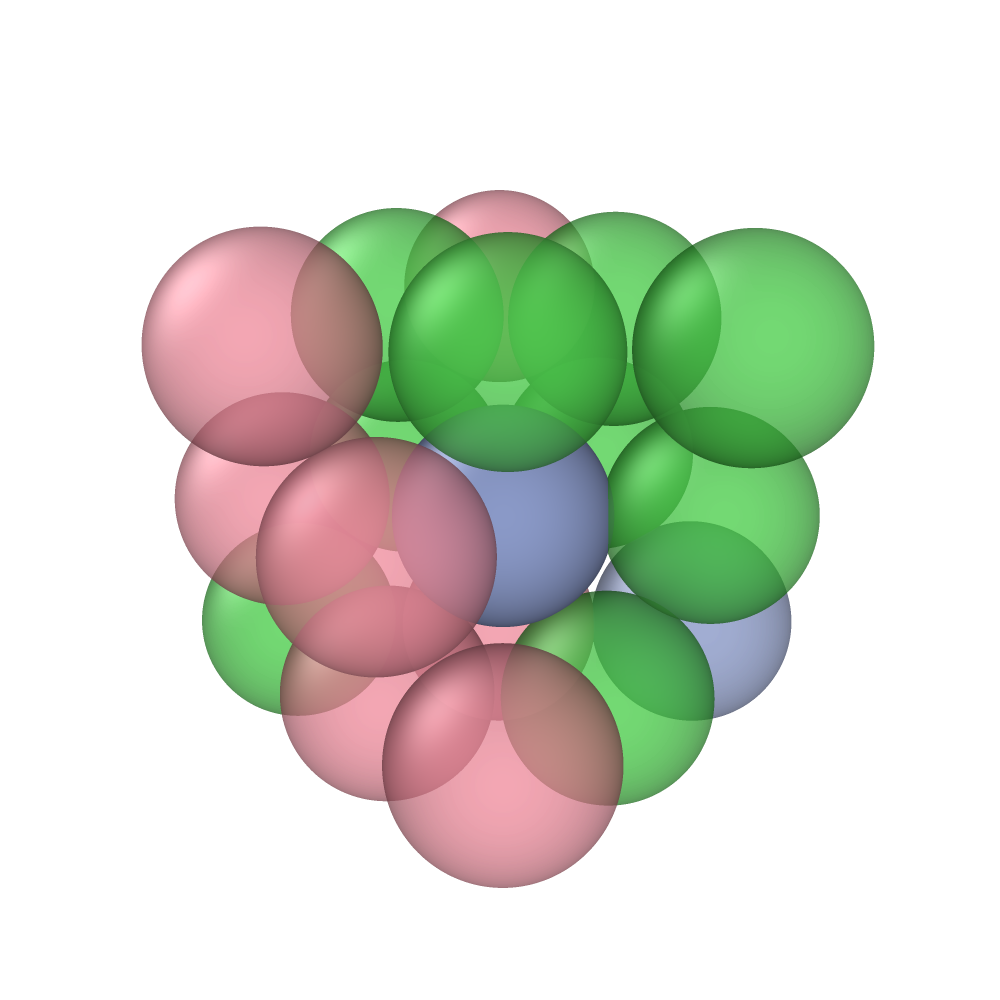}}
        \put(124,-7){\includegraphics[width=0.1\textwidth]{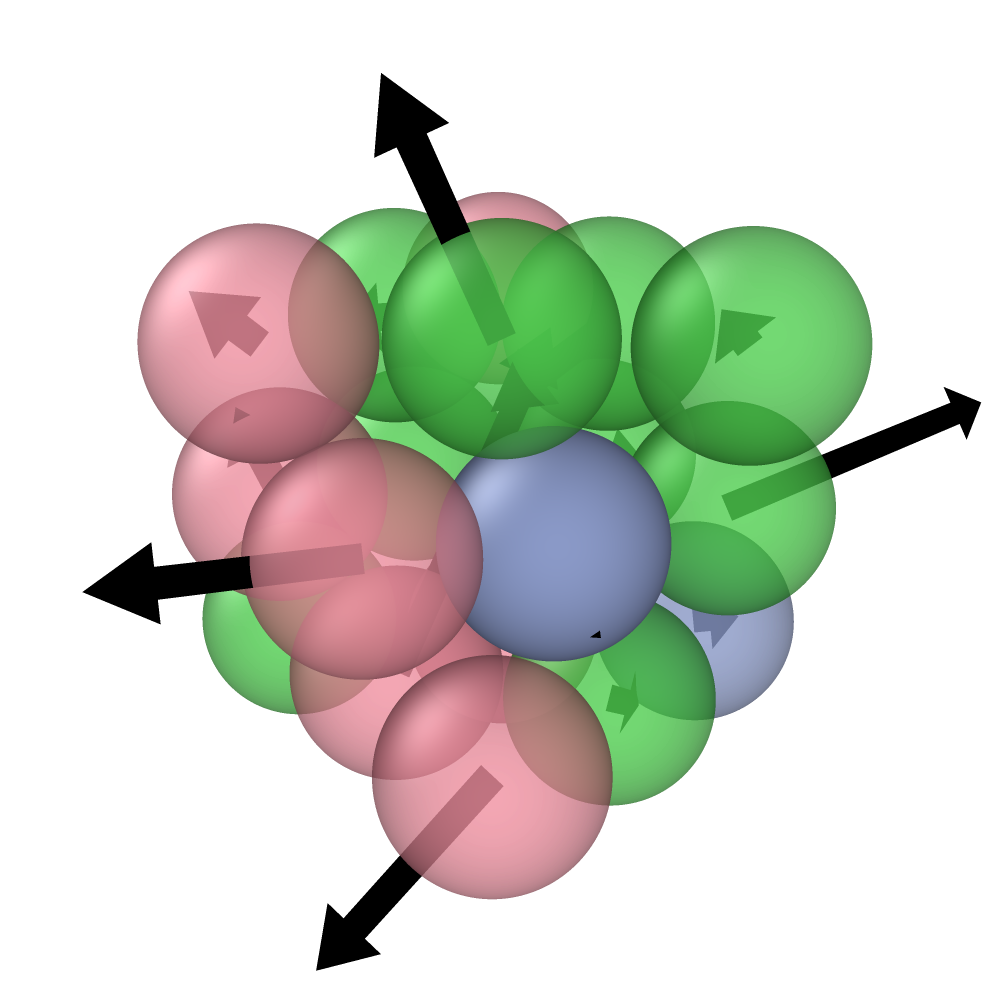}}
        \put(158,-7){\includegraphics[width=0.1\textwidth]{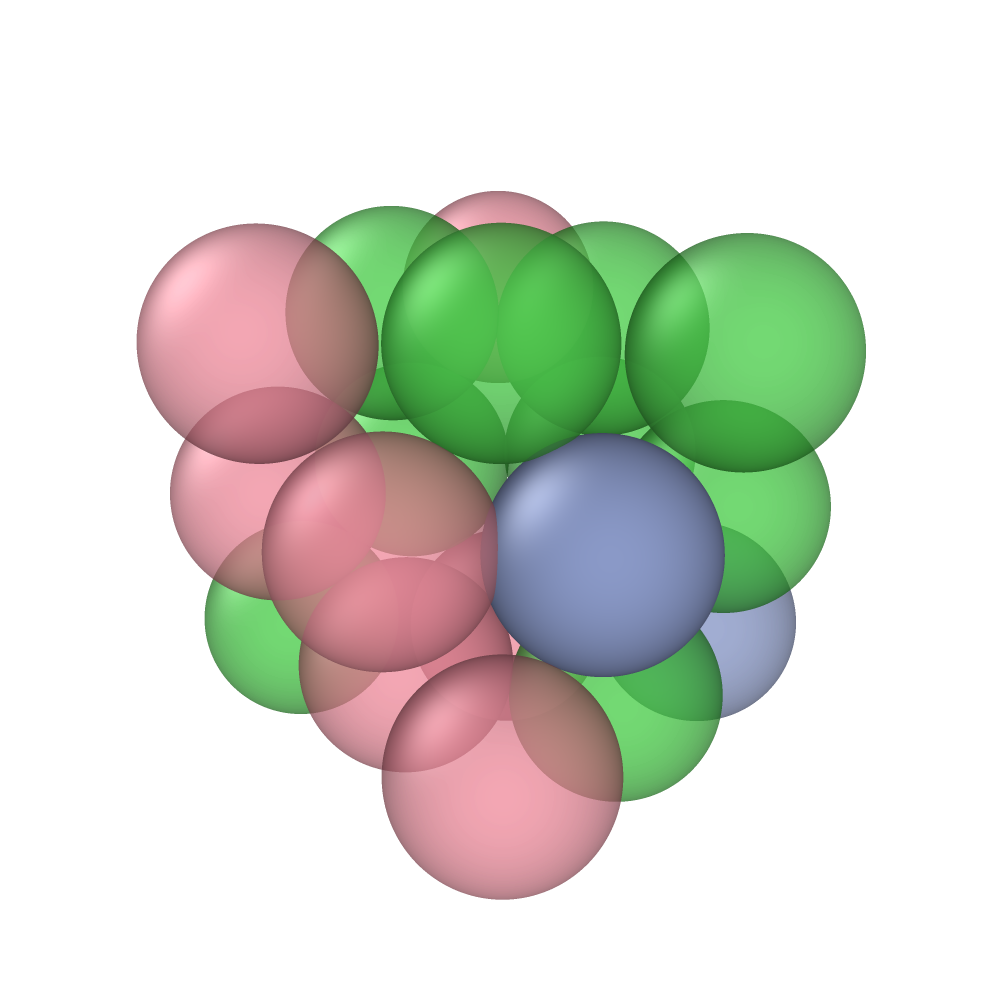}}
        \LabelFig{98}{25}{\scriptsize \romnum{1}} 
        \LabelFig{131}{25}{\scriptsize \romnum{2}}
        \LabelFig{166}{25}{\scriptsize \romnum{3}}
        
       \begin{tikzpicture}
             \coordSys{0.8}{1.4}{.55}{1}{-0.4}{-1}{-.6}{-0.0}{1} 
             \arrow{4.8}{3.7}{2.0}{-0.2}{1.0}{}
             \lline{4.8}{3.7}{.4}{1.0}{0.9}{}
             \arrow{4.8}{3.7}{3.3}{0.4}{-.3}{}	
        \end{tikzpicture}

    %
    \end{overpic}
    \caption{Thermally-assisted vacancy migration in \glsix at $T=1400$ K. The line segments in \textbf{a}) indicate defect trajectories over order 100 Monte Carlo moves of size $1$ \r{A}. Each Monte Carlo step in \textbf{b}) involves a center (blue) atom in \romnum{1} diffusing to a neighboring vacant site as in \romnum{3} through a saddle-point configuration of \romnum{2}. The arrows in \romnum{2} denote atomic displacements relative to \romnum{1}. The migration barrier is determined based on the energy cost between \romnum{2} and \romnum{1} on the energy hyper-surface along the reaction coordinate.  During the course of each simulation, we typically observe $12$ different sets of noncrystalline topology as in \romnum{1}.}
    \label{fig:randomWalk}
    \vspace{-.7cm}
\end{figure}
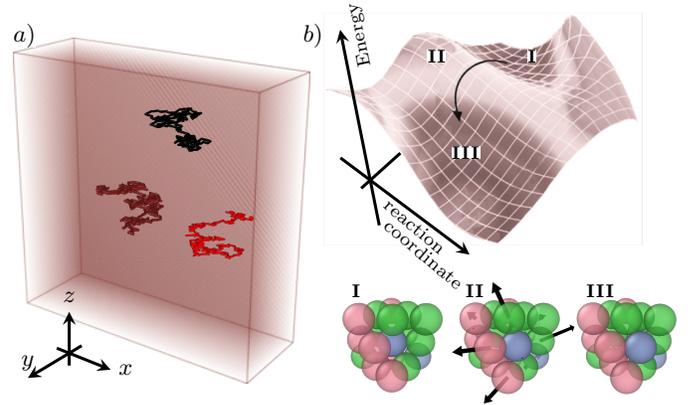

Atomic-scale transport properties in complex concentrated alloys (CCAs) have long been hypothesized to be characterized by comparatively slow kinetics, as opposed to pure metals and conventional alloys, hence the term \emph{sluggish} diffusion \cite{shang2021mechanical,jien2006recent}.
Together with high entropy of mixing, severe lattice distortion, and also, the cocktail effect, these so-called ``core effects" are commonly identified as the principal sources of exceptional CCA thermo-mechanical properties ({\it e.g.} single-phase thermodynamic stability \cite{zhang2008solid}, creep resistance \cite{li2018creep}, and high-temperature strength \cite{chen2018review}).
Sluggishness of diffusion dynamics, in particular, connects to apparent compositional and underlying atomic structure complexities \cite{li2019mechanical}, but its demonstration and connection to multi-principal element alloys' outstanding properties have been challenging \cite{miracle2017critical}. In this Letter, we demonstrate in molecular simulations (\cf Fig.~\ref{fig:randomWalk}) the sluggishness of vacancies in CCAs,  and further model it in terms of subdiffusive fractional Brownian dynamics. We investigate two alloys with exceptional mechanical properties (equiatomic FCC NiCoCr and NiCoCrFeMn \cite{li2019mechanical}), and we develop connections of the subdiffusive vacancy dynamics to underlying crystal lattice distortions.

 Diffusion sluggishness in CCAs has been evasive, given that typical measurements of elemental tracer diffusivities display conventional metallic behavior~\cite{DABROWA2019193,miracle2017critical}. Instead, it appears that slow diffusion kinetics may not be a robust feature of CCAs in an absolute temperature sense~\cite{tsai2013sluggish,jien2006recent,miracle2017critical} but can be perceived within a reduced temperature scale (\ie with melting temperature $T_m$ as the scaling factor) when compared with pure metals and/or conventional alloys~\cite{beke2016diffusion,vaidya2016ni,DABROWA2019193}.
\add[KK]{This is based on a heuristic argument that alloys' diffusion coefficients $D$ at $T_m$ are almost independent of specific chemical compositions but tend to show variations with the crystal structure \cite{miracle2017critical}. 
It follows that $D\propto e^{-\Delta E_*(1/T_*-1)}$ with reduced units $T_*=T/T_m$ and $\Delta E_* = \Delta E/k_BT_m$.
Here $k_B$ is the Boltzmann constant.}
This suggestion was further strengthened by high rescaled activation energies $\Delta E_*$ of interdiffusion, due to inherent ruggedness in potential energy landscape, from severe lattice distortions~\cite{xi2022mechanism,dkabrowa2020state}, confirming the importance of rescaling, for tracer(inter) diffusivities.  
In contrast, CCA modeling efforts to extract atomic-level transport properties have been mainly centered on coarse-grained meso-scale modeling of vacancy-driven diffusivity under thermal activation \cite{osetsky2016specific,kottke2020experimental} and/or irradiation conditions \cite{yang2018irradiation}, without explicit links to atomistic, compositional complexities \cite{ponga2022effects,zhou2022vacancy,sugita2022vacancy,wang2022disentangling}.
It is currently clear that the characterization of kinetic sluggishness requires the  thorough understanding of composition-dependent atomistic features. 

This work reports on a novel atomistic-based simulation investigation of diffusion properties in two key complex concentrated alloys that have consistently displayed excellent mechanical properties~\cite{shang2021mechanical}. We investigate vacancy migration in single-phase FCC equiatomic \glsix and \glseven solid solutions using molecular simulations, and we identify \emph{anomalous} dynamics akin to (self-)diffusion in other complex systems with heterogeneous substructure \cite{metzler2000random}. 
We find that these complex concentrated alloys are characterized by vacancy sub-diffusion, in contrast to single-element metals, through a fractional Brownian process \cite{mandelbrot1968fractional}, due to severe lattice distortions, that governs long-term kinetics of thermally-assisted defect motion.
We investigate a fairly broad range of simulation timescales through a kinetic Monte-Carlo sampling framework that is implemented, using the kinetic Activation-Relaxation Technique ($k$-ART) \cite{el2008kinetic}, that detects vacancy energetics, as illustrated in Fig.~\ref{fig:randomWalk}.
The short-term dynamics is resolved by directly probing statistics of vacancy hopping, in terms of rest times, and pure FCC Ni behavior is used as a benchmark.
We argue that long waiting periods, statistically characterized by broad \emph{non}-exponential temporal distributions, can best describe sluggishness in metal diffusion.
The former can be described as a direct consequence of barrier energy scales and their broad spectrum owing to atomic-level chemical complexities.\\


\begin{figure}[t] 
    \centering
    \begin{overpic}[width=0.49\columnwidth]{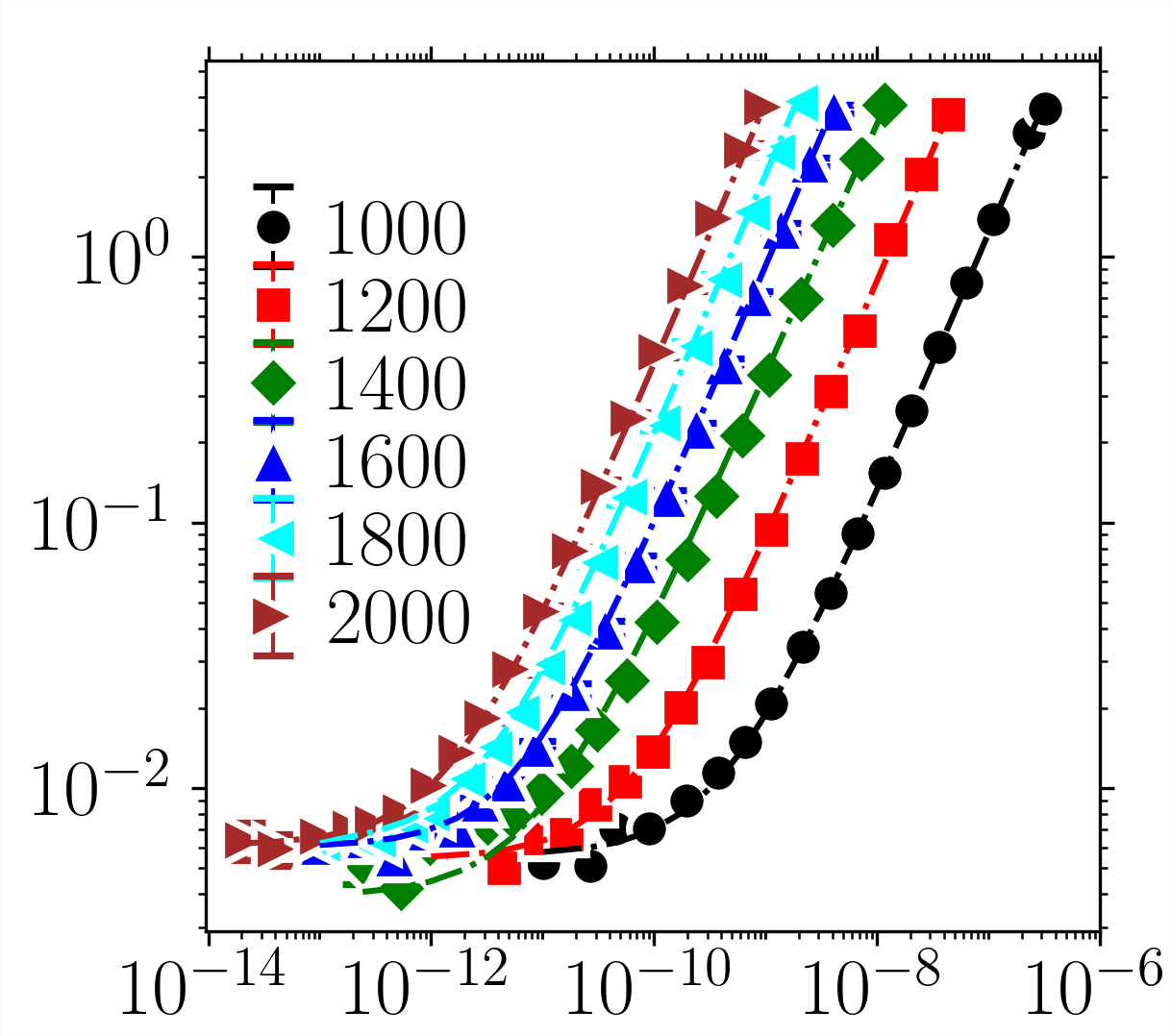}
        \LabelFig{16}{87}{$a)$ \scriptsize Ni}
        \Labelxy{50}{-6}{0}{$t~\text{(s)}$}
        \Labelxy{-7}{37}{90}{msd~(\r{A}$^2$)}
        \Labelxy{24}{75}{0}{\scriptsize $T(\text{K})$}
	\end{overpic}
    \begin{overpic}[width=0.49\columnwidth]{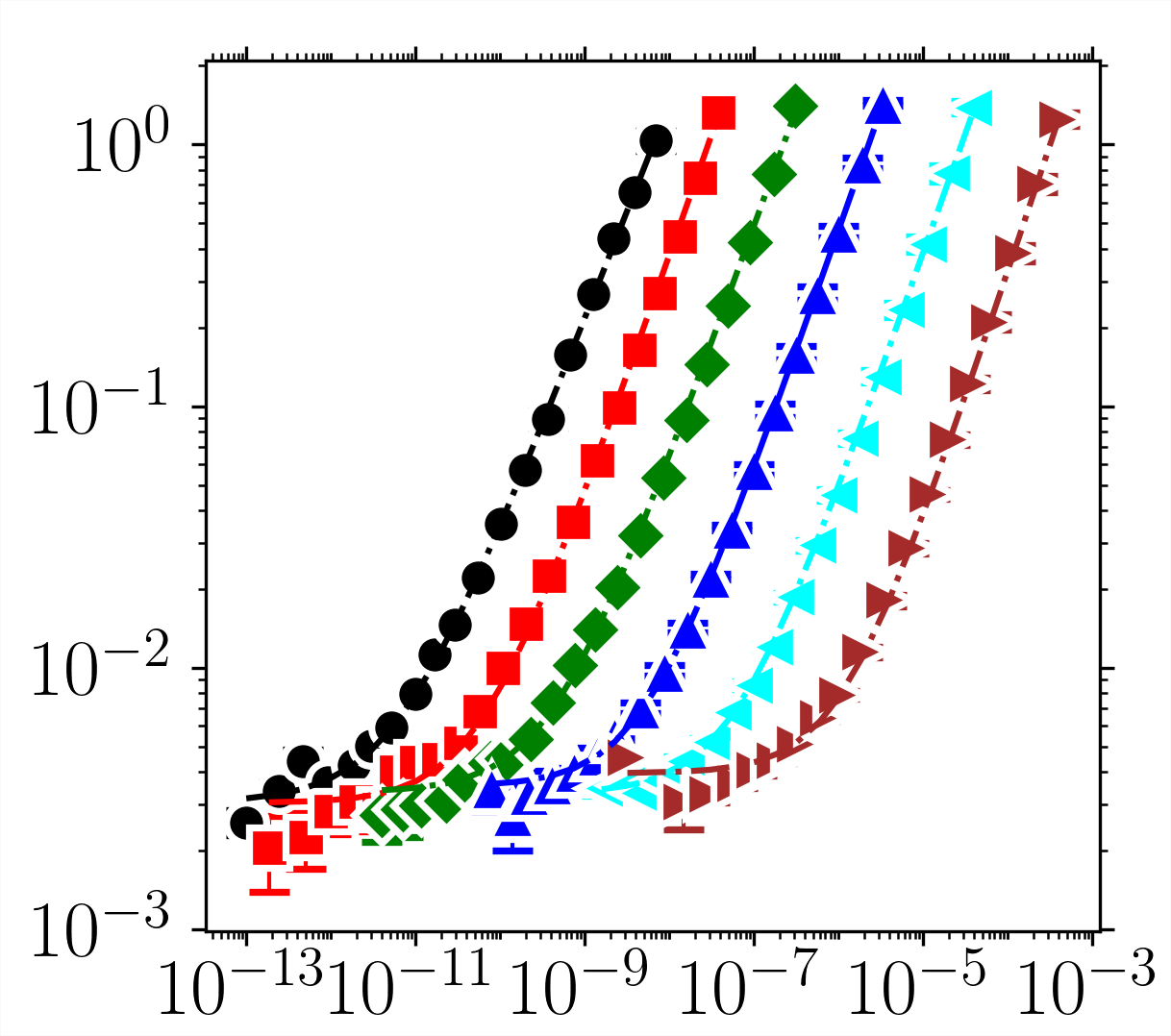}
        \LabelFig{16}{87}{$b)$  \scriptsize \glsix}
        \Labelxy{50}{-6}{0}{$t~\text{(s)}$}
        \Labelxy{-7}{37}{90}{msd~(\r{A}$^2$)}
	\end{overpic}
	\vspace{8pt}
 
    \begin{overpic}[width=0.49\columnwidth]{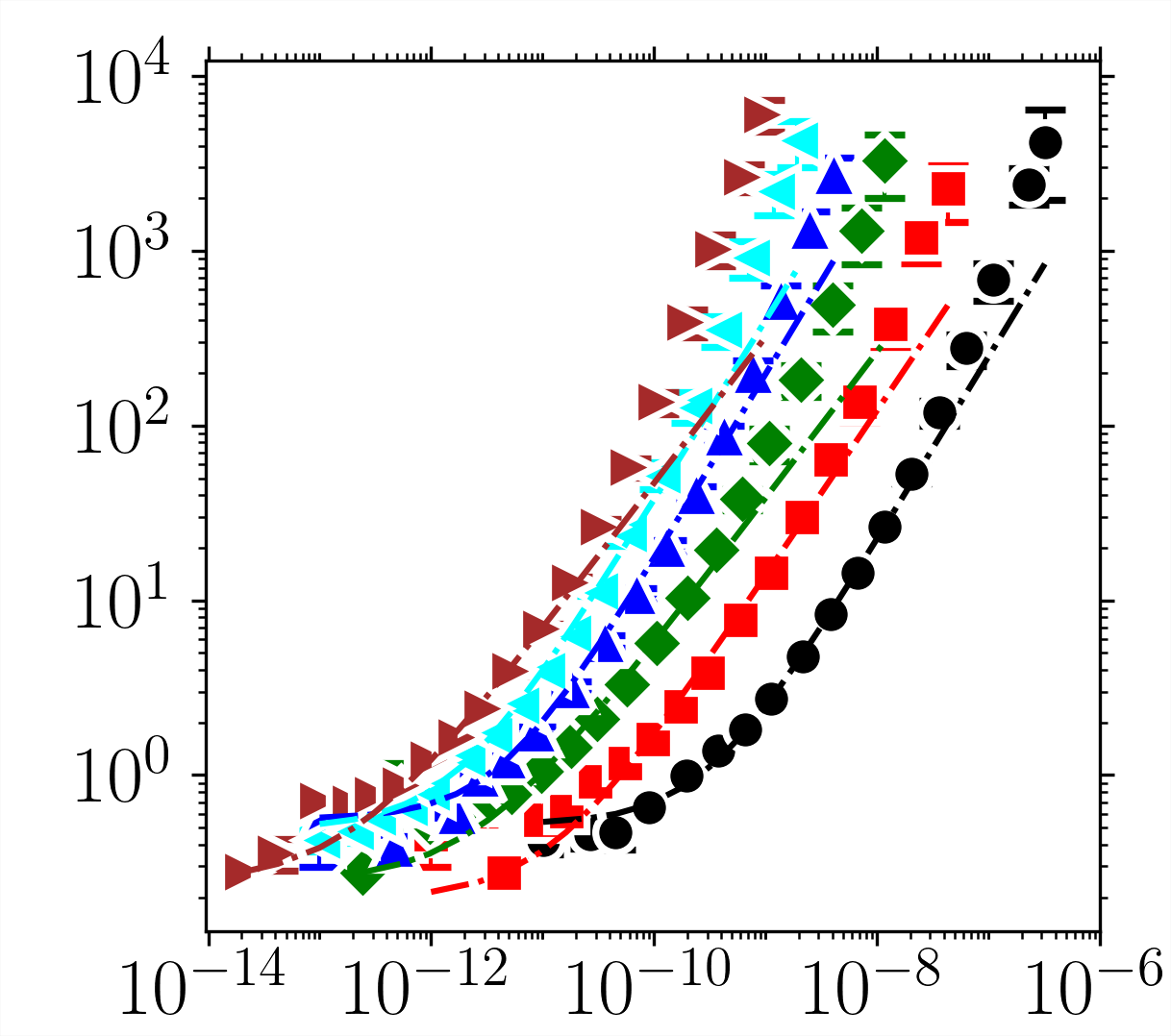}
        \LabelFig{16}{87}{$c)$}
        \Labelxy{50}{-6}{0}{$t~\text{(s)}$}
        \Labelxy{-7}{37}{90}{$\text{msd}_v$~(\r{A}$^2$)}
	\end{overpic}
    \begin{overpic}[width=0.49\columnwidth]{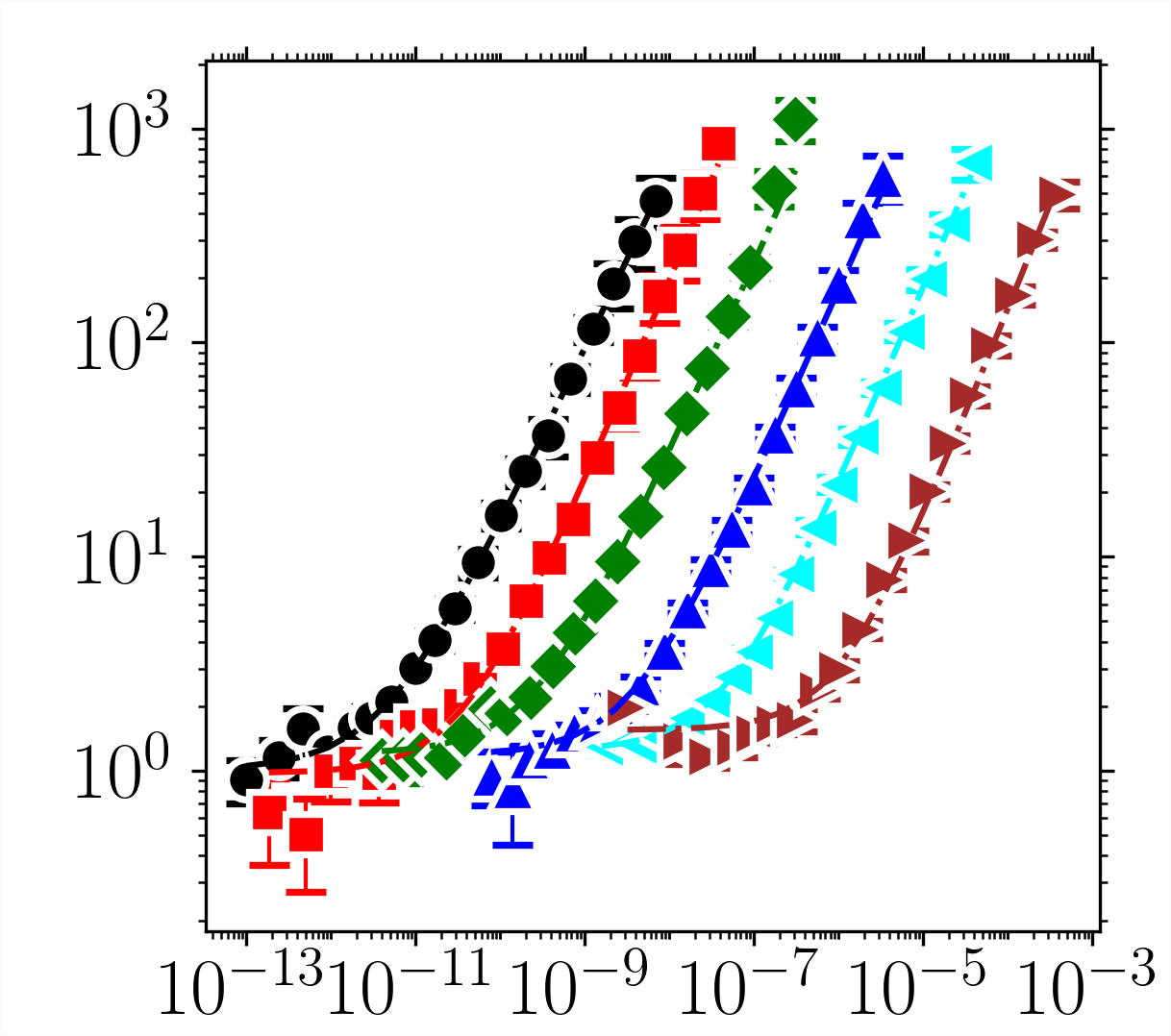}
        \LabelFig{16}{87}{$d)$}
        \Labelxy{50}{-6}{0}{$t~\text{(s)}$}
        \Labelxy{-7}{37}{90}{$\text{msd}_v$~(\r{A}$^2$)}
	\end{overpic}
	\vspace{4pt}
    \caption{Total mean-squared displacements (msd) as a function of time $t$ corresponding to \textbf{a}) pure Ni \textbf{b}) \glsix alloy at different temperatures. Panels \textbf{c}) and \textbf{d}) are the same as \textbf{a}) and \textbf{b}) but plot the vacancy mean-squared displacements $\text{msd}_v(t)$.
    Dashdotted curves indicate fitting curves $\text{msd}(t) \propto t^{2H}$ at $t\rightarrow \infty$.
    The curves in \textbf{b}) and \textbf{d}) are shifted horizontally for the better clarity. \note[KK]{data in c)?}
    }
    \label{fig:msdTemp}
\end{figure}

\noindent\emph{Methods---}
Model Ni and \glsix alloys were implemented as systems of $N= 1370$ atoms within cubic boxes with dimension $L= 26.0$~\r{A} in a three-dimensional ($d=3$) periodic setup.
The interatomic forces were derived from the embedded-atom method potential developed recently by Ma et al. \cite{Li2019}.
To test the robustness of our findings with respect to interatomic details, we also made use of the \emph{modified} embedded-atom (m-eam) framework proposed by Choi et al. \cite{choi2018understanding} which was successfully applied in the context of Cantor alloys.
The \glseven model alloys consist of order $N= 13,500$ atoms within periodic cubes of size $L= 54.0$~\r{A}.
Defect-free crystalline structures were prepared by performing energy minimization in LAMMPS \cite{plimpton1995fast} and were further relaxed upon the insertion of a point defect (single vacancy in this case).
We also checked that prepared multi-component alloys closely resemble random solid solutions with no or negligible chemical ordering effects. 

To probe the vacancy dynamics under thermal effects, we make use of the $k$-ART software \cite{el2008kinetic} based on the initial structures prepared at zero temperature.
We opted not to feed thermalized samples at finite temperatures to $k$-ART in order to magnify the effects of lattice distortions~\cite{el2008kinetic,kart}.
To investigate vacancy-driven diffusivity in metals, we perform between $10^2-10^3$ Monte Carlo steps within the temperature range $1000-2000$ K.
Figure~\ref{fig:randomWalk} shows defect trajectories in \glsix as well as typical atomic rearrangements next to a vacant site at $T=1400$ K. 
The vacancy dynamics and associated hopping closely resembles a random walk in three dimensions with temporally-uncorrelated increments in space.\\

\begin{figure}[t]
    \centering
    \begin{overpic}[width=0.49\columnwidth]{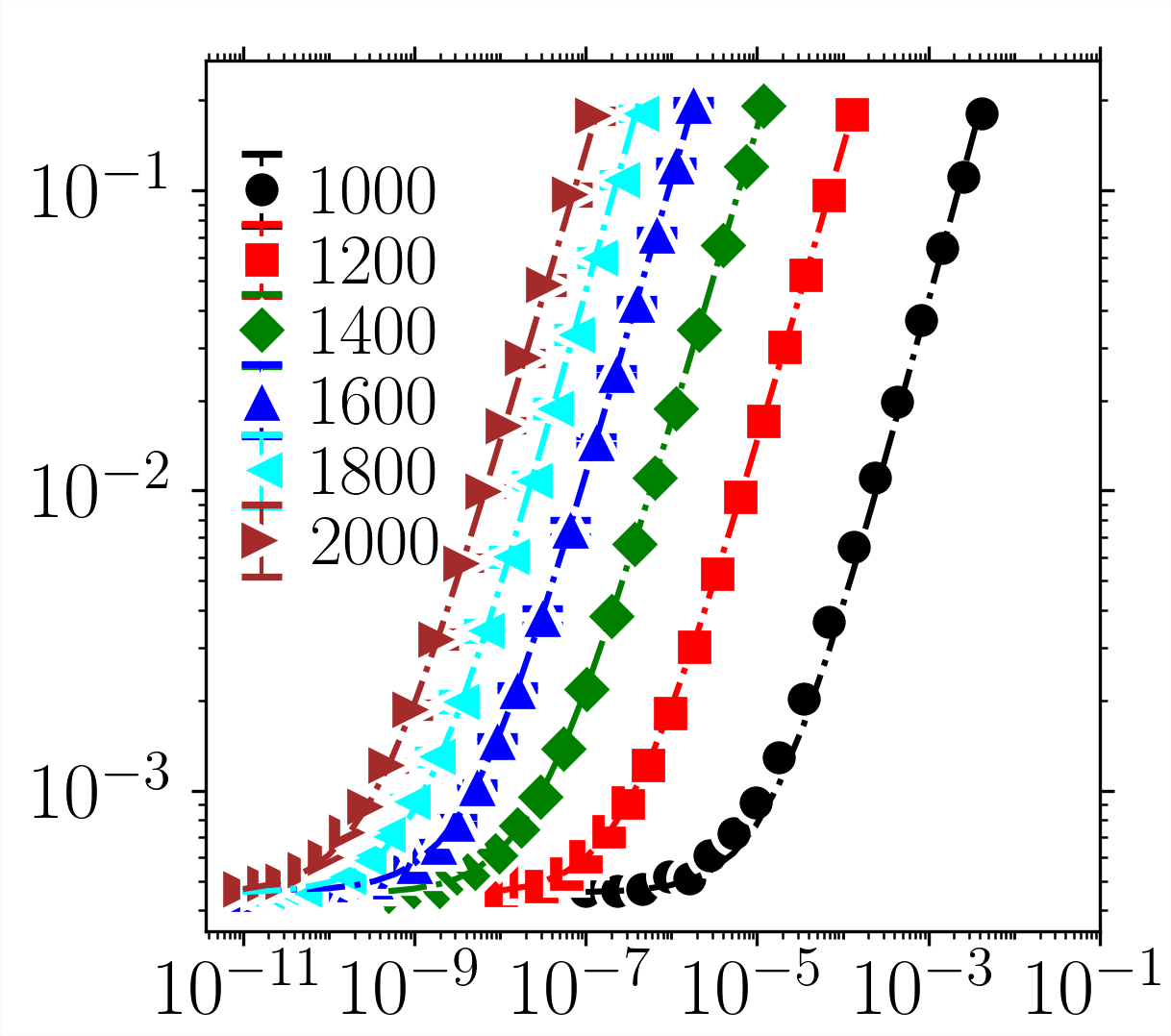}
        \LabelFig{16}{87}{$a)$ \scriptsize Ni}
        \Labelxy{50}{-6}{0}{$t~\text{(s)}$}
        \Labelxy{-7}{37}{90}{msd~(\r{A}$^2$)}
        \Labelxy{24}{77}{0}{\scriptsize $T(\text{K})$}
	\end{overpic}
    \begin{overpic}[width=0.49\columnwidth]{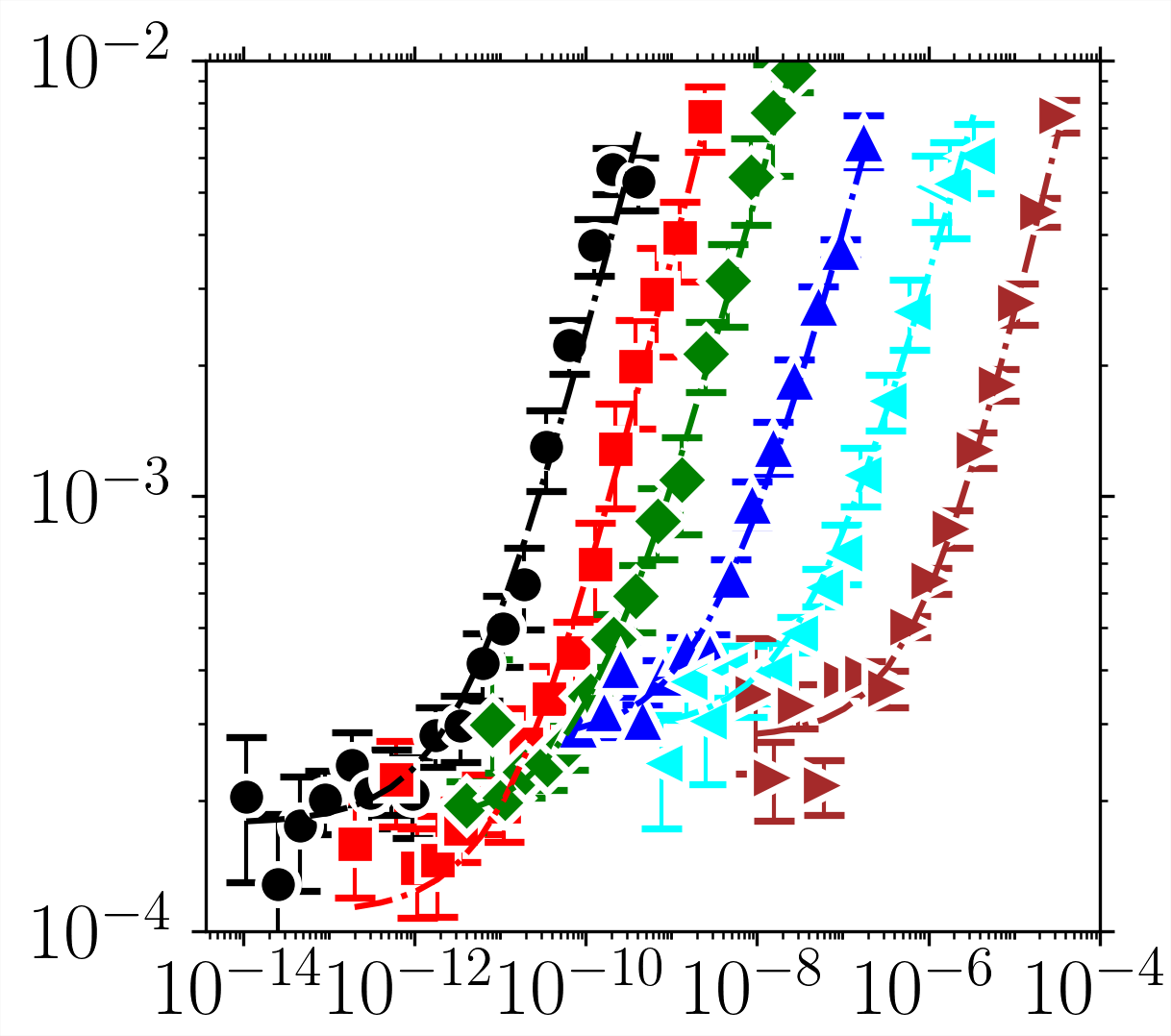}
        \LabelFig{16}{87}{$b)$  \scriptsize \glseven}
        \Labelxy{50}{-6}{0}{$t~\text{(s)}$}
        \Labelxy{-7}{37}{90}{msd~(\r{A}$^2$)}
	\end{overpic}
	\vspace{4pt}
 
    \begin{overpic}[width=0.49\columnwidth]{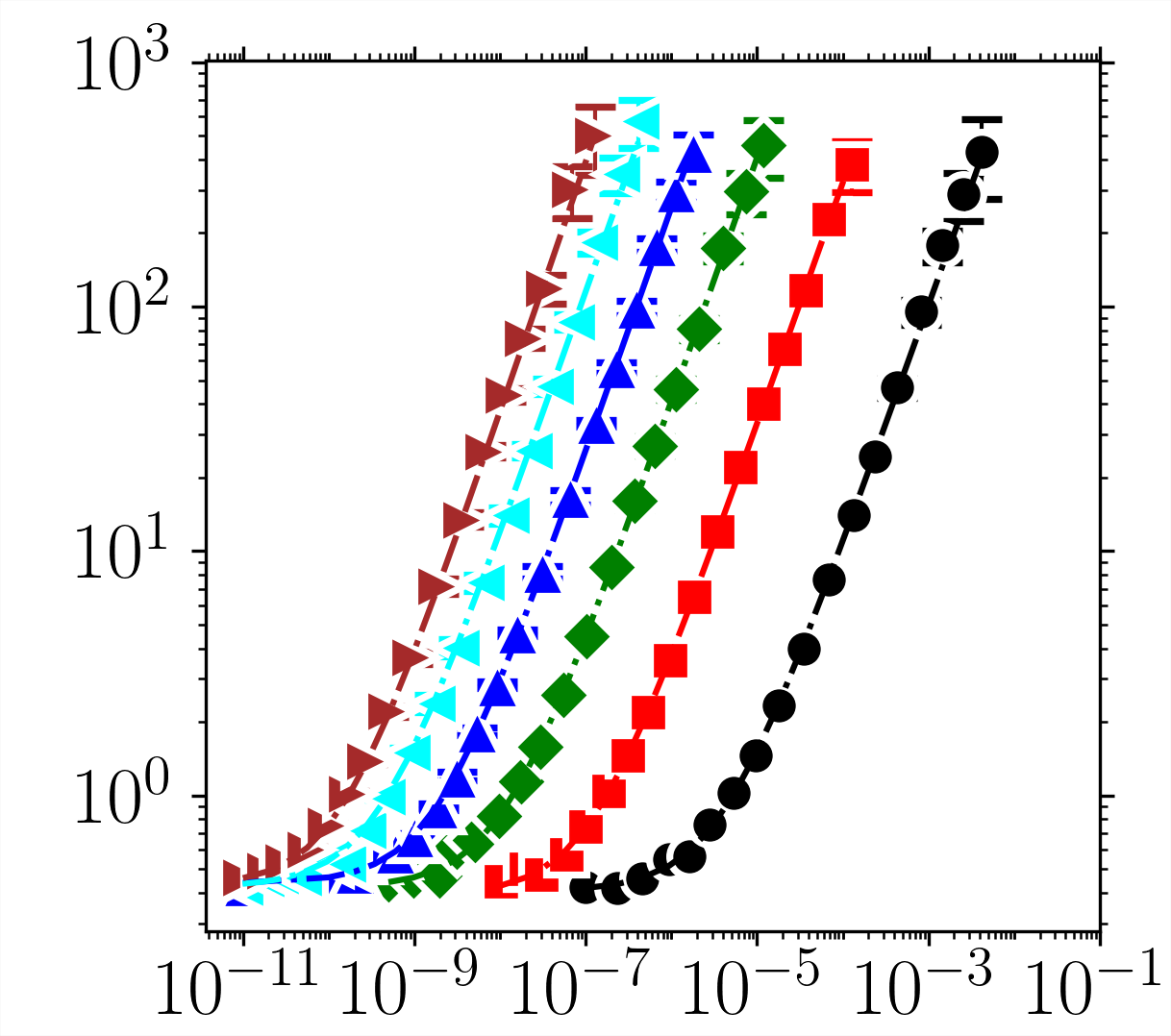}
        \LabelFig{16}{87}{$c)$}
        \Labelxy{50}{-6}{0}{$t~\text{(s)}$}
        \Labelxy{-7}{37}{90}{$\text{msd}_v$~(\r{A}$^2$)}
	\end{overpic}
    \begin{overpic}[width=0.49\columnwidth]{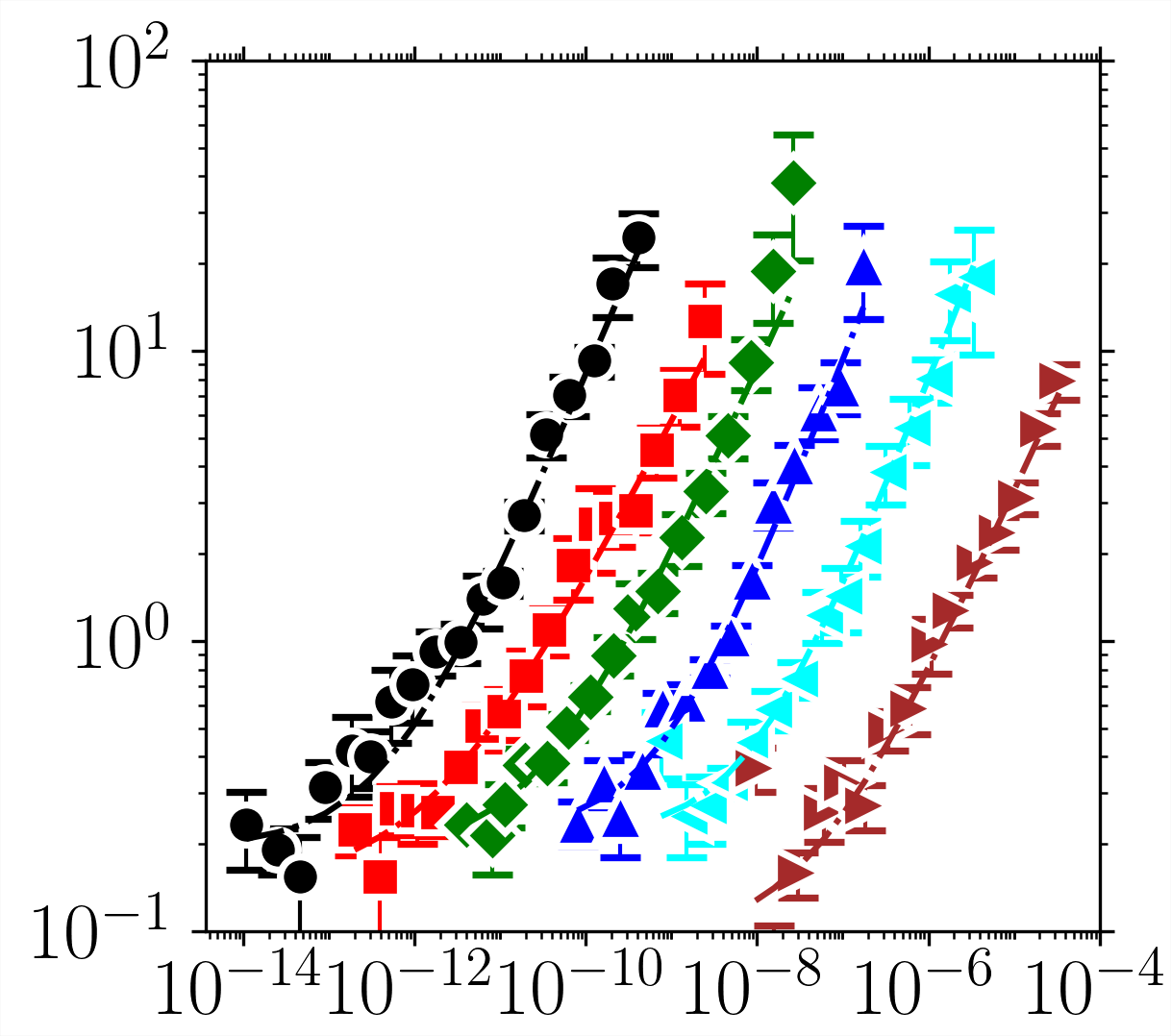}
        \LabelFig{16}{87}{$d)$}
        \Labelxy{50}{-6}{0}{$t~\text{(s)}$}
        \Labelxy{-7}{37}{90}{$\text{msd}_v$~(\r{A}$^2$)}
	\end{overpic}
	\vspace{8pt}

    \caption{Total mean-squared displacements msd(\r{A}$^2$) as a function of time $t\text{(s)}$ corresponding to model \textbf{a}) Ni \textbf{b}) \glseven alloy using the m-eam potential function. Panels \textbf{c}) and \textbf{d}) are the same as \textbf{a}) and \textbf{b}) but plot the vacancy mean-squared displacements $\text{msd}_v(t)$.
    Dashdotted curves indicate fitting curves $\text{msd}(t) \propto t^{2H}$ at $t\rightarrow \infty$.
    The curves in \textbf{b}) and \textbf{d}) are shifted horizontally for the better clarity.}
    \label{fig:msdTempSMcantor}
\end{figure}



\noindent\emph{Anomalous Diffusion---}
We calculate the mean-squared displacement of atoms including all effects due to atoms crossing periodic boundaries.
The displacement vector $u_{i\alpha}(t^\prime,t)=\vec{r}_{i}(t^\prime+t)-\vec{r}_{i}(t^\prime)$ is defined per atom $i=1...N$ given a reference time $t^\prime$ and over duration $t$.
Here $\vec{r}_{i}(t)$ denotes the position of atom $i$ at time $t$. 
Squared displacements are summed and averaged over atoms $i$ and different reference times $t^\prime$ to obtain the displacement variance as a function of duration $t$, \ie $\text{msd}(t)=\langle~\vec{u}_{i}(t^\prime,t)~.~\vec{u}_{i}(t^\prime,t)~\rangle_{i,t^\prime}$. 
The msd associated with the single vacancy $\text{msd}_v(t)$ is defined in a similar manner but including motion of a subset of atoms in the nearest neighborhood of the vacant site (see Fig.~\ref{fig:randomWalk}).
To improve collected statistics in the random solid solution alloy, we consider ensembles of eight different realizations associated with each temperature of interest.
The temporal evolution of $\text{msd}(t)$ and $\text{msd}_v(t)$ corresponding to pure Ni and \glsix alloy are shown in Fig.~\ref{fig:msdTemp}(a-d) at various temperatures.
Both sets of curves mark the cross-over from an initial plateau regime, as a signature of solid-like behavior, at short time-scales to a diffusive regime due to relaxations at later times.
Similar trends can be also seen in Fig.~\ref{fig:msdTempSMcantor}(a-d) corresponding to pure Ni and \glseven based on the the m-eam potential function.

\begin{figure}[b]
    \centering
    \begin{overpic}[width=0.49\columnwidth]{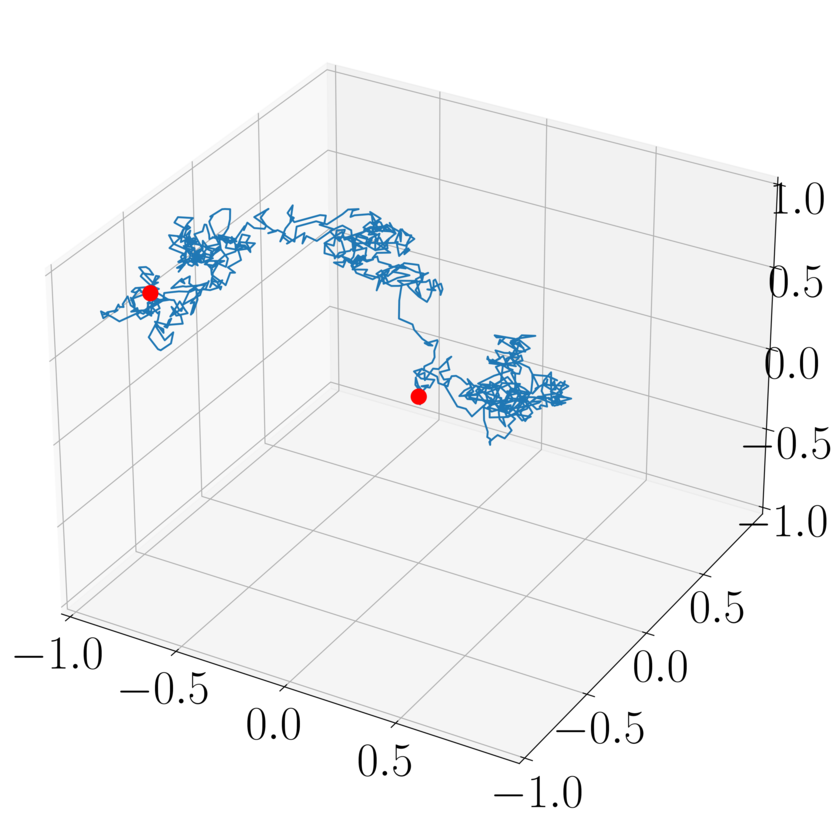}
        \LabelFig{16}{87}{$a)~H=0.5$}
        \Labelxy{80}{10}{45}{$x$~\scriptsize(\r{A})}
        \Labelxy{26}{4}{-18}{$y$~\scriptsize(\r{A})}
        \Labelxy{-2}{40}{90}{$z$~\scriptsize(\r{A})}
	\end{overpic}
    \begin{overpic}[width=0.49\columnwidth]{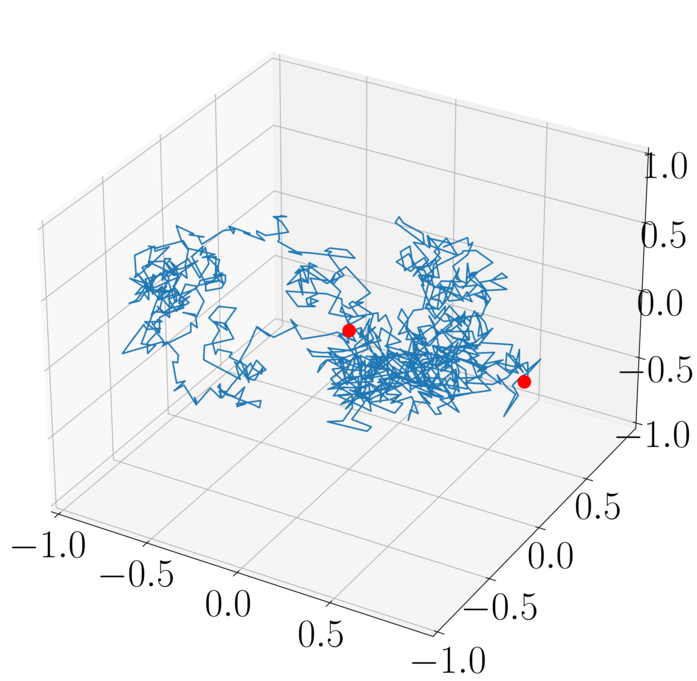}
        \LabelFig{16}{87}{$b)~H=0.4$}
        \Labelxy{80}{10}{45}{$x$~\scriptsize(\r{A})}
        \Labelxy{26}{4}{-18}{$y$~\scriptsize(\r{A})}
        \Labelxy{-2}{40}{90}{$z$~\scriptsize(\r{A})}
	\end{overpic}
        \vspace{11pt}

    \begin{overpic}[width=0.49\columnwidth]{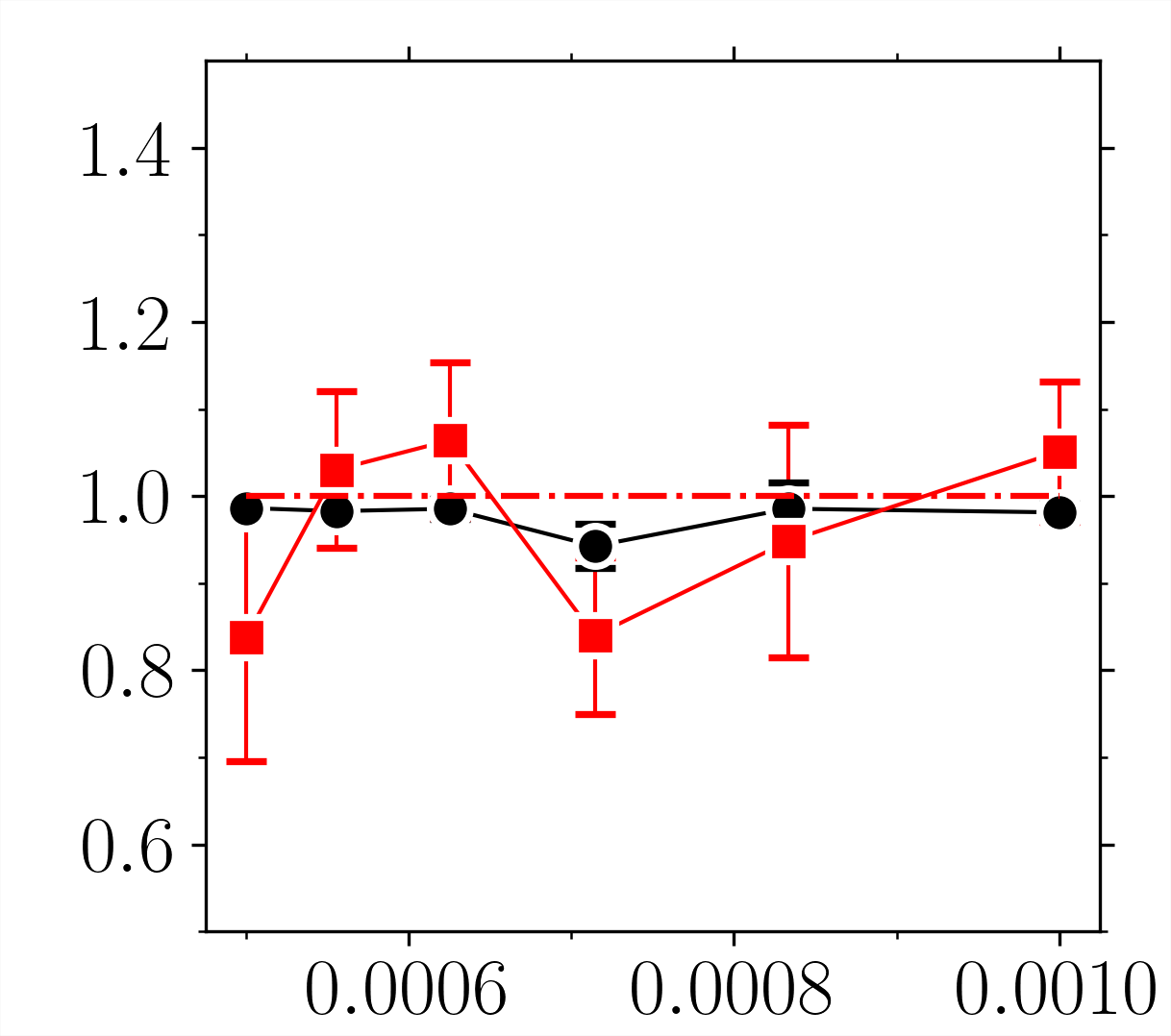}
        %
        \LabelFig{16}{87}{$c)$ Ni}
        \Labelxy{50}{-6}{0}{$1/T~(\text{K}^{-1})$}
        \Labelxy{-3}{42}{90}{$2H$}
	\end{overpic}
    \begin{overpic}[width=0.49\columnwidth]{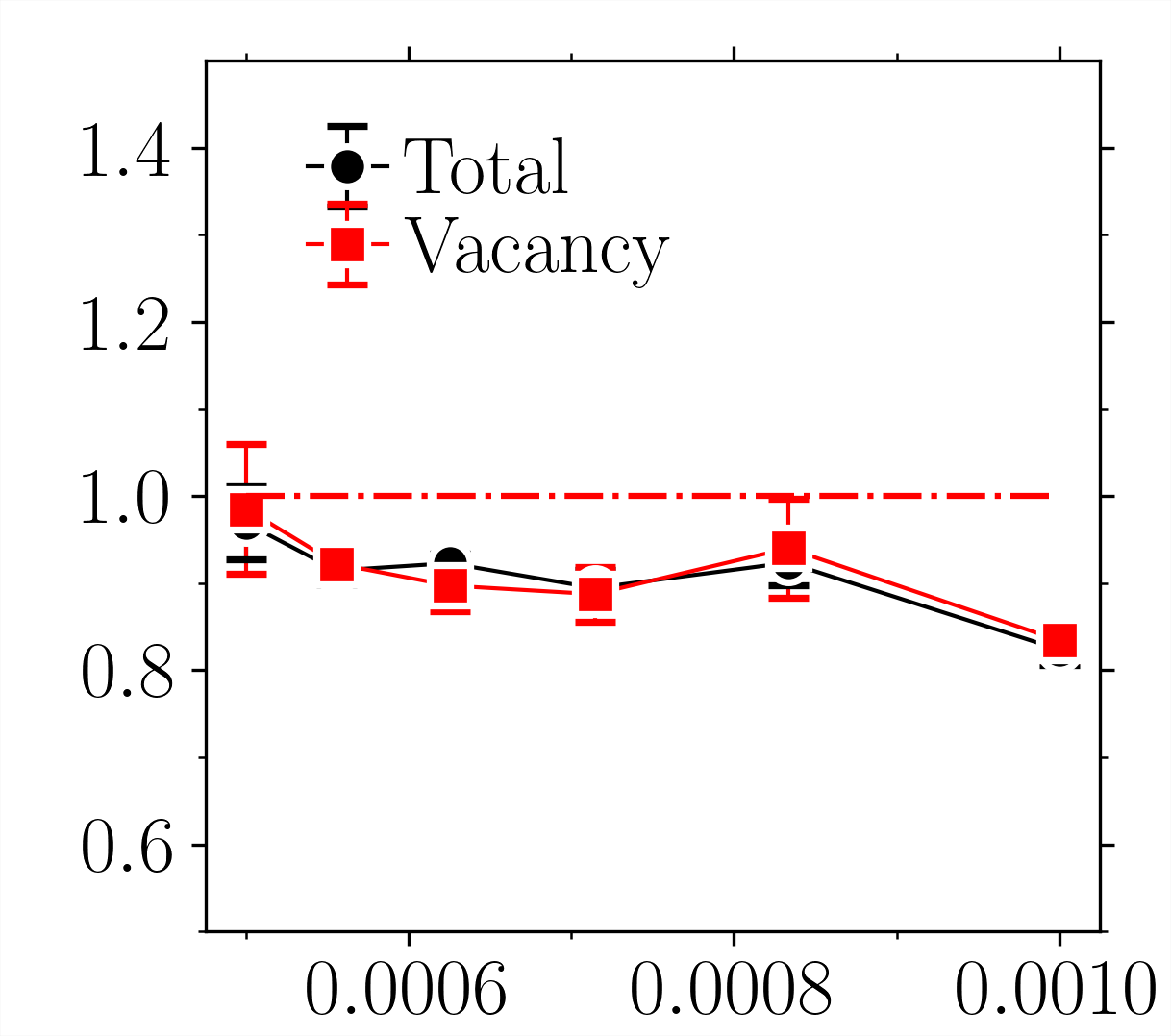}
        %
        \LabelFig{16}{87}{$d)$ \glsix}
        \Labelxy{50}{-6}{0}{$1/T~(\text{K}^{-1})$}
        \Labelxy{-3}{42}{90}{$2H$}
	\end{overpic}
    \vspace{5pt}

     \begin{overpic}[width=0.49\columnwidth]{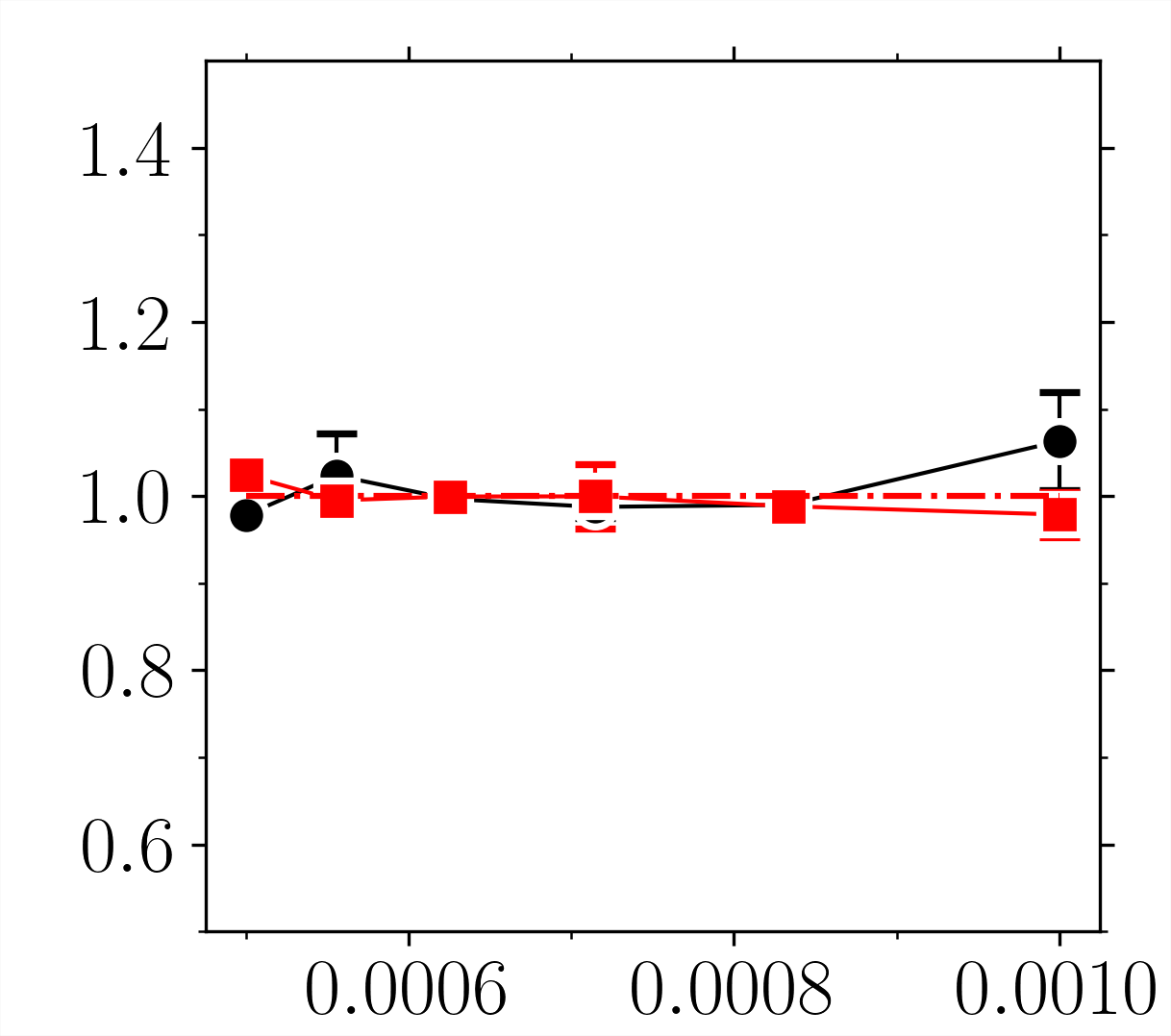}
        \LabelFig{16}{87}{$e)$~Ni}
        \Labelxy{50}{-6}{0}{$1/T~(\text{K}^{-1})$}
        \Labelxy{-7}{42}{90}{$2H$}
	\end{overpic}
    \begin{overpic}[width=0.49\columnwidth]{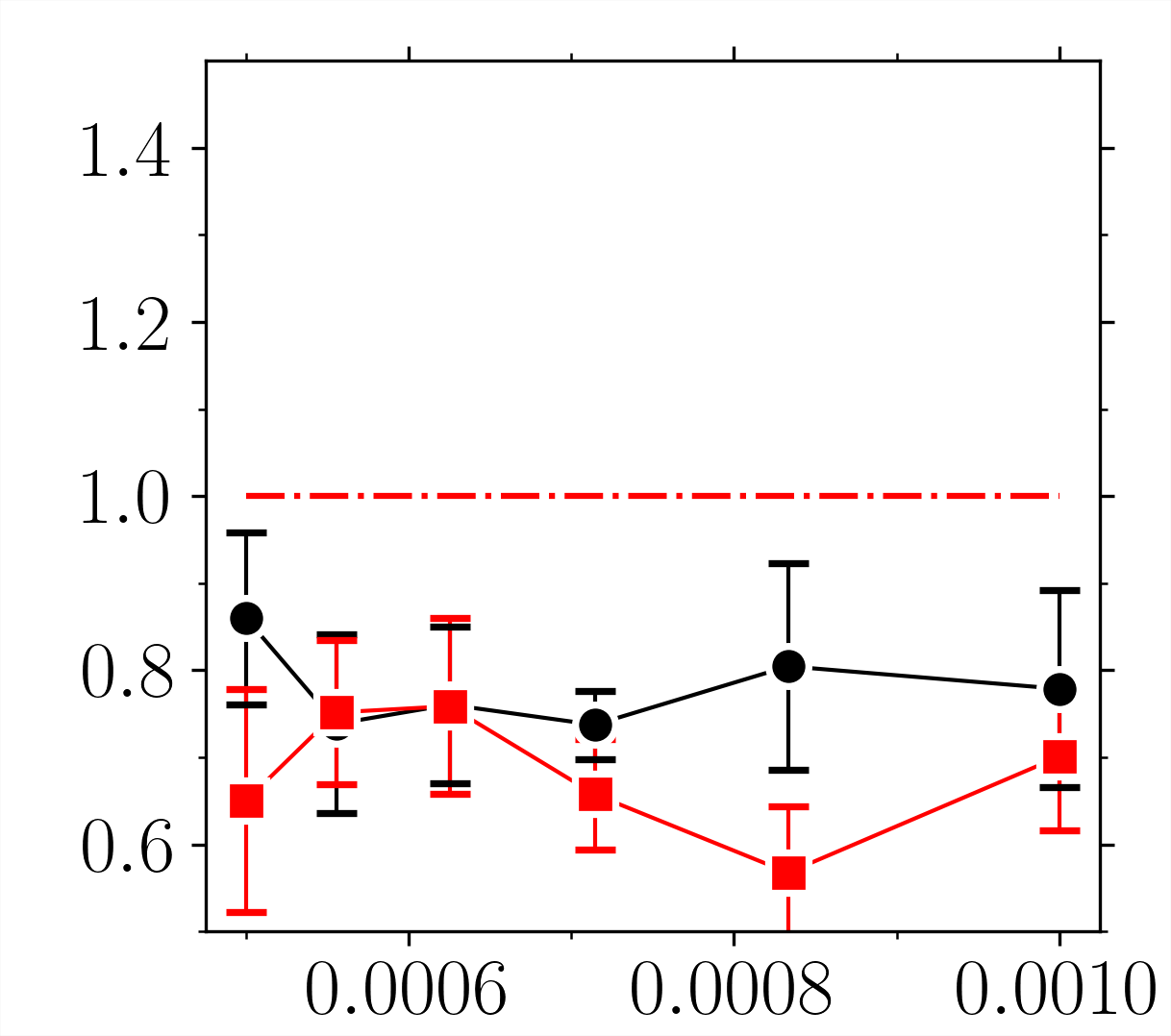}
        \LabelFig{16}{87}{$f)$~\glseven}
        \Labelxy{50}{-6}{0}{$1/T~(\text{K}^{-1})$}
        \Labelxy{-7}{42}{90}{$2H$}
	\end{overpic}
        \caption{Model fractional Brownian walks associated with Hurst exponents \textbf{a}) $H=0.5$ \textbf{b}) $H=0.4$, each showing $1,000$ discrete increments with zero mean and standard deviation \textbf{a}) $0.03$ and \textbf{b}) $0.06$ \r{A}. Estimated Hurst exponents $H$ based on the msd data presented in Fig.~\ref{fig:msdTemp} and \ref{fig:msdTempSMcantor} correspond to \textbf{c}) pure Ni \textbf{d}) \glsix alloy \textbf{e}) m-eam based Ni \textbf{f}) m-eam based \glseven alloy. The dashdotted lines indicate $H=1/2$. The (red) markers indicate the start points at $(0,0,0)$ and the end points.}
    \label{fig:hurst}
\end{figure}

To describe the observed cross-over, we fit a nonlinear model, $\text{msd}(t)=\langle u^2 \rangle +Kt^{2H}$, to the msd data by the least-squares regression, giving the results shown as dashdotted curves in Fig.~\ref{fig:msdTemp}(a-d).
Here $H$ is the Hurst exponent \cite{kantz2004nonlinear} and $K$ and $\langle u^2 \rangle$ are the fit parameters.
In the long-time limit $t\rightarrow\infty$, the variance scales like a power-law with the time lag as $\text{msd}(t)\propto t^{2H}$.
In Fig.~\ref{fig:hurst}(c), we recover the square-root dependence of displacements with time, \ie $H=1/2$, over the range of studied temperatures for pure Ni obeying a standard diffusion process. 
In this case, the slope of the msd curves versus time--- that is, $K$--- is equivalent to the diffusion coefficient.
Interestingly, the Hurst exponents fitted to the \glsix data in Fig.~\ref{fig:hurst}(d) suggest a subdiffusive behavior with $0< H < 1/2$ \cite{metzler2000random} showing an overall growth toward $H=1/2$ with increasing temperature $T$.
The emerging subdiffusion seems to be also relevant in \glseven alloys as in Fig.~\ref{fig:hurst}(f), as opposed to pure Ni in Fig.~\ref{fig:hurst}(e), but our data indicate almost no (meaningful) temperature-dependence associated with exponents $H$.

We may now proceed with the hypothesis of anti-correlations between successive increments and thus fractional Brownian motion as a potential source of subdiffusion \cite{mandelbrot1968fractional}.
This is shown by a generated Brownian path in Fig.~\ref{fig:hurst}(b) corresponding to $H<1/2$ where the random walker tends to be self-trapped within certain cages at short/intermediate timescales.
One can also show that a standard Brownian motion (with $H=1/2$) in Fig.~\ref{fig:hurst}(a) may not possess this caging property.
The characteristic scale associated with such traps may be inferred from the initial plateau regions within the $\text{msd}_v(t)$ fits in Fig.~\ref{fig:msdTemp}(d) with $\langle u^2 \rangle^{1/2}=1.1-1.5$ \r{A}.
This length should correspond to almost half the mean vacancy hopping distance (e.g. mean nearest-neighbor distance between atoms) as sketched in Fig.~\ref{fig:randomWalk}.
The time lag marking the cross-over from the plateau regime to anomalous diffusion should also set a (temperature-dependent) caging timescale. 

If we are to assume uncorrelated increments within a Markov process, another plausible explanation for suppressed diffusion is the presence of long rest times (with independent but bounded increments) which could be understood in the framework of continuous-time random walks \cite{klages2008anomalous}.
In what follows, we describe the observed subdiffusive trends based on underlying waiting time distributions.\\

\begin{figure}[t] 
    \centering
    \begin{overpic}[width=0.49\columnwidth]{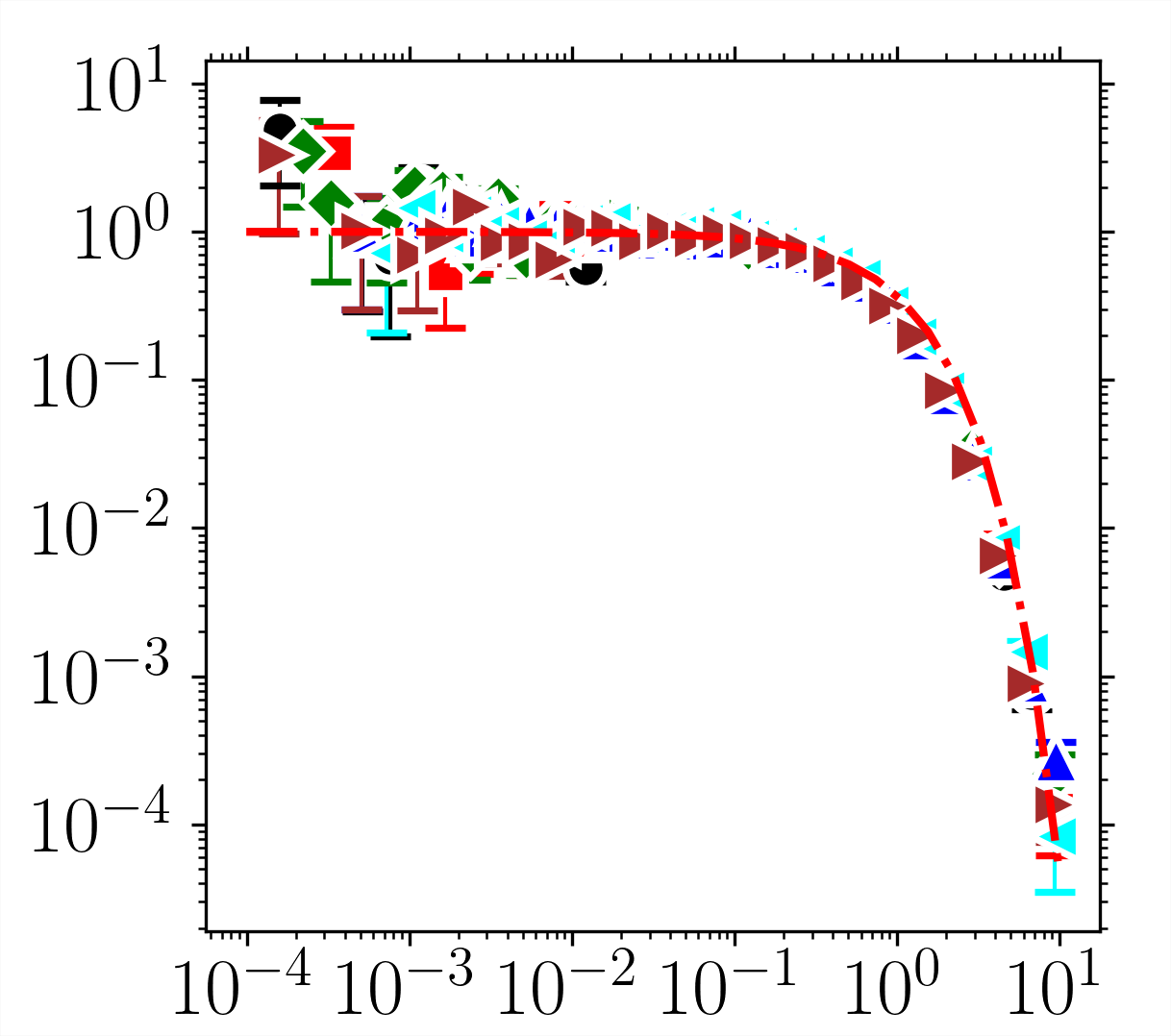}
        \put(19,10){\includegraphics[width=0.1\textwidth]{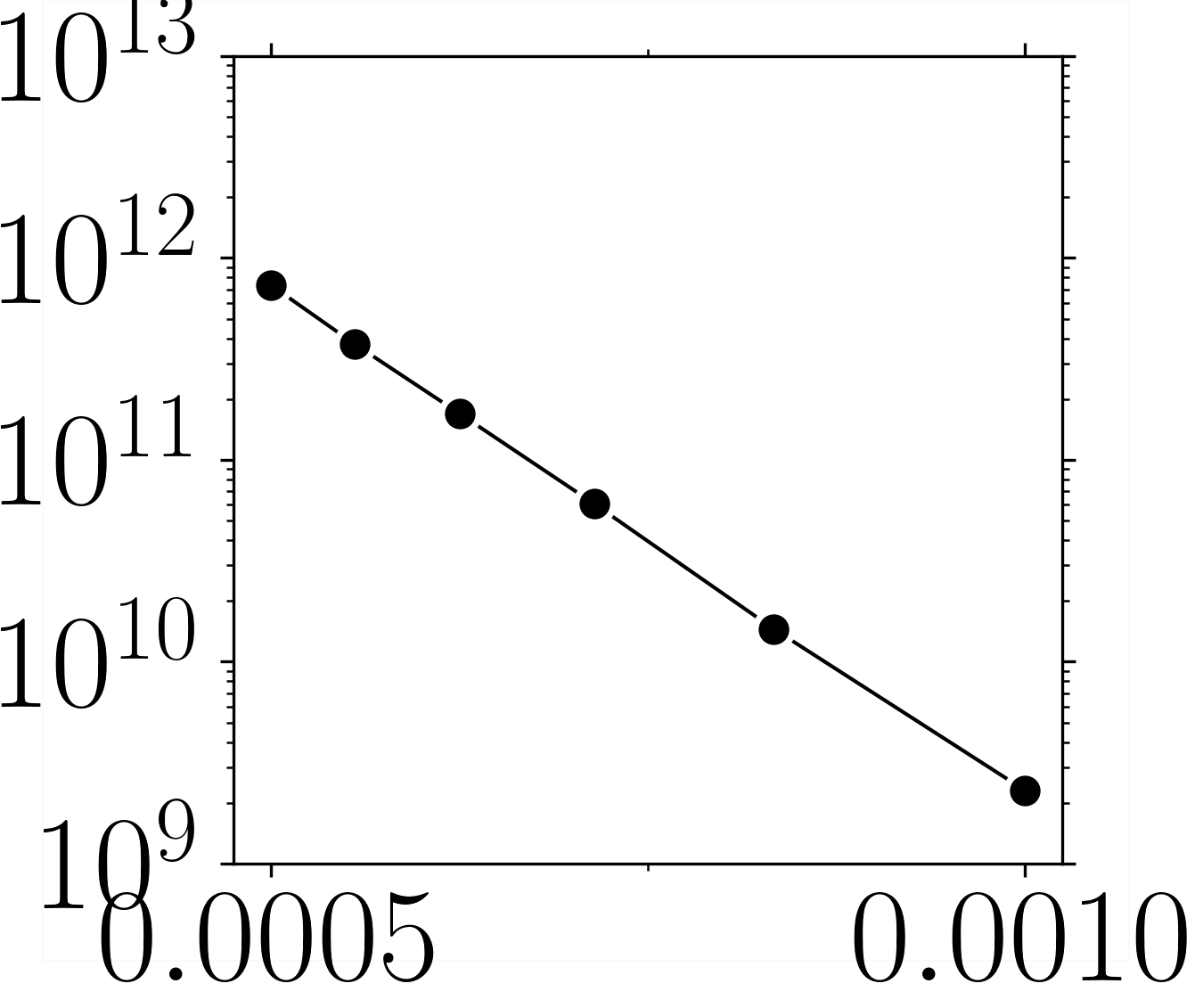}}
        \LabelFig{16}{87}{$a)$ \scriptsize Ni}
        \Labelxy{50}{-6}{0}{$\lambda t_w$}
        \Labelxy{-7}{37}{90}{$\lambda^{-1}p(t_w)$}
        \Labelxy{34}{17}{0}{\tiny $1/T$}
        \Labelxy{59}{22}{90}{\tiny $\lambda~(\text{s}^{-1})$}
	\end{overpic}
    \begin{overpic}[width=0.49\columnwidth]{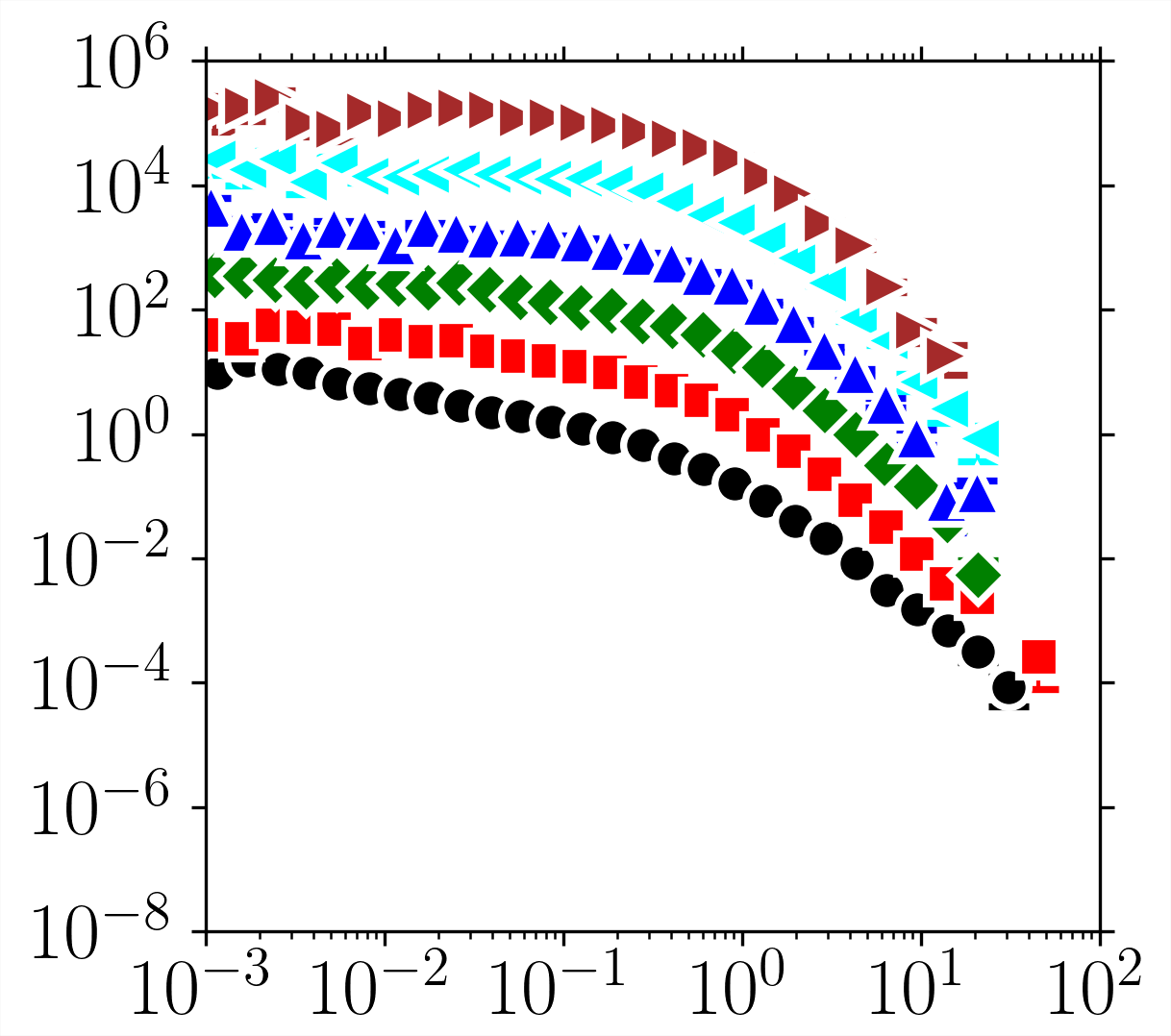}
        \put(19,10){\includegraphics[width=0.1\textwidth]{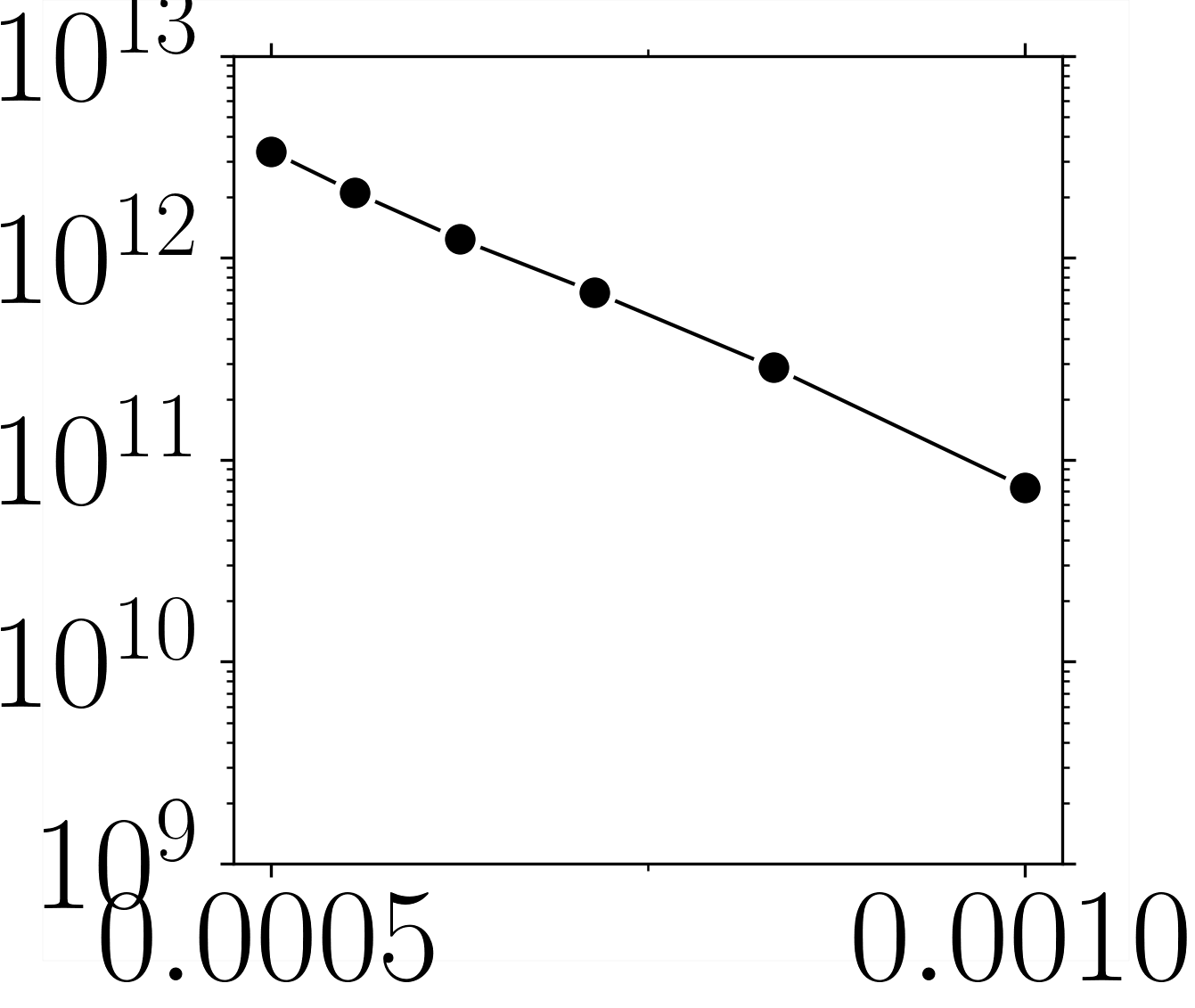}}
        \LabelFig{16}{87}{$b)$  \scriptsize \glsix}
        \Labelxy{50}{-6}{0}{$\lambda t_w$}
        \Labelxy{-7}{37}{90}{$p(t_w)$}
        \Labelxy{34}{17}{0}{\tiny $1/T$}
        \Labelxy{59}{21}{90}{\tiny $\lambda~(\text{s}^{-1})$}
	\end{overpic}
	\vspace{8pt}

    \begin{overpic}[width=0.49\columnwidth]{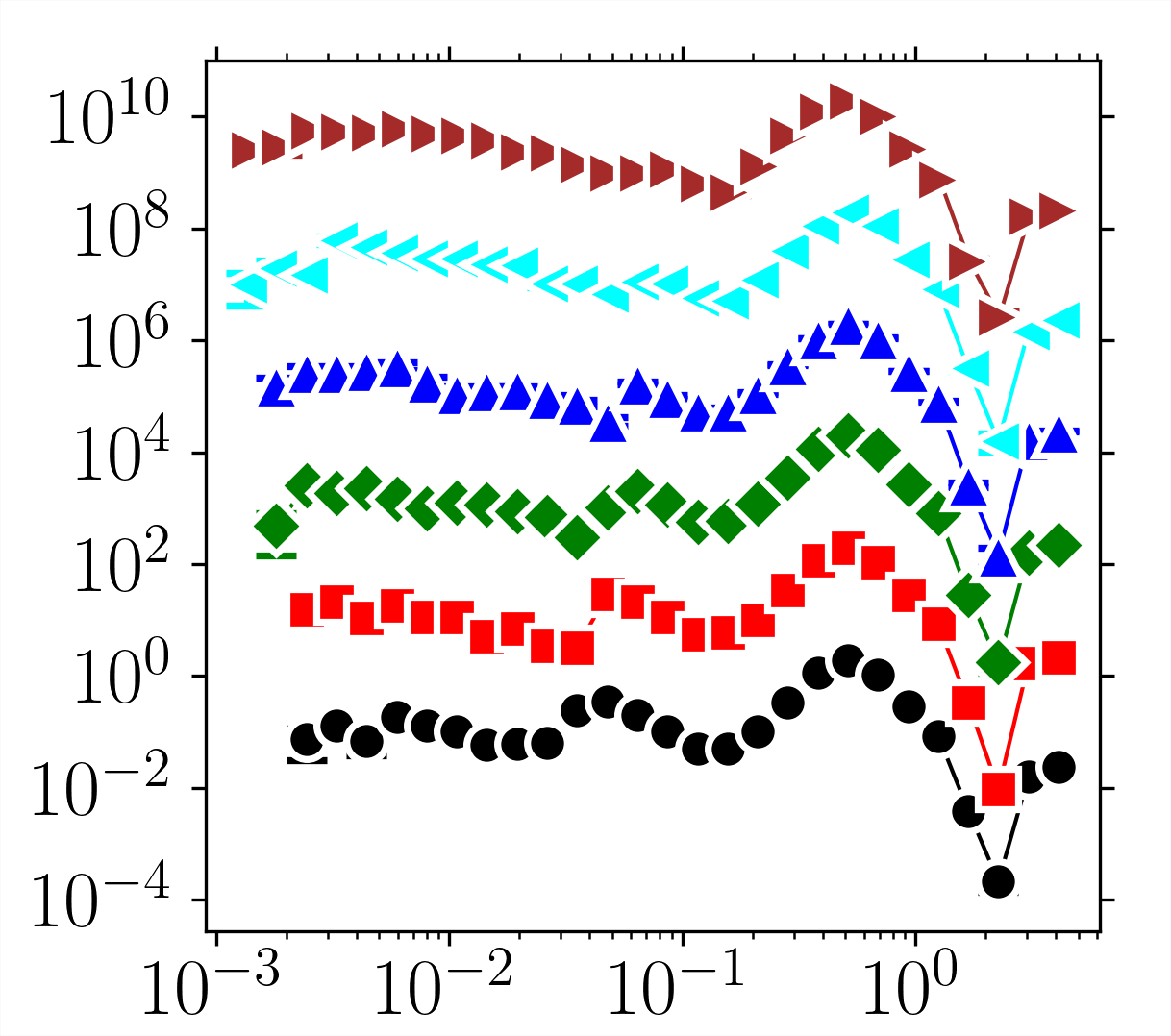}
        %
        %
        \LabelFig{16}{87}{$c)$}
        \Labelxy{50}{-6}{0}{$\Delta E$~(eV)}
        \Labelxy{-6}{46}{90}{$p(\Delta E)$}
    \end{overpic}
    \begin{overpic}[width=0.49\columnwidth]{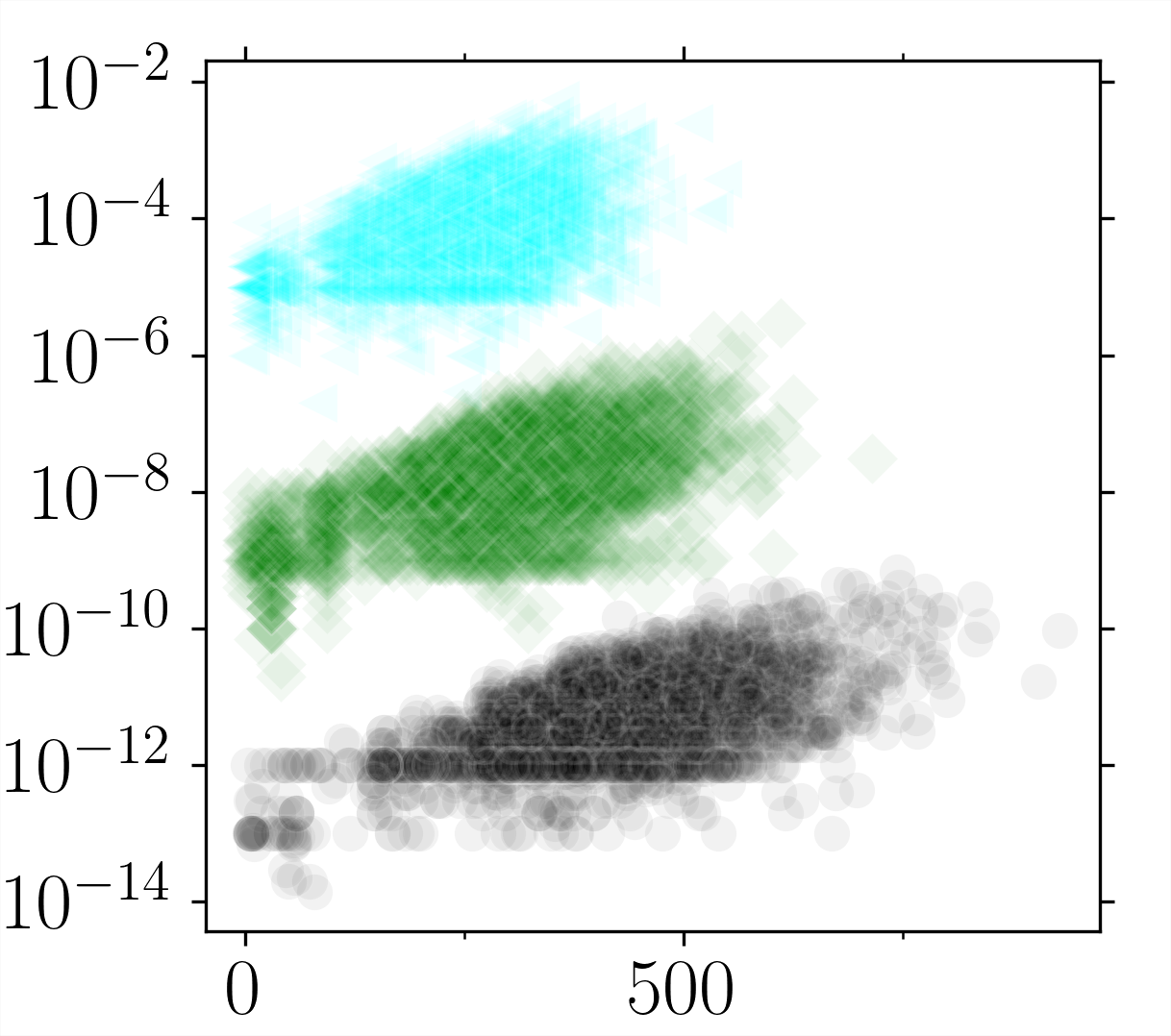}
        \LabelFig{16}{87}{$d)$}
        \Labelxy{50}{-6}{0}{$\Delta E/k_BT$}
        \Labelxy{-6}{46}{90}{$t_w$~(s)}
    \end{overpic}
    \caption{Atom-scale statistics at various temperatures $T=1000$ K (\protect\circTxtFill{0}{0}{black}), $1200$ K (\protect\legSqTxt{0}{0}{red}), $1400$ K (\protect\legDiamondTxt{0}{0}{darkspringgreen}), $1600$ K (\protect\legTriangleTxt{0}{0}{blue}), $1800$ K (\protect\leftTriang{0}{0}{cyan}), $2000$ K (\protect\rightTriang{0}{0}{brown}). Waiting time distributions $\lambda^{-1}p(t_w)$ scaled by mean activity rate $\lambda(\text{s}^{-1})$ corresponding to \textbf{a}) pure Ni \textbf{b}) \glsix alloy. Statistics of energy barriers corresponding to \glsix \textbf{c}) statistical distributions $p(\Delta E)$  \textbf{d}) scatter plot of waiting times $t_w$(s) and scaled energy barriers $\Delta E/k_BT$ at $T=1000$, $1400$, and $1800$ K.
    The dashdotted (red) curve denotes $\text{exp}(-\lambda t_w)$.
    The insets in \textbf{a}) and \textbf{b}) plot activation rate $\lambda$ against inverse temperature $1/T$.
    The curves in \textbf{b}) are shifted vertically for the sake of clarity. The data points in \textbf{d}) are shifted vertically for a better clarity.
    }
    \label{fig:waitTimes}
\end{figure}

\noindent\emph{Waiting Times and Energy Barriers---}
For a (homogeneous) Poisson process with independent events of constant rate $\lambda$, the wait time statistics should obey an exponential distribution $p(t_w)=\lambda~\text{exp}(-\lambda t_w)$.
The relevance of such dynamics for pure Ni is demonstrated in Fig.~\ref{fig:waitTimes}(a) where the rescaled distributions $\lambda^{-1} p(t_w)$ are plotted against scaled waiting times $\lambda t_w$.

We observe significant deviations from the hypothesis of a Poisson process, particularly at low $T$, as demonstrated in Fig.~\ref{fig:waitTimes}(b) corresponding to the \glsix alloy.
At $T=1000$ K, the rescaled distribution in the main plot is characterized by a fairly shallow power-law crossing over to a steeper decay that extends for almost two decades in $t_w$.
As $T$ is increased toward $2000$ K, we see a gradual transition to exponential-like decays which is consistent with our msd data indicating a subdiffusive-to-diffusive cross-over at elevated temperatures (\cf Fig.~\ref{fig:hurst}(d)). 
At $T=1000$ K, we find $P(t_w)\propto t_w^{-(1+\alpha)}$ with $\alpha = 1.0$ asymptotically for $\lambda t_w>1$ indicating long rest periods.
The waiting time distributions associated with the Cantor alloy indicate nearly the same scaling properties (data not shown).
We note that broad $t_w$ distributions with diverging mean times (\ie $0<\alpha<1$) but with a bounded jump length variance corresponds to a subdiffusive Brownian process \cite{metzler2000random}.
In that case, $\text{msd}(t) \propto t^\alpha$ and, therefore, $H=\alpha/2$ . 

The inset of Fig.~\ref{fig:waitTimes}(a) and (b) also show the relevance of Arrhenius-based activation with $\lambda \propto \text{exp}(-\Delta E_\text{eff}/k_BT)$ with \emph{effective} barriers $\Delta E^\text{ni}_\text{eff}= 1.0$ eV and $\Delta E^\text{nicocr}_\text{eff}= 0.65$ eV.
Given the melting temperatures $T^\text{ni}_m=2100$ K \footnote{Here the melting point was determined, using molecular dynamics simulations, by considering the volume change during the melt process that is known to develop a discontinuity across $T_m$.} and $T^\text{nicocr}_m=1650$ K \cite{Li2019}, it follows that $\Delta E^\text{ni}_* > \Delta E^\text{nicocr}_* $ with $\Delta E^\text{ni}_*=5.5$ and $\Delta E^\text{nicocr}_*=4.6$.
This, however, does not agree with most empirical observations that, in general, CCAs tend to have a higher scaled energy barrier than conventional alloys.  
We remark that $\Delta E_*$ is typically inferred from the Arrhenius-like dependence of the diffusion coefficients on (reduced) temperature and that the latter quantity is mathematically ill-defined in our case owing to the anomalous diffusion behavior associated with \glsix\!\!\!.
Furthermore, the inferred slope associated with NiCoCr may show variations depending on the selected range of $T$ in our regression analysis. 
Due to a slightly negative curvature at lower temperatures, \ie non-Arrhenius behavior as in the inset of Fig.~\ref{fig:waitTimes}(b), one might infer larger effective energies leading to a better agreement with experimental findings.
That said, the Hurst exponent should be viewed as the robust measure of sluggishness in our study as opposed to the scaled activation energies.

Given the above estimates for exponent $\alpha$, the scaling law predicts $H=0.5$ (\ie the standard diffusion) which contradicts our observations based on Fig.~\ref{fig:hurst}(d) at low temperatures. 
A plausible explanation could be given when one considers the bi-linear form for waiting time distributions (on logarithmic scales) in Fig.~\ref{fig:waitTimes}(b) with a relatively shallow slope $\alpha=0.4$ corresponding to $\lambda t_w<1$. 
This could yield a Hurst exponent $H < 1/2$ in a rough agreement with the observed trends in Fig.~\ref{fig:hurst}(d).
\remove[KK]{We also allude to statistics of jump lengths and corresponding \emph{truncated} power-laws that might rule out the L\'evy flight hypothesis.} 
On a different note, we observed power-law decays associated with jump size distributions $p(\Delta x)$ for \glsix as shown in Supplementary Materials (SM).
In the context of complex disordered alloys, the observed power-law behavior could be understood theoretically in terms of both local atomic misfits and vacancy hopping that induce long-range residual strains within the embedding elastic medium \cite{geslin2021microelasticity,geslin2021microelasticity_b}.
In the large $\Delta x$ limit, the theory further predicts $p(\Delta x)\propto |\Delta x|^{-(1+\mu)}$ with $\mu=d/(d-1)$ \cite{karimi2019self} in fair agreement with the observed scaling behavior for \glsix (see Fig.~S6(b) in SM).

Dynamics of the vacancy hopping is governed by local barriers that, in the context of concentrated solid solutions, are expected to have a broad distribution of energy scales owing to heterogeneities in local chemical environments.
This is evidenced Fig.~\ref{fig:waitTimes}(c) showing statistical distributions of local energy barriers $p(\Delta E)$ in \glsix spanning at least three decades in $\Delta E$ with mean energy $\langle \Delta E \rangle = 0.8-0.9$ eV and characteristic peaks around $\Delta E = 0.5$ eV at all temperatures.
We remark that pure Ni possess only one energy scale ($\Delta E = 1.0$ eV) due to the tranlational symmetry of the crystal.
We also note the appearance of smaller peaks at about $\Delta E =  0.05$ eV that tend to become suppressed as $T$ is increased toward $2000$ K.
One might naively attribute the emerging characteristic energy scales to the different constituent elements in \glsix on accounts that each species presumably have a typical chemical environment surrounding it.
However, elemental energy distributions (\ie $p(\Delta E)$ conditioned based on atomic types) do not support this hypothesis (data not shown).
As shown in the scatter plots of Fig.~\ref{fig:waitTimes}(d), the exponential dependence of (mean) waiting times $t_w$ on the normalized energy barrier $\Delta E/k_BT$ validates the relevance of the Arrhenius-like activation.\\


\noindent\emph{Conclusions \& Discussions---}
We have presented direct numerical evidence that thermally-driven dynamics of single atomic vacancies in model FCC CCAs tends to obey a subdiffusive behavior, and therefore sluggish diffusion, over sufficiently long timescales.
This observation is in stark contrast to pure single-element metals in which the vacancy migration typically features a normal diffusive process. 
We have argued that short-time dynamics has a strong bearing on the observed long-term kinetics and is well-described asymptotically by a set of scale-free characteristics and critical scaling exponents.
In this context, we have reported the relevance of fractional Brownian motion as a potential origin of suppressed diffusion possibly due to (anti-)correlation effects.
As for the emerging sluggishness, another plausible scenario seems to be the presence of long rest periods and the property that vacancy migration energies possess wide statistical distributions, covering almost four orders of magnitude. 
Broad energy-scales have their root in underlying chemical/structural disorder and such microstructure-dynamics correlations have important implications in alloys' property prediction and design.\\

\noindent\emph{Acknowledgments---}
We wish to express our gratitude to N. Mousseau for sharing the $k$-ART code and insightful discussions. This research was funded by the European Union Horizon 2020 research and innovation program under grant agreement no. 857470 and from the European Regional Development Fund via Foundation for Polish Science International Research Agenda PLUS program grant no. MAB PLUS/2018/8.

\bibliography{references}

\end{document}


\title{Multiscale modeling of kinetic sluggishness in equiatomic NiCoCr and NiCoCrFeMn single-phase solid solutions}

\author{Kamran Karimi$^1$}
\author{Stefanos Papanikolaou$^1$}
\affiliation{%
 $^1$ NOMATEN Centre of Excellence, National Center for Nuclear Research, ul. A. Sołtana 7, 05-400 Swierk/Otwock, Poland\\
}%

\maketitle

\section*{Supplementary Materials}

\section{$k$-ART algorithm}
Kinetic Activation-Relaxation Technique ($k$-ART) constructs an exhaustive catalog of atomically-resolved topology and relevant energy barriers to be used for \emph{event} sampling (i.e. vacancy hopping) within the Arrhenius-based activation process \cite{voter2007introduction}. 
In this context, the hopping rate obeys the following dynamics $\Omega_j = \Omega_0~\text{exp}(-\Delta E_j/k_BT)$ where index $j\in \mathcal{T}(i)$ refers to a distinct topology identity classified based on the local environment of a center atom $i$ with respect to its nearest neighbors within the cutoff radius $r_c=4.0$~\r{A}.
Here $k_B$ is the Boltzmann constant. 
Energy barriers $\Delta E_j$ are determined based on force fields implemented in LAMMPS by associating each topology index $j$ with a nearest saddle point and next minimum in the energy hypersurface within the activation-relaxation framework (see \cite{el2008kinetic} and references therein).
We also set $\Omega_0=10^{13}~\text{s}^{-1}$ as a typical vibrational frequency although a recent work by Mousseaua et al. \cite{sauve2022unexpected} reported non-negligible variations of this prefactor in concentrated solid solutions.
Apart from being a computational convenience, our choice for a fixed $\Omega_0$ can still capture the underlying hopping dynamics which is strongly governed by chemical heterogeneities in $\Delta E_j$.   
Upon each activation process, the topology of evolved atoms corresponding to the new relaxed configuration will be updated to be used within the next Monte Carlo sampling.

\begin{figure*}[t]
    \centering
	%
    \begin{overpic}[width=0.23\textwidth]{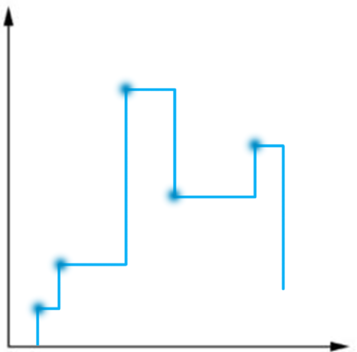}
        %
        \Labelxy{100}{-6}{0}{$t$}
        \Labelxy{30}{-6}{0}{$t_{k}$}
        \Labelxy{50}{-6}{0}{$t_{k+1}$}
        \Labelxy{-8}{92}{90}{$x_i(t)$}
        \Labelxy{-10}{40}{90}{$x_i^{(k+1)}$}
        \Labelxy{-10}{64}{90}{$x_i^{(k)}$}
        \begin{tikzpicture}
             \coordinate (a) at (0,0); 
             \node[white] at (a) {\tiny.};                 
             \lline{1.3}{0}{.1}{0}{0.8}{}
             \lline{1.9}{0}{.1}{0}{0.8}{}
             %
             \lline{0}{1.7}{.1}{1.0}{0.0}{}
             \lline{0}{2.9}{.1}{1.0}{0.0}{}
        \end{tikzpicture}
        
	\end{overpic}
	\vspace{8pt}
    \caption{Sketch of a displacement timeseries $x(t)$ associated with atom $i$ in one dimension with jump size $\Delta x_i=x_i^{(k+1)}-x_i^{(k)}$ and waiting times $t_w=t_{k+1}-t_{k}$ between event $k$ and $k+1$ as denoted by the symbols.}
    \label{fig:hoppingSM}
\end{figure*}

Based on the above framework, waiting (or rest) times refer to time intervals between two consecutive events defined as $t_w=t_{k+1}-t_{k}$ in Fig.~\ref{fig:hoppingSM}.
Here $k$ denotes the event index. 
After the elapsed time $t_w$, atom $i$ at position $\vec{r}_i^{\hspace{1pt}(k)}$ moves to position $\vec{r}_i^{\hspace{1pt}(k)}+\Delta \vec{r}_i$ at $t_{k+1}$ with increments $\Delta \vec{r}_i=(\Delta x_i,\Delta y_i, \Delta z_i)$.
We note that the individual components $\Delta x_i$, $\Delta y_i$, and $\Delta z_i$ are statistically equivalent as the activation kinetics has no angular preference (up to local crystal symmetry).

\section*{Robustness of $H$ exponent}
\begin{figure*}[t]
    \centering
    \begin{overpic}[width=0.23\textwidth]{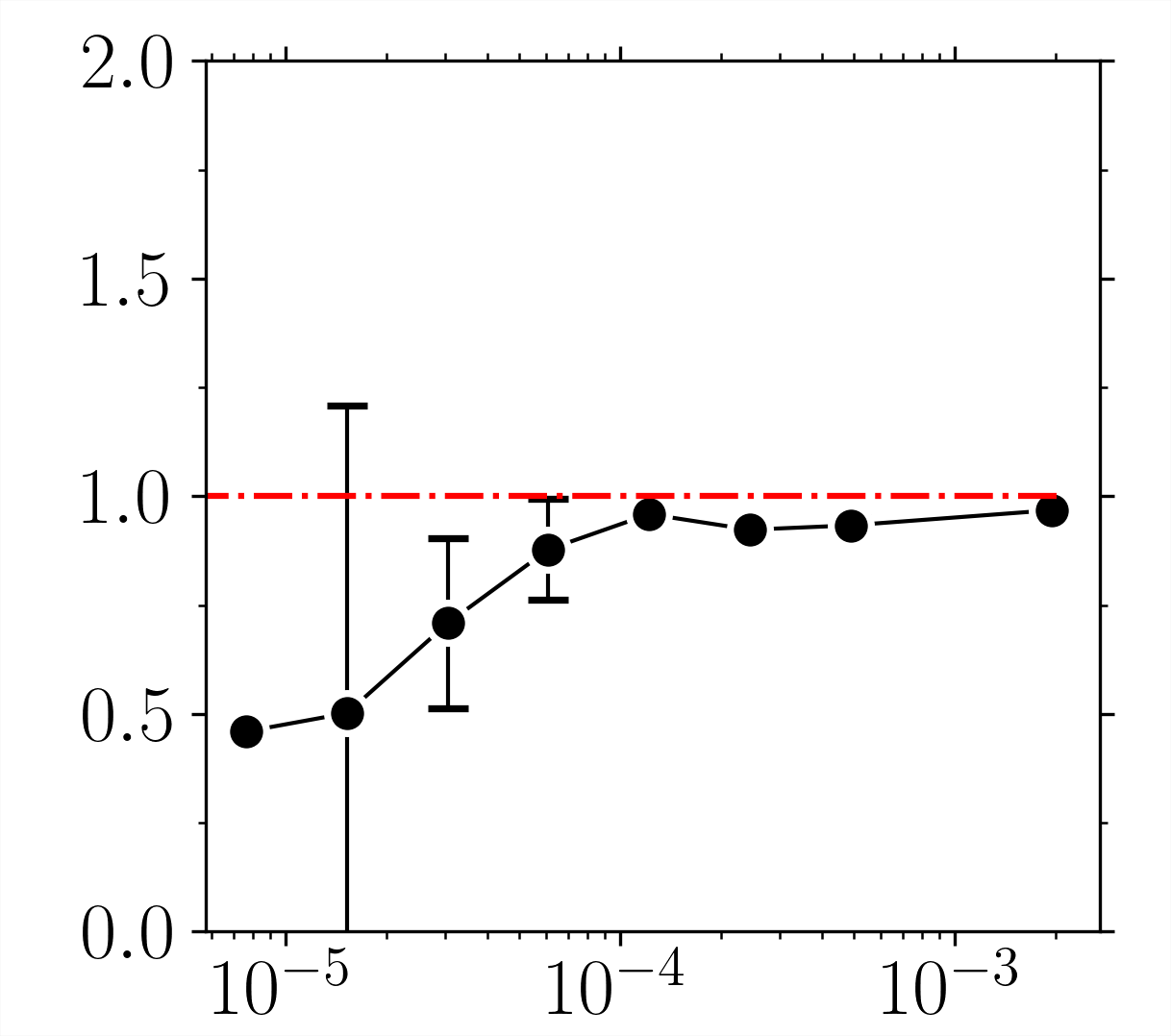}
        \LabelFig{16}{87}{$a)~ T=1000$ K}
        \Labelxy{-3}{42}{90}{$2H$}
	\end{overpic}
    \begin{overpic}[width=0.23\textwidth]{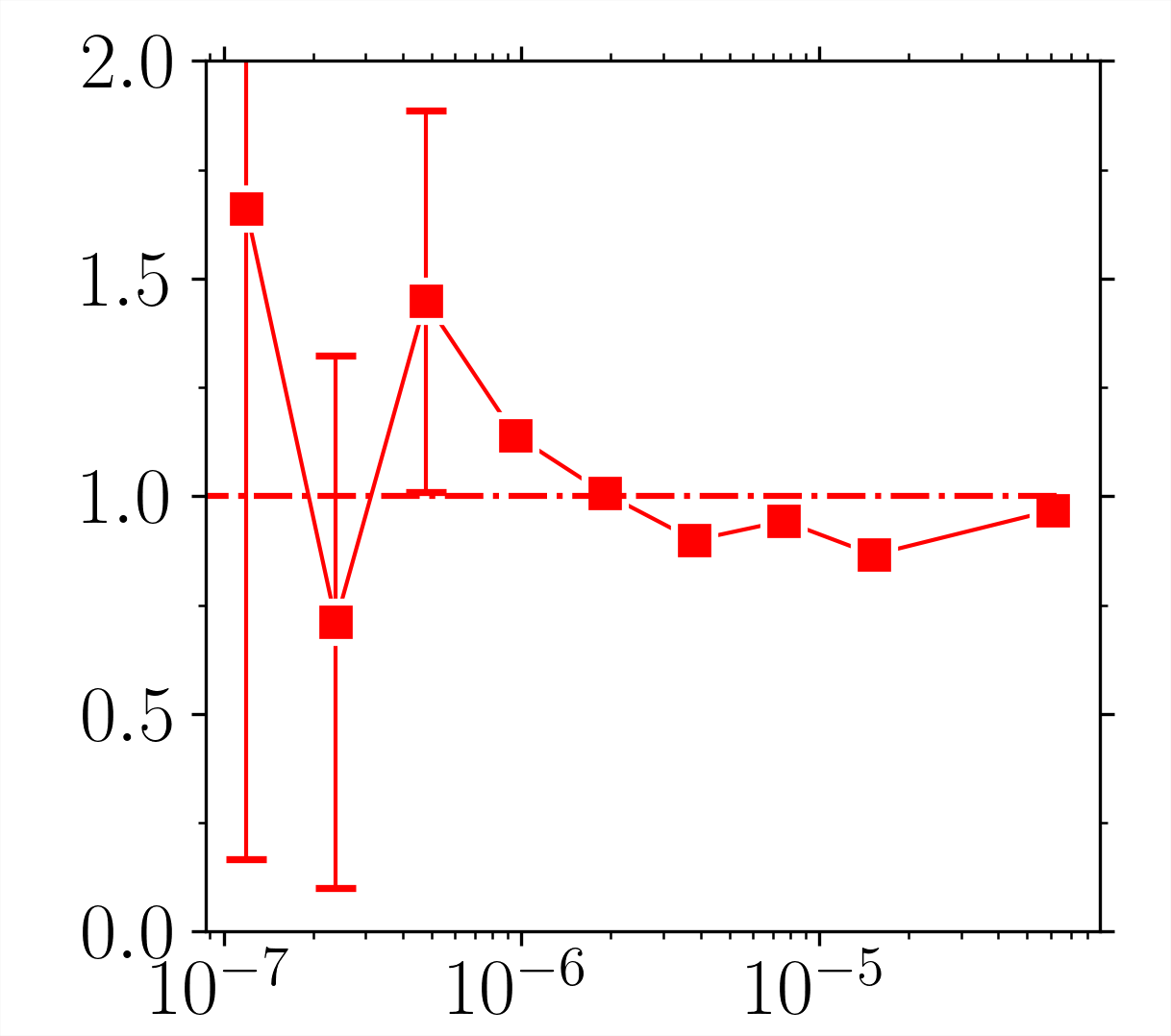}
        \LabelFig{16}{87}{$b)~ T=1200$ K}
	\end{overpic}
    \begin{overpic}[width=0.23\textwidth]{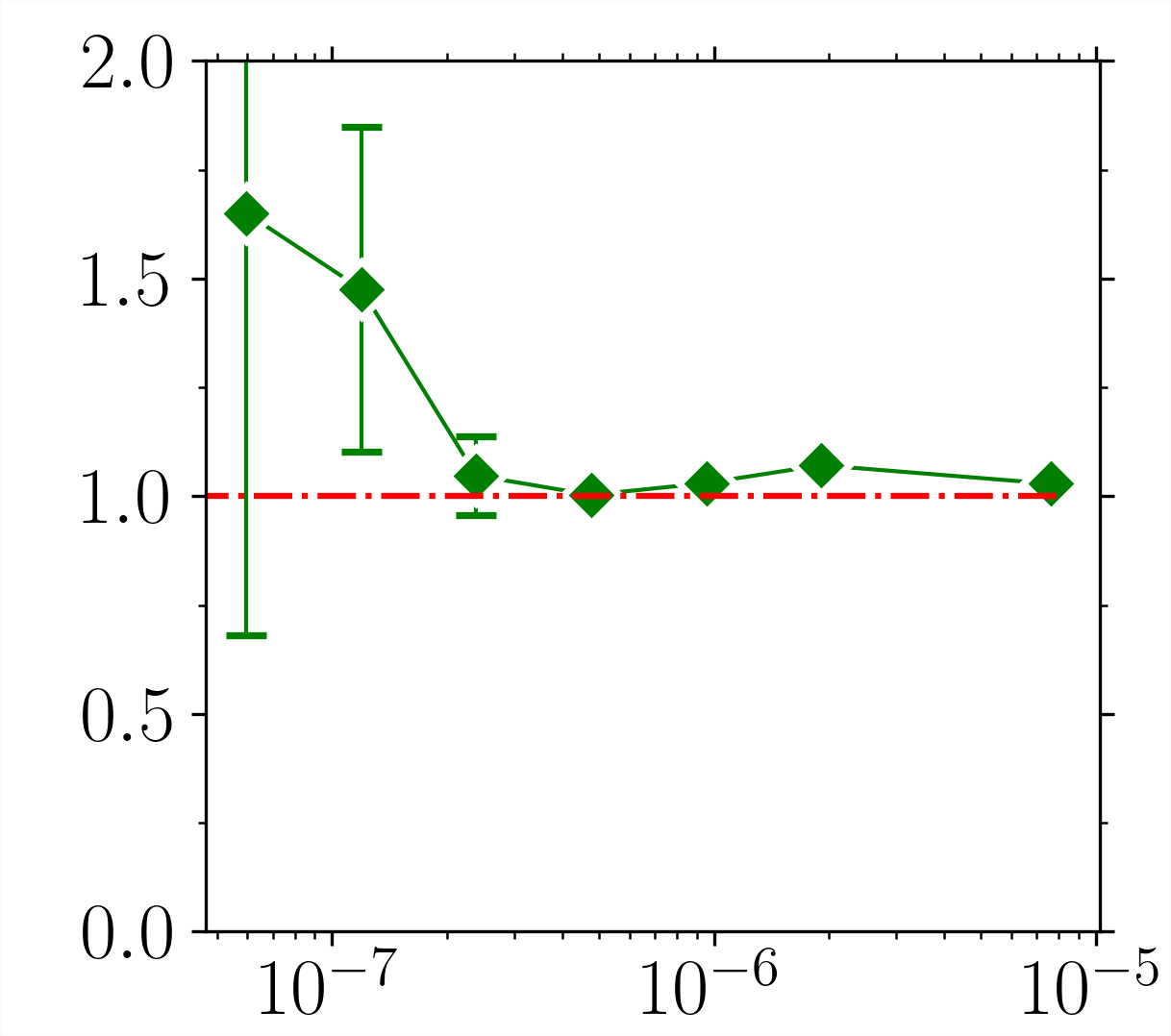}
        \LabelFig{16}{87}{$c)~ T=1400$ K}
	\end{overpic}

    \begin{overpic}[width=0.23\textwidth]{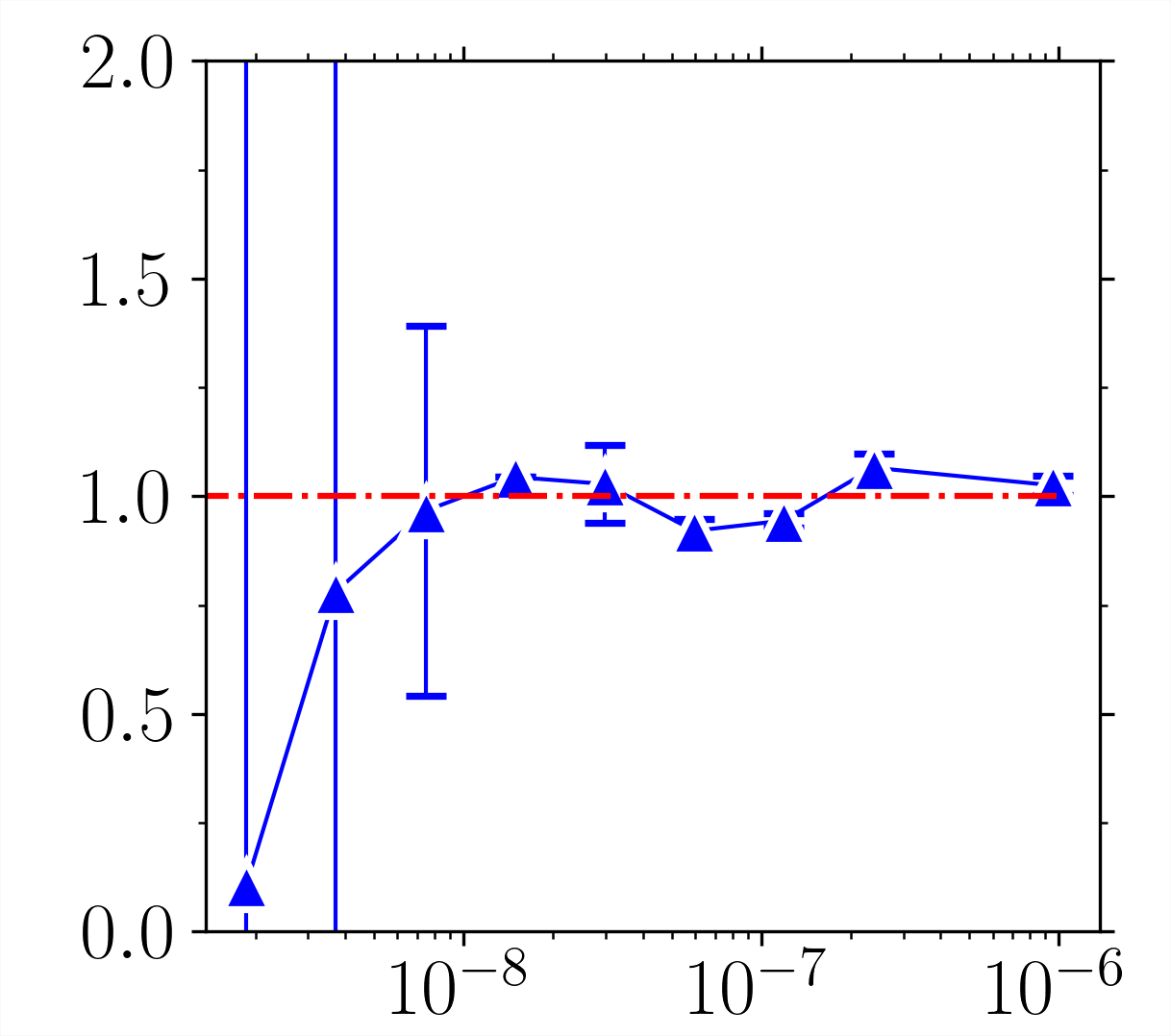}
        \LabelFig{19}{14}{$d)~ T=1600$ K}
        \Labelxy{50}{-6}{0}{$t_c$(s)}
        \Labelxy{-3}{42}{90}{$2H$}
	\end{overpic}
    \begin{overpic}[width=0.23\textwidth]{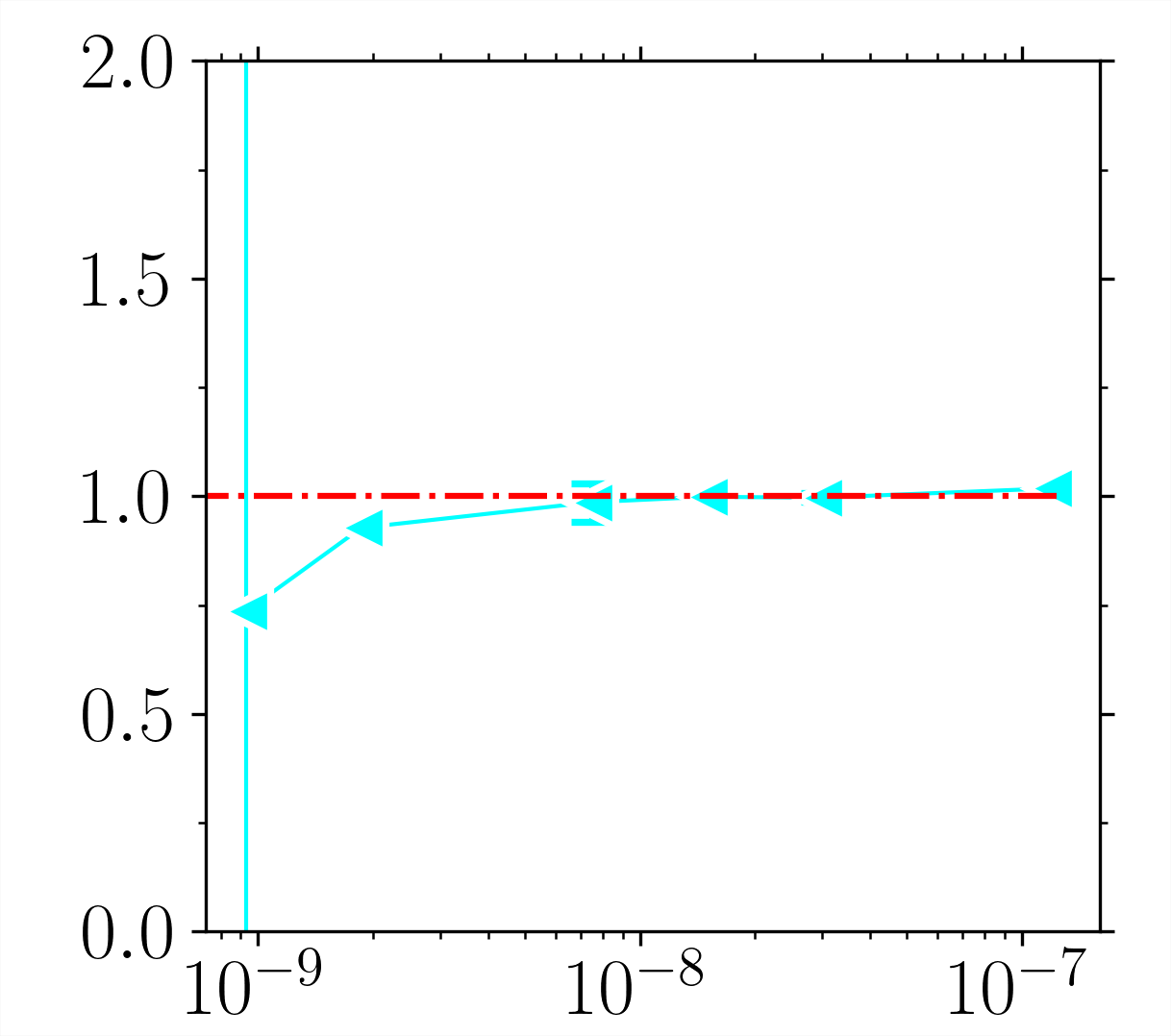}
        \LabelFig{19}{14}{$e)~ T=1800$ K}
        \Labelxy{50}{-6}{0}{$t_c$(s)}
	\end{overpic}
    \begin{overpic}[width=0.23\textwidth]{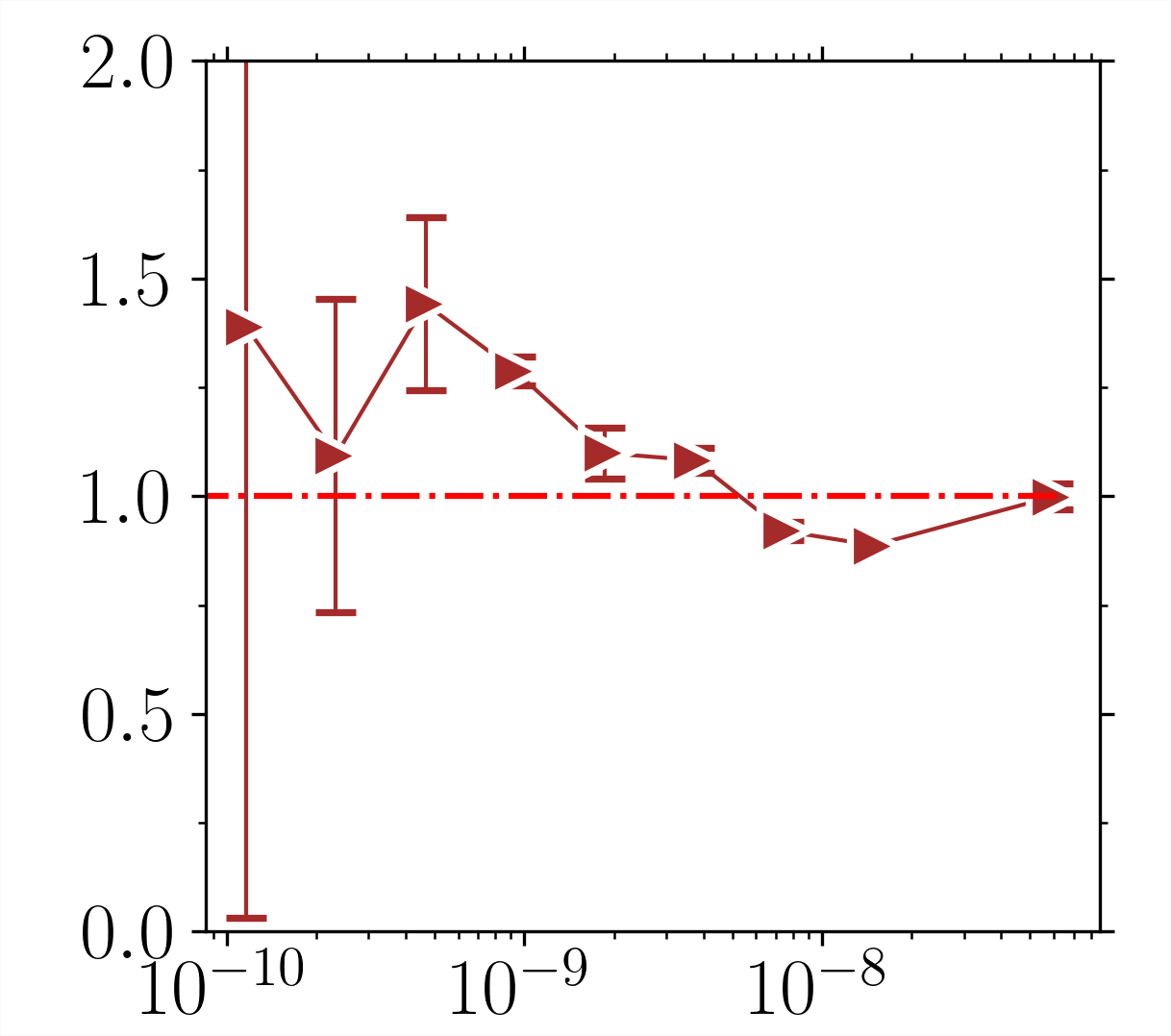}
        \LabelFig{19}{14}{$f)~ T=2000$ K}
        \Labelxy{50}{-6}{0}{$t_c$(s)}
	\end{overpic}
    \caption{Estimated exponent $H$ plotted against $t_c$ associated with pure Ni at various temperatures $T$. The (red) dashdotted lines indicate $H = 1/2$.}
    \label{fig:robustHNi}
\end{figure*}

\begin{figure*}[t]
    \centering
    \begin{overpic}[width=0.23\textwidth]{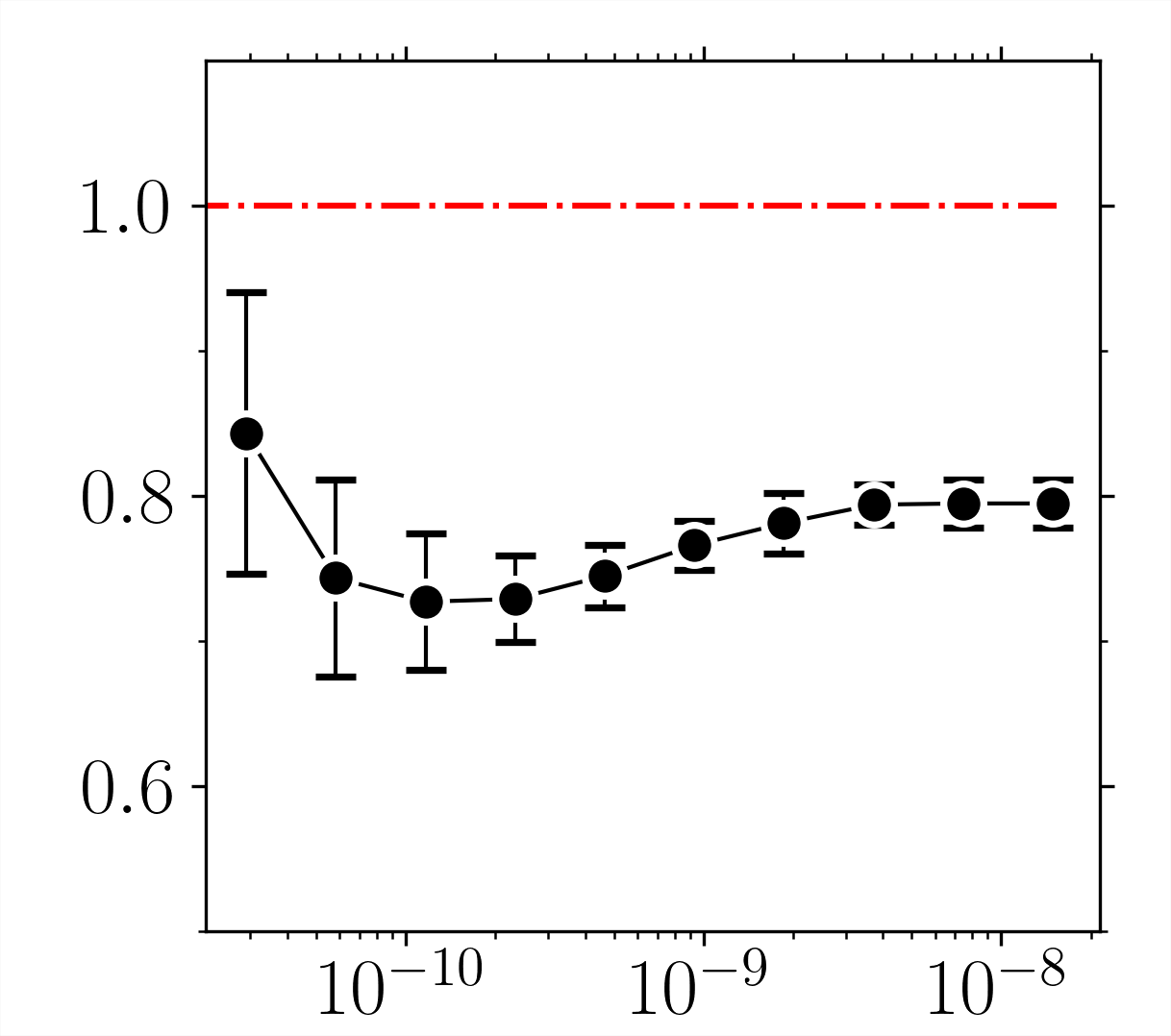}
        \LabelFig{16}{87}{$a)~ T=1000$ K}
        \Labelxy{-3}{42}{90}{$2H$}
	\end{overpic}
    \begin{overpic}[width=0.23\textwidth]{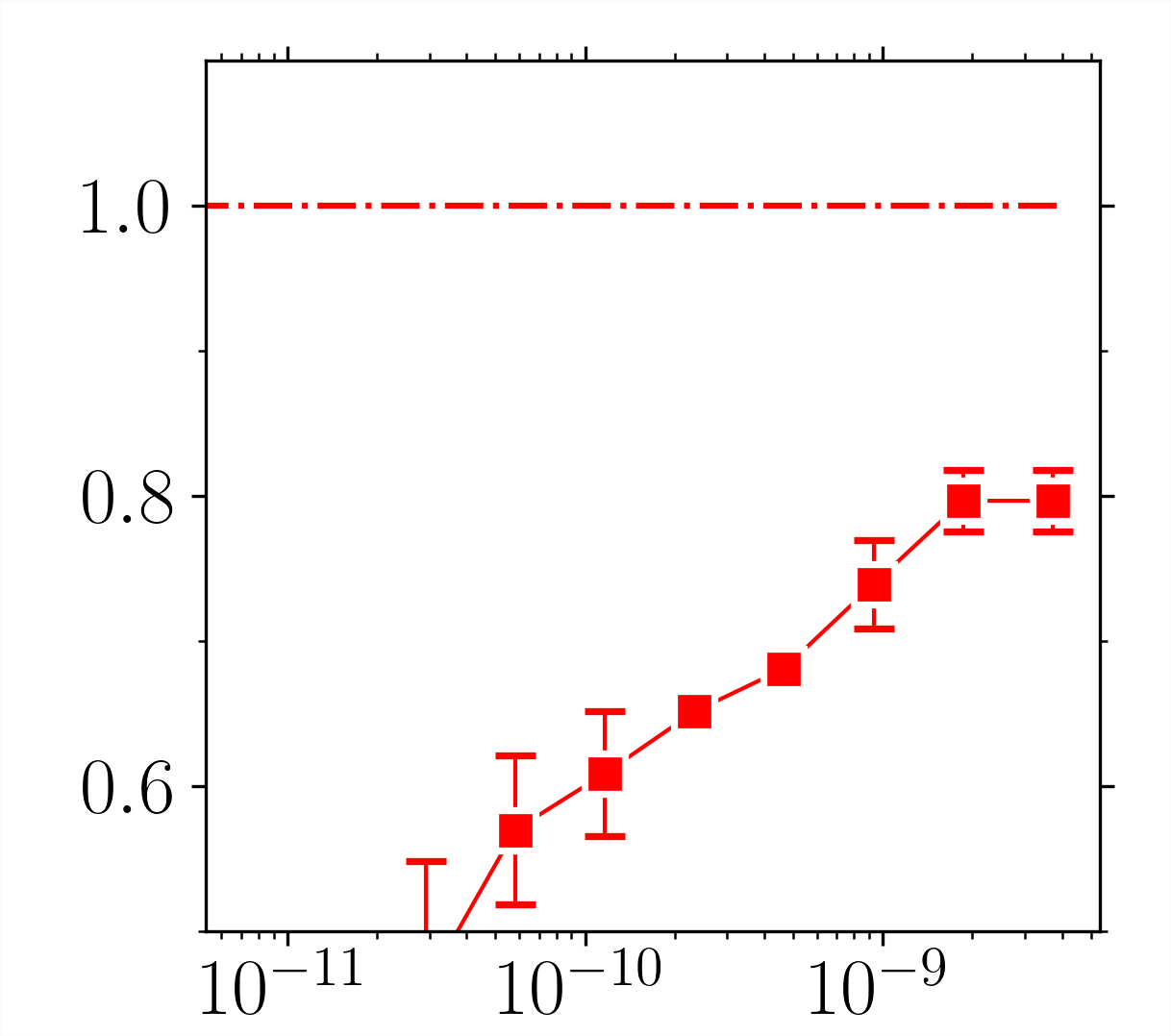}
        \LabelFig{16}{87}{$b)~ T=1200$ K}
	\end{overpic}
    \begin{overpic}[width=0.23\textwidth]{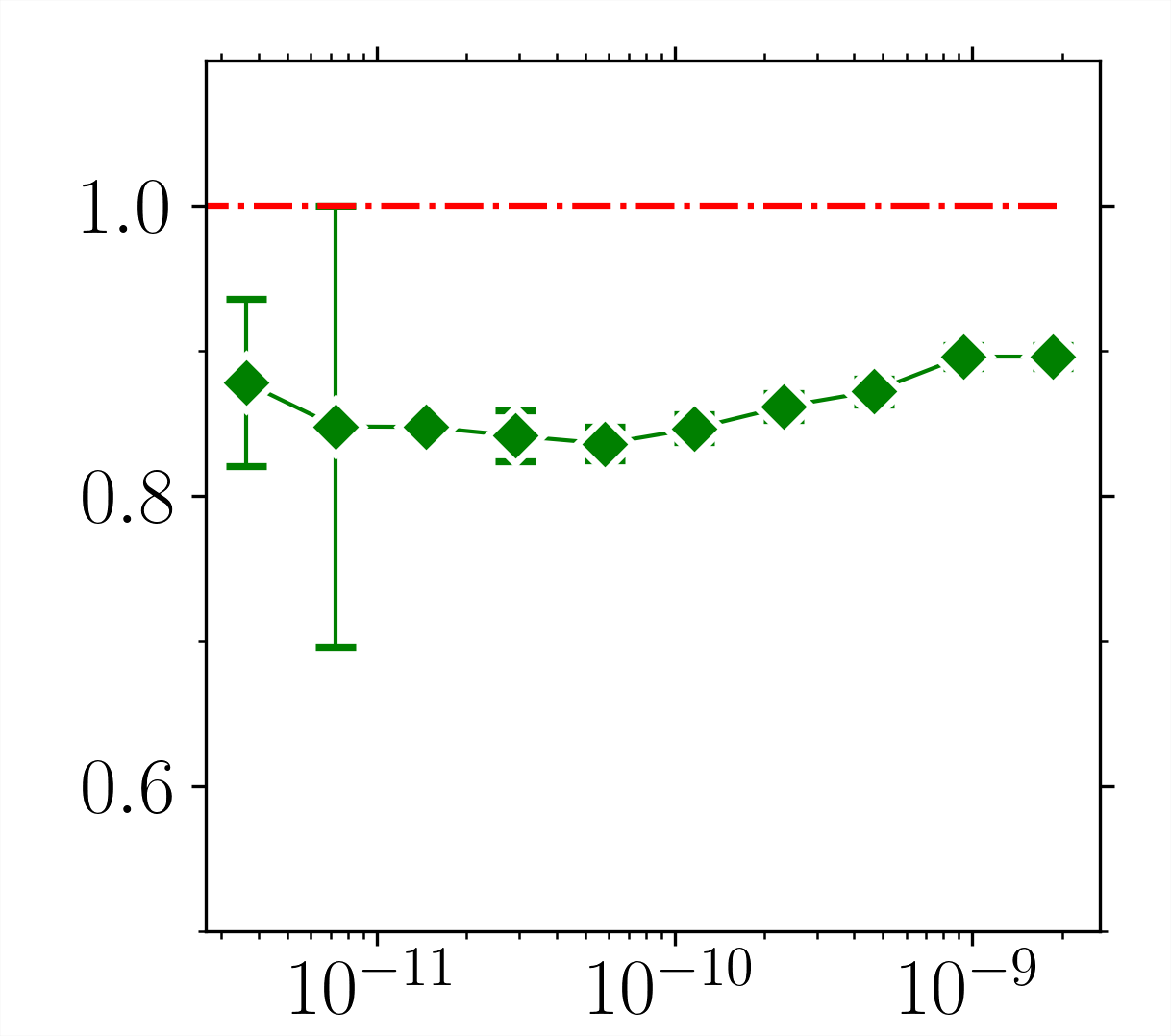}
        \LabelFig{16}{87}{$c)~ T=1400$ K}
	\end{overpic}

    \begin{overpic}[width=0.23\textwidth]{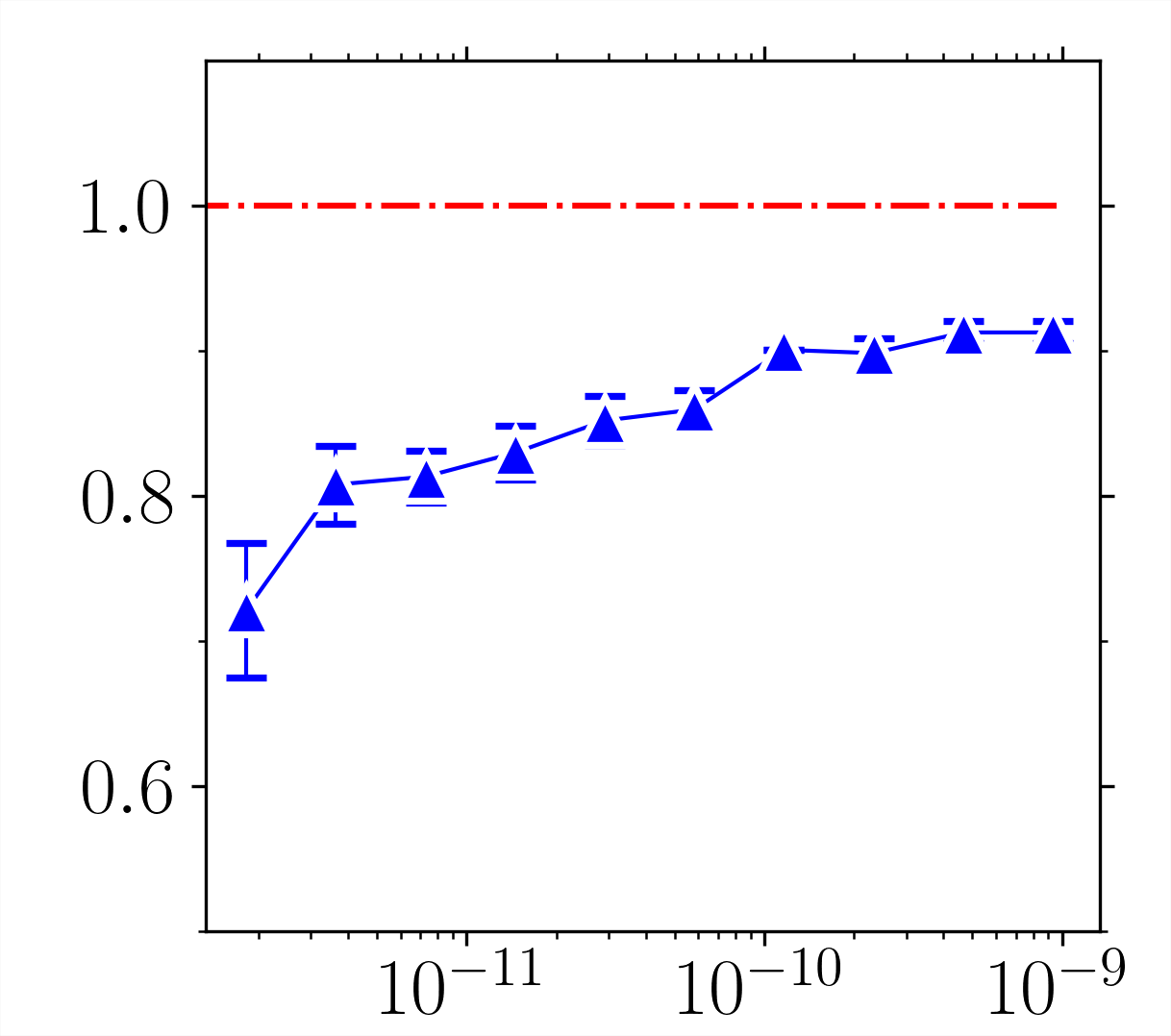}
        \LabelFig{19}{14}{$d)~ T=1600$ K}
        \Labelxy{50}{-6}{0}{$t_c$(s)}
        \Labelxy{-3}{42}{90}{$2H$}
	\end{overpic}
    \begin{overpic}[width=0.23\textwidth]{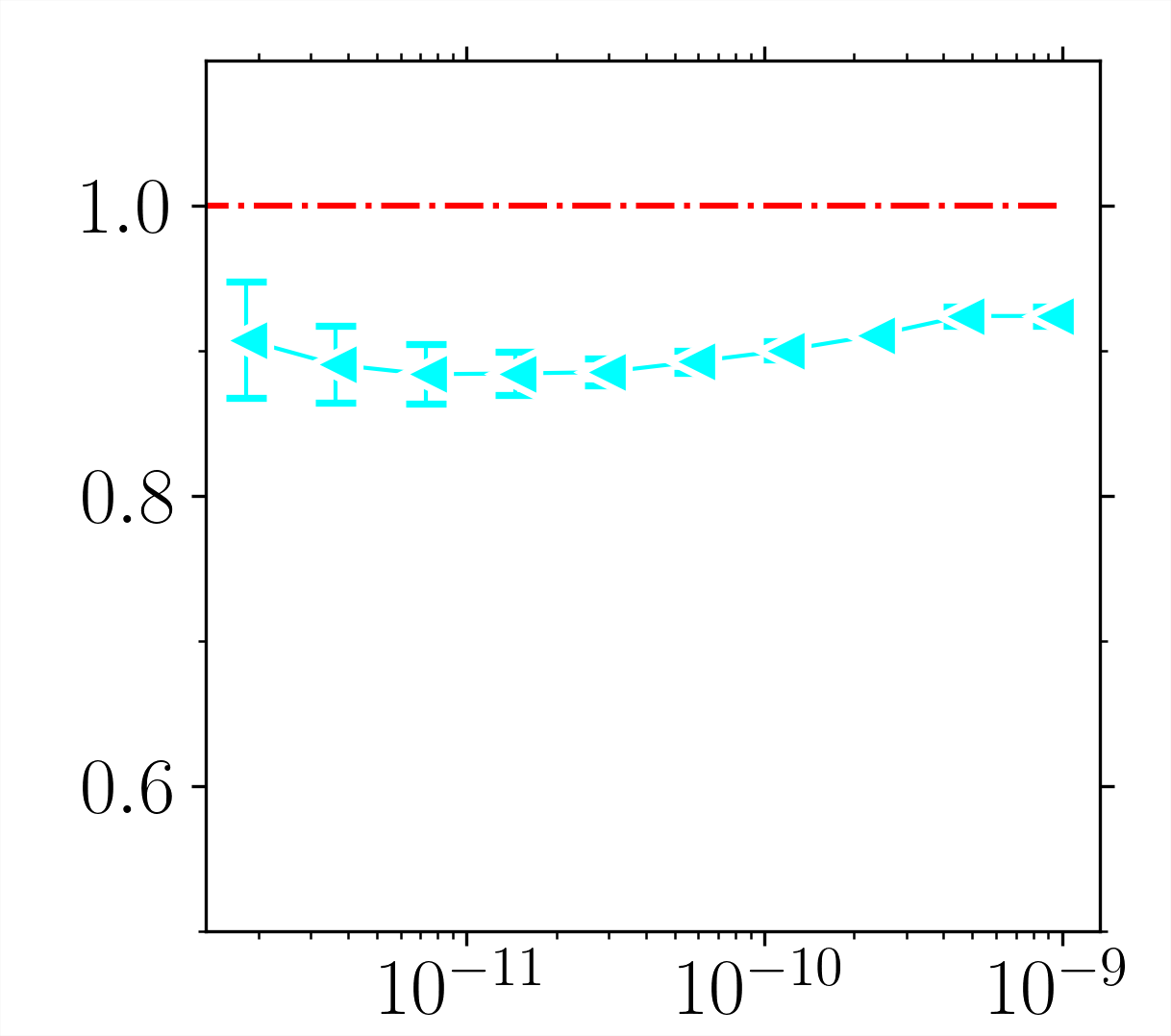}
        \LabelFig{19}{14}{$e)~ T=1800$ K}
        \Labelxy{50}{-6}{0}{$t_c$(s)}
	\end{overpic}
    \begin{overpic}[width=0.23\textwidth]{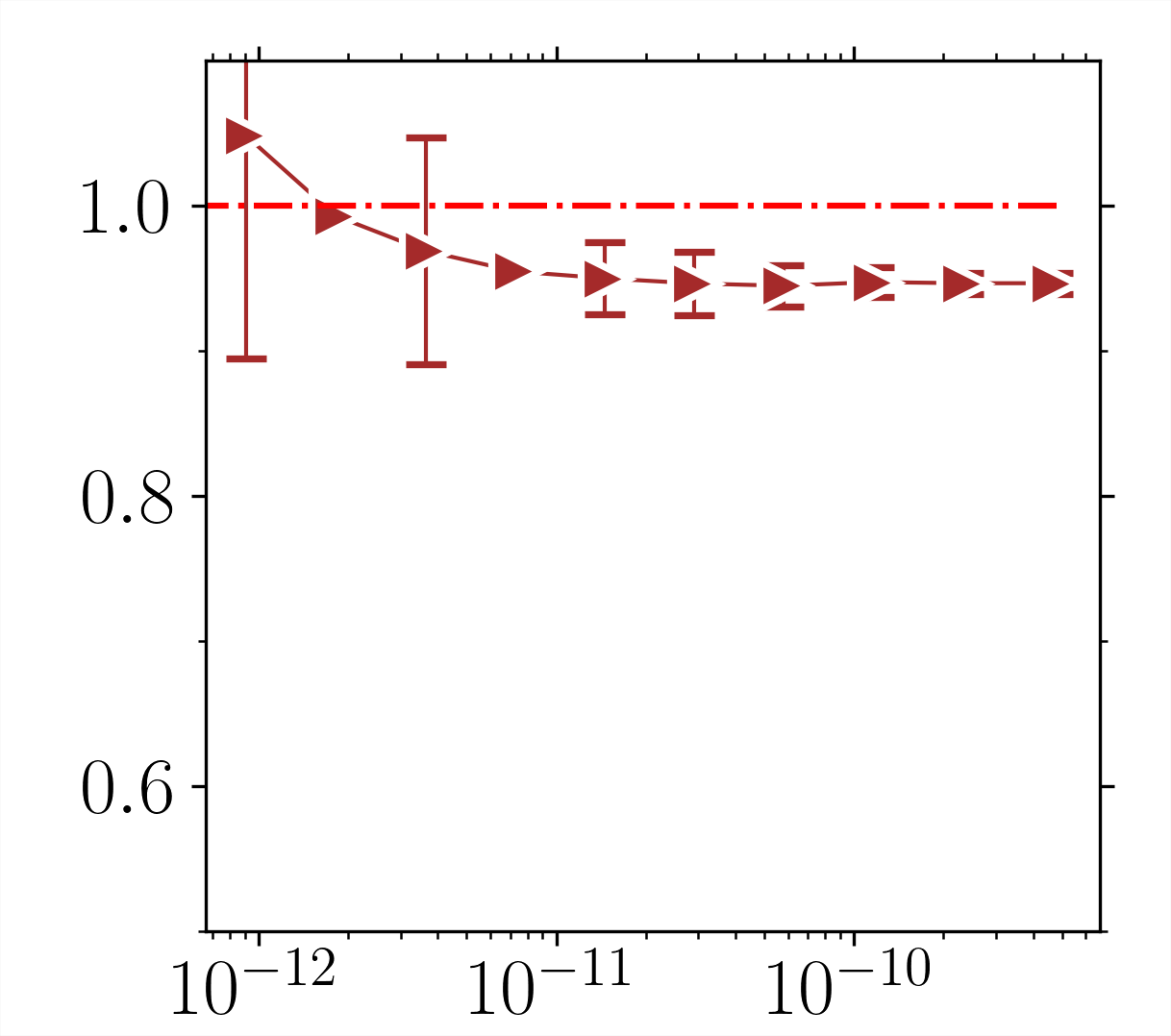}
        \LabelFig{19}{14}{$f)~ T=2000$ K}
        \Labelxy{50}{-6}{0}{$t_c$(s)}
	\end{overpic}
    \caption{Estimated exponent $H$ plotted against $t_c$ associated with \glsix at various temperatures $T$. The (red) dashdotted lines indicate $H = 1/2$.}
    \label{fig:robustHNiCoCr}
\end{figure*}

As a robustness analysis, we vary $t_c$ systematically and repeat the regression analysis by including the mean-square displacement (msd) data in Fig.~2(a) and (b) within the temporal window $0< t < t_c$ to estimate the Hurst exponents $H$. 
The results are shown in Fig.~\ref{fig:robustHNi}(a-f) and Fig.~\ref{fig:robustHNiCoCr}(a-f) corresponding to pure Ni and NiCoCr at all temperatures.
We obtain fairly robust estimates of $H$ featuring a progressive evolution toward well-defined asymptotes at sufficiently large values of $t_c$.

\section*{fractional Brownian motion}
\begin{figure*} 
    \centering
    \begin{overpic}[width=0.32\textwidth]{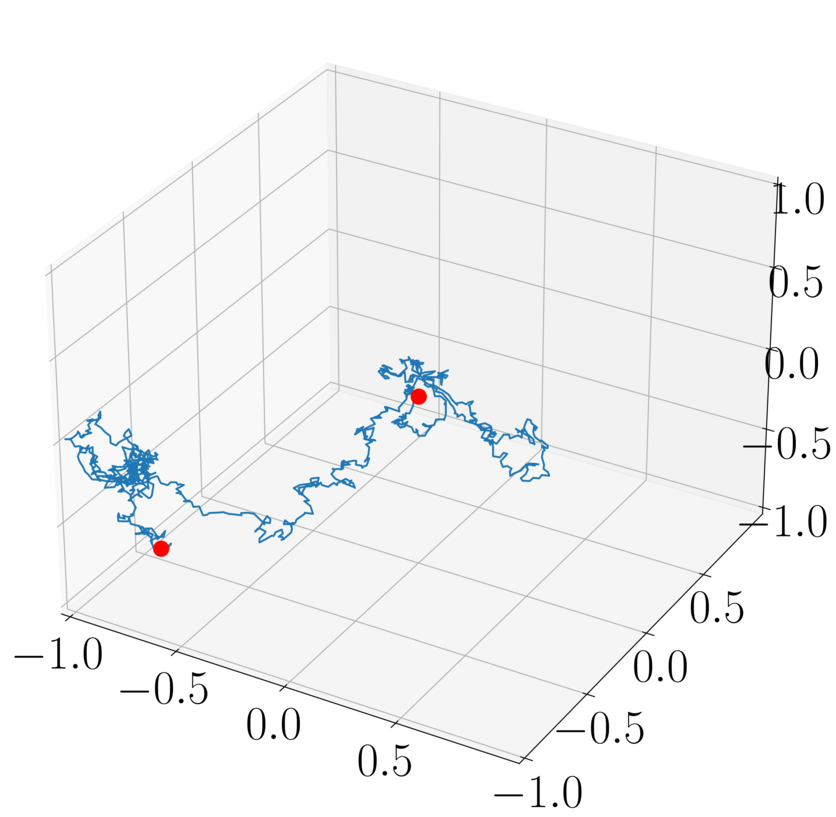}
        \LabelFig{16}{87}{$a)~H=0.6$}
        \Labelxy{80}{10}{45}{$x$\scriptsize(\r{A})}
        \Labelxy{26}{4}{-18}{$y$\scriptsize(\r{A})}
        \Labelxy{-2}{40}{90}{$z$\scriptsize(\r{A})}
	\end{overpic}
    \begin{overpic}[width=0.32\textwidth]{Figs/Diffusive_reduced.png}
        \LabelFig{16}{87}{$b)~H=0.5$}
        \Labelxy{80}{10}{45}{$x$\scriptsize(\r{A})}
        \Labelxy{26}{4}{-18}{$y$\scriptsize(\r{A})}
        \Labelxy{-2}{40}{90}{$z$\scriptsize(\r{A})}
	\end{overpic}
    \begin{overpic}[width=0.32\textwidth]{Figs/subDiffusive_reduced.png}
        \LabelFig{16}{87}{$c)~H=0.4$}
        \Labelxy{80}{10}{45}{$x$\scriptsize(\r{A})}
        \Labelxy{26}{4}{-18}{$y$\scriptsize(\r{A})}
        \Labelxy{-2}{40}{90}{$z$\scriptsize(\r{A})}
	\end{overpic}
    \caption{Model fractional Brownian walks associated with Hurst exponents \textbf{a}) $H=0.6$ \textbf{b}) $H=0.5$ \textbf{c}) $H=0.4$. The (red) markers indicate the start points at $(0,0,0)$ and the end points. 
    }
    \label{fig:fbm}
\end{figure*}

A fractal Brownian motion (fBm) is a stochastic process where the fBm increments are not necessarily independent (as in classical Brownian motion) and can instead be correlated in time with a slowly-decaying memory \cite{mandelbrot1968fractional}.
If the increments are positively correlated, then $H > 1/2$ and diffusion is enhanced as shown in Fig.~\ref{fig:fbm}(a). Anti-correlated increments leads to $H < 1/2$ implying subdiffusion as in Fig.~\ref{fig:fbm}(c). For uncorrelated increments, one recovers ordinary diffusion with $H=1/2$ as in Fig.~\ref{fig:fbm}(b).
Here the fractional Brownian paths were implemented based on the Davies-Harte Method \cite{fbm}.

\begin{figure*}[b]
    \centering
    \begin{overpic}[width=0.23\textwidth]{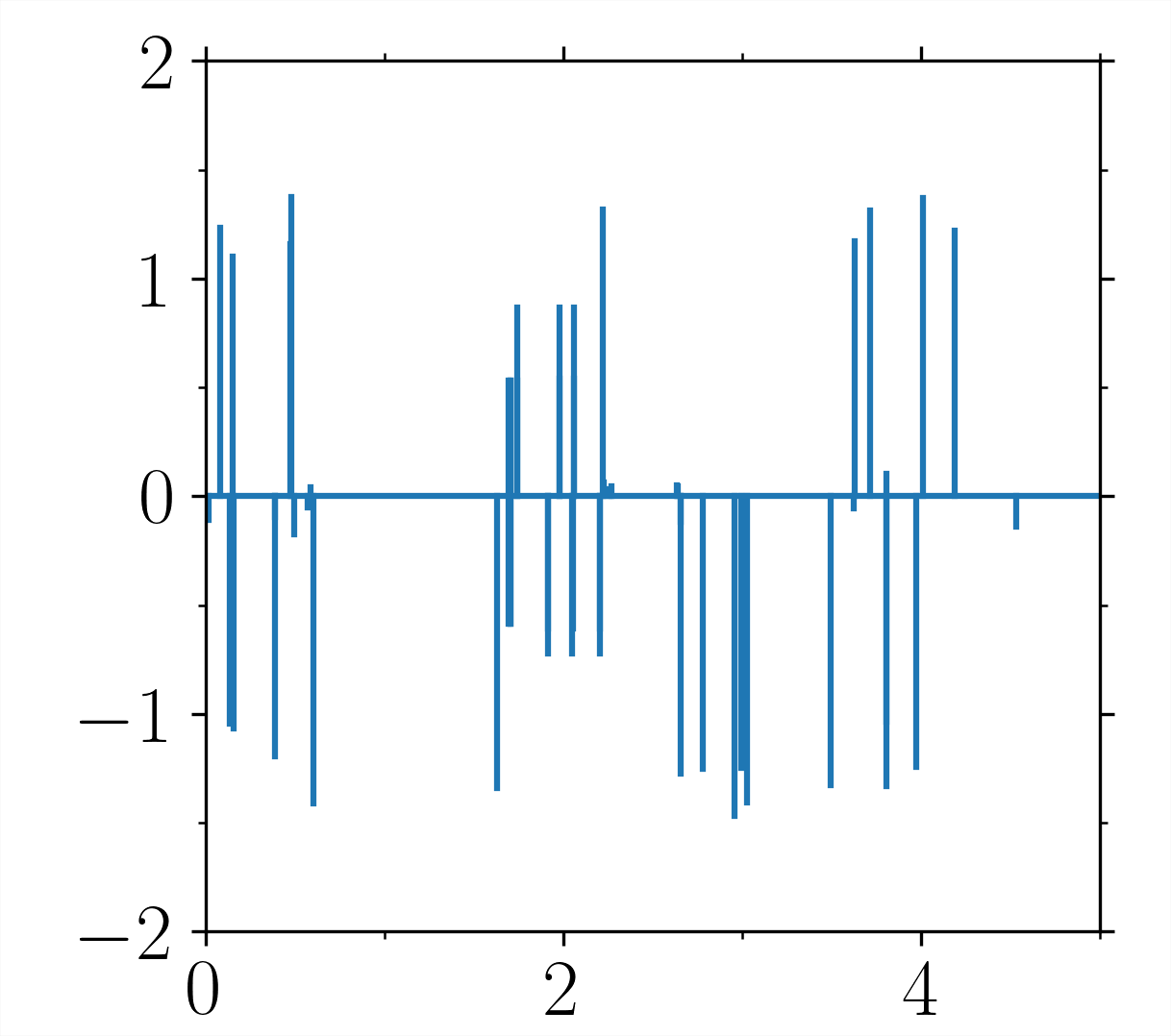}
        \LabelFig{16}{87}{$a)$}
        \Labelxy{50}{-6}{0}{$t(\times 10^{-10}~\text{s})$}
        \Labelxy{-4}{37}{90}{$\Delta x$(\r{A})}
	\end{overpic}
    \begin{overpic}[width=0.23\textwidth]{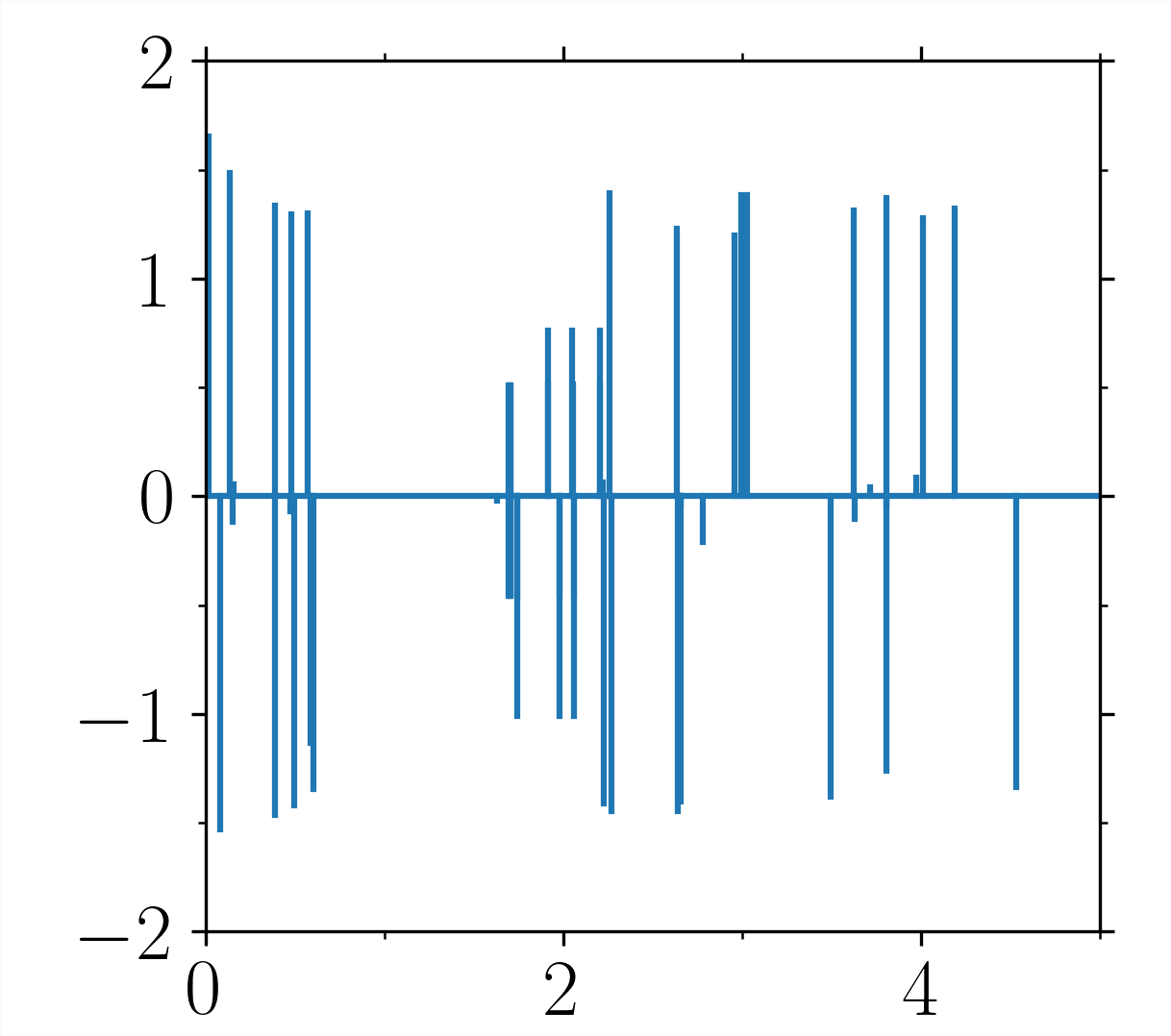}
        \LabelFig{16}{87}{$b)$}
        \Labelxy{50}{-6}{0}{$t(\times 10^{-10}~\text{s})$}
        \Labelxy{-4}{37}{90}{$\Delta y$(\r{A})}
	\end{overpic}
	\vspace{8pt}
    \begin{overpic}[width=0.23\textwidth]{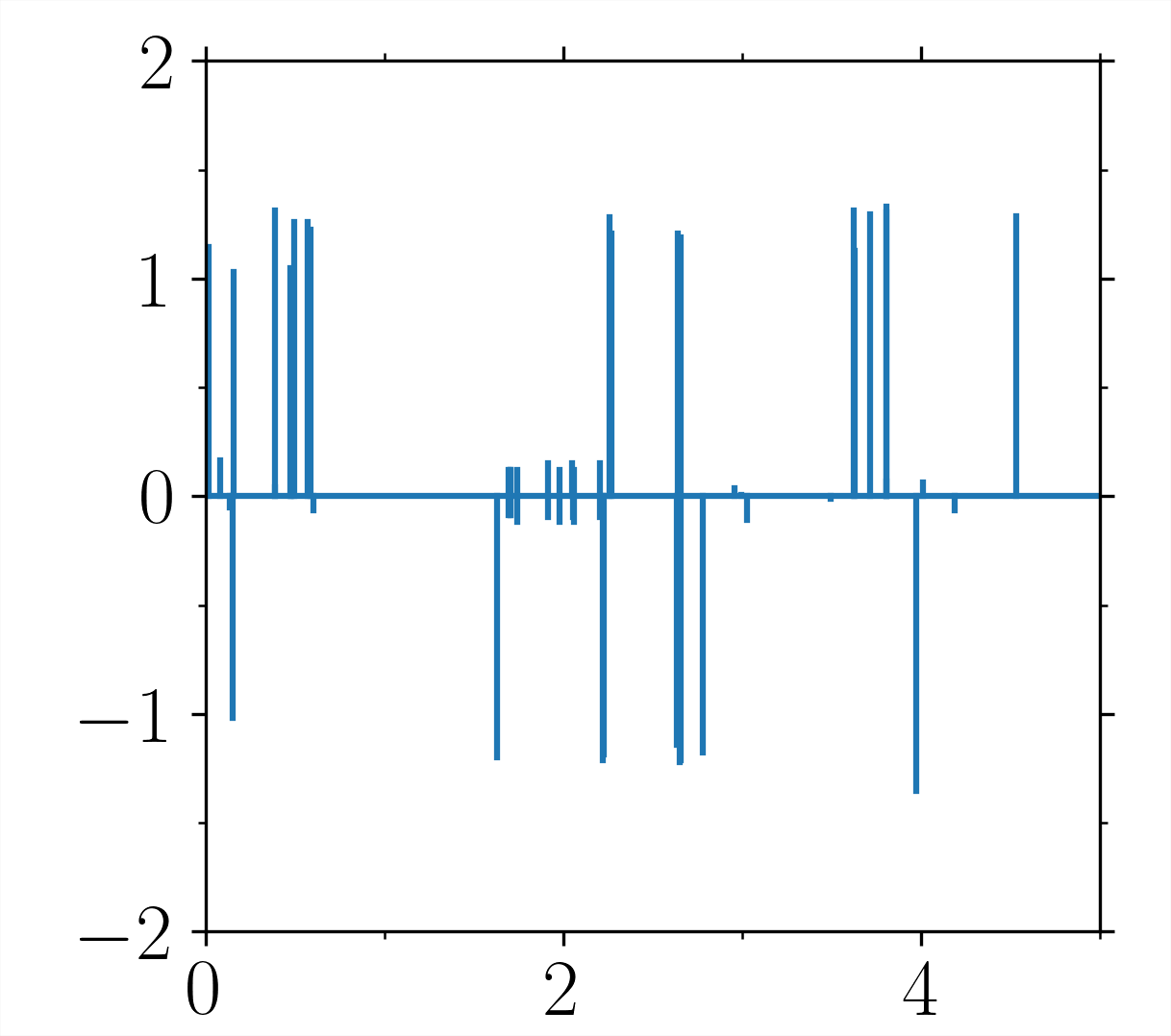}
        \LabelFig{16}{87}{$c)$}
        \Labelxy{50}{-6}{0}{$t(\times 10^{-10}~\text{s})$}
        \Labelxy{-4}{37}{90}{$\Delta z$(\r{A})}
	\end{overpic}

     \begin{overpic}[width=0.23\textwidth]{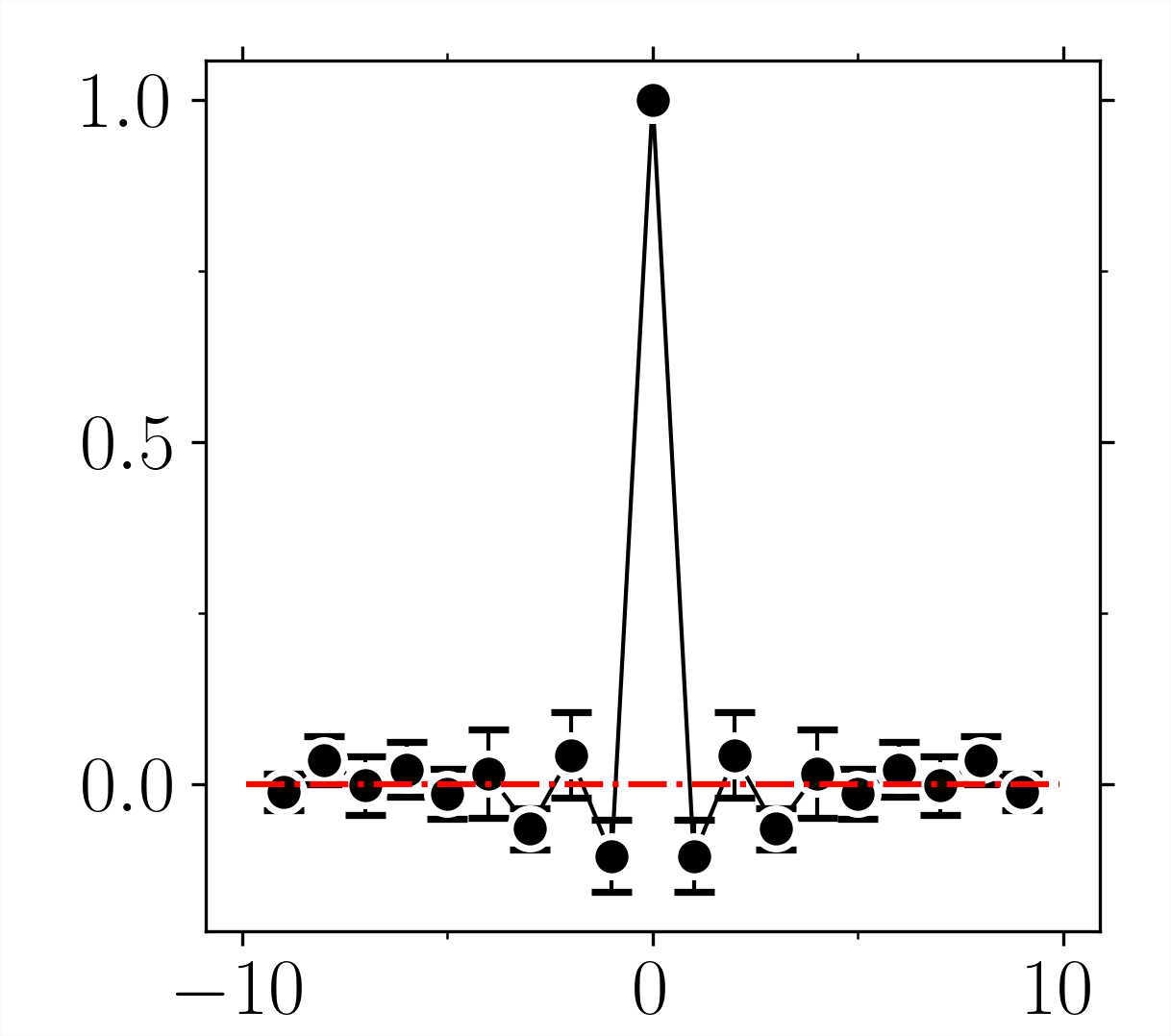}
        \LabelFig{16}{87}{$d)$}
        \Labelxy{50}{-6}{0}{$n$}
        \Labelxy{-4}{37}{90}{$c_{xx}$}
	\end{overpic}
    \begin{overpic}[width=0.23\textwidth]{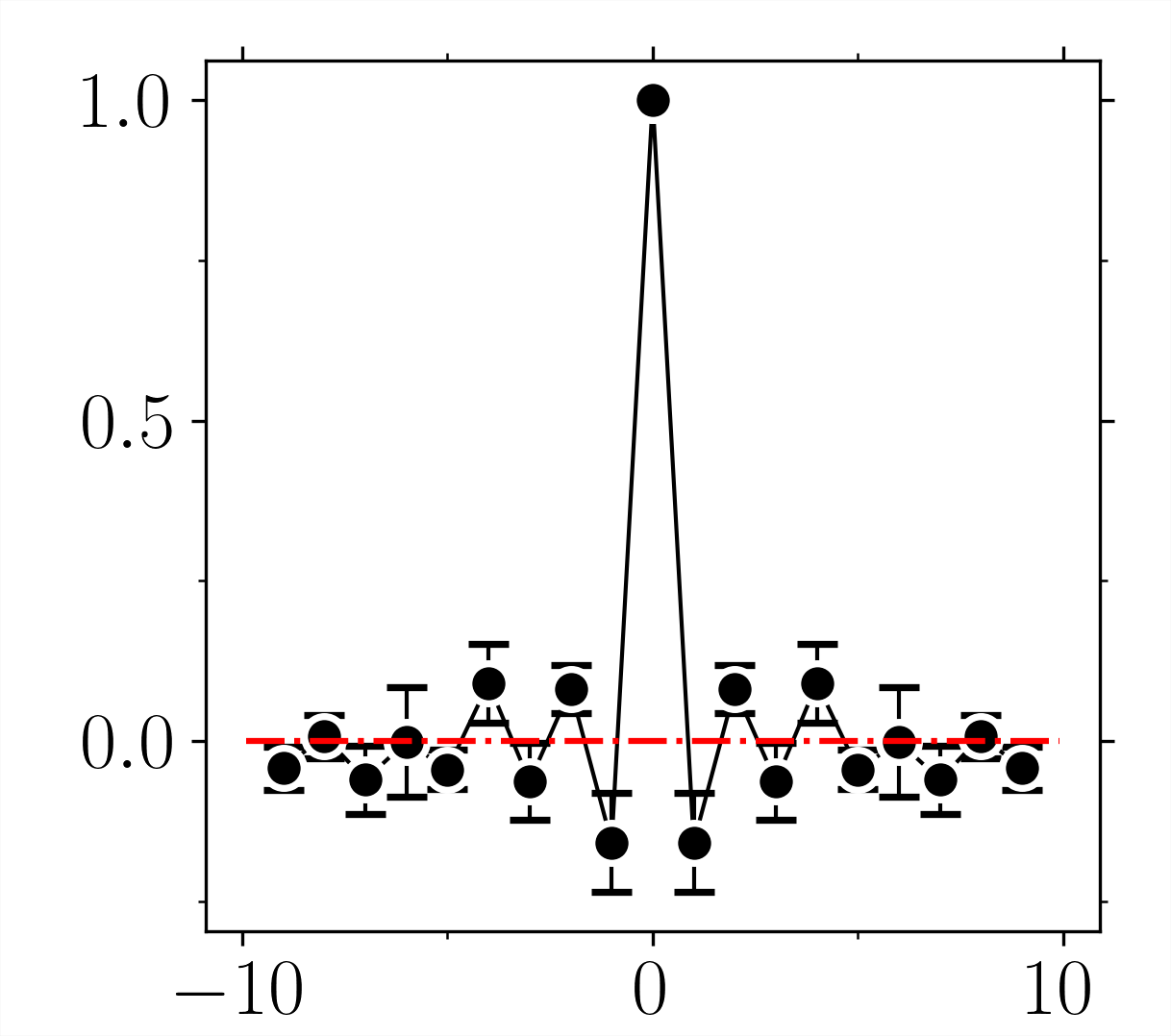}
        \LabelFig{16}{87}{$e)$}
        \Labelxy{50}{-6}{0}{$n$}
        \Labelxy{-4}{37}{90}{$c_{yy}$}
	\end{overpic}
	\vspace{8pt}
    \begin{overpic}[width=0.23\textwidth]{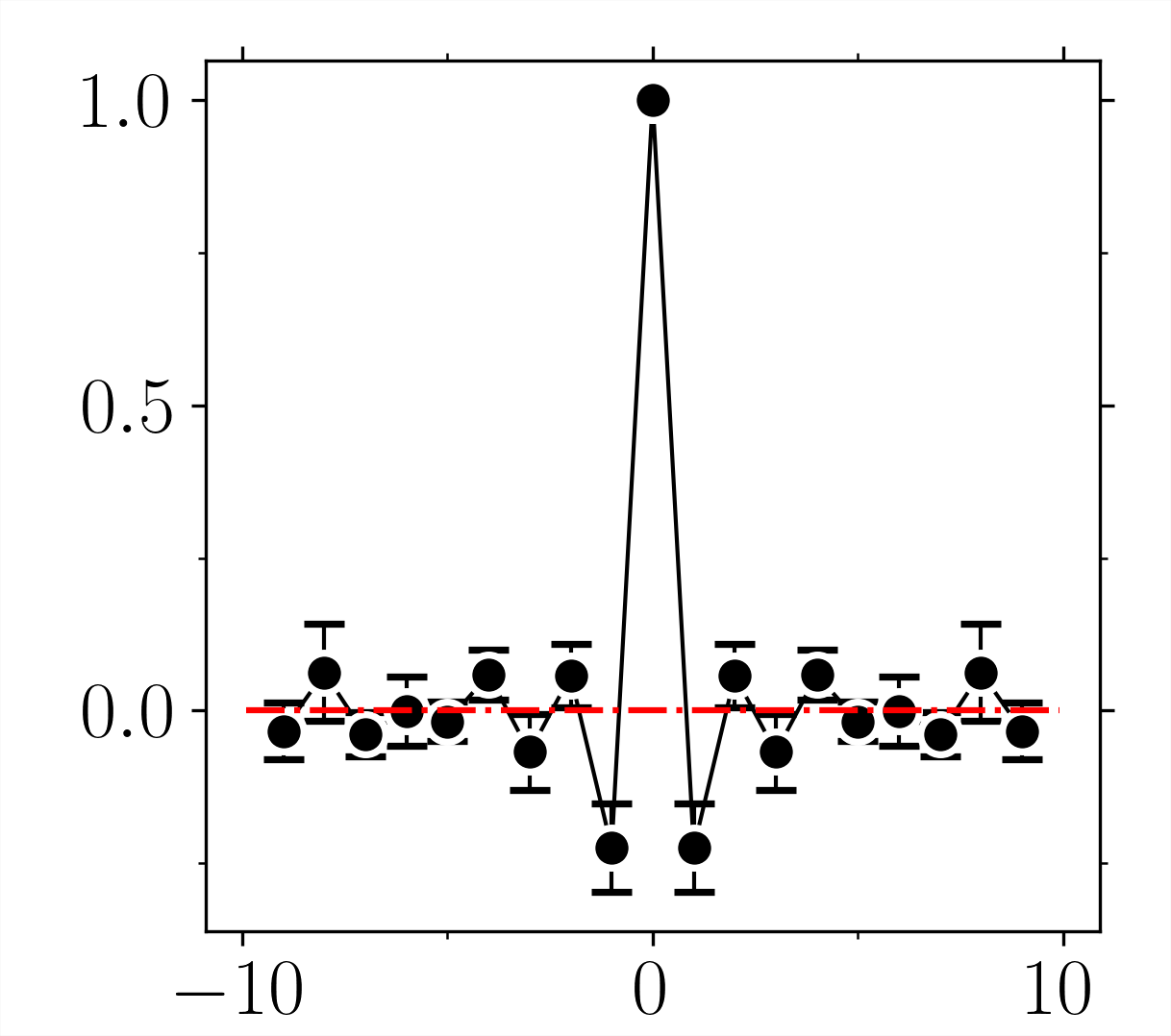}
        \LabelFig{16}{87}{$f)$}
        \Labelxy{50}{-6}{0}{$n$}
        \Labelxy{-4}{37}{90}{$c_{zz}$}
	\end{overpic}

     \begin{overpic}[width=0.23\textwidth]{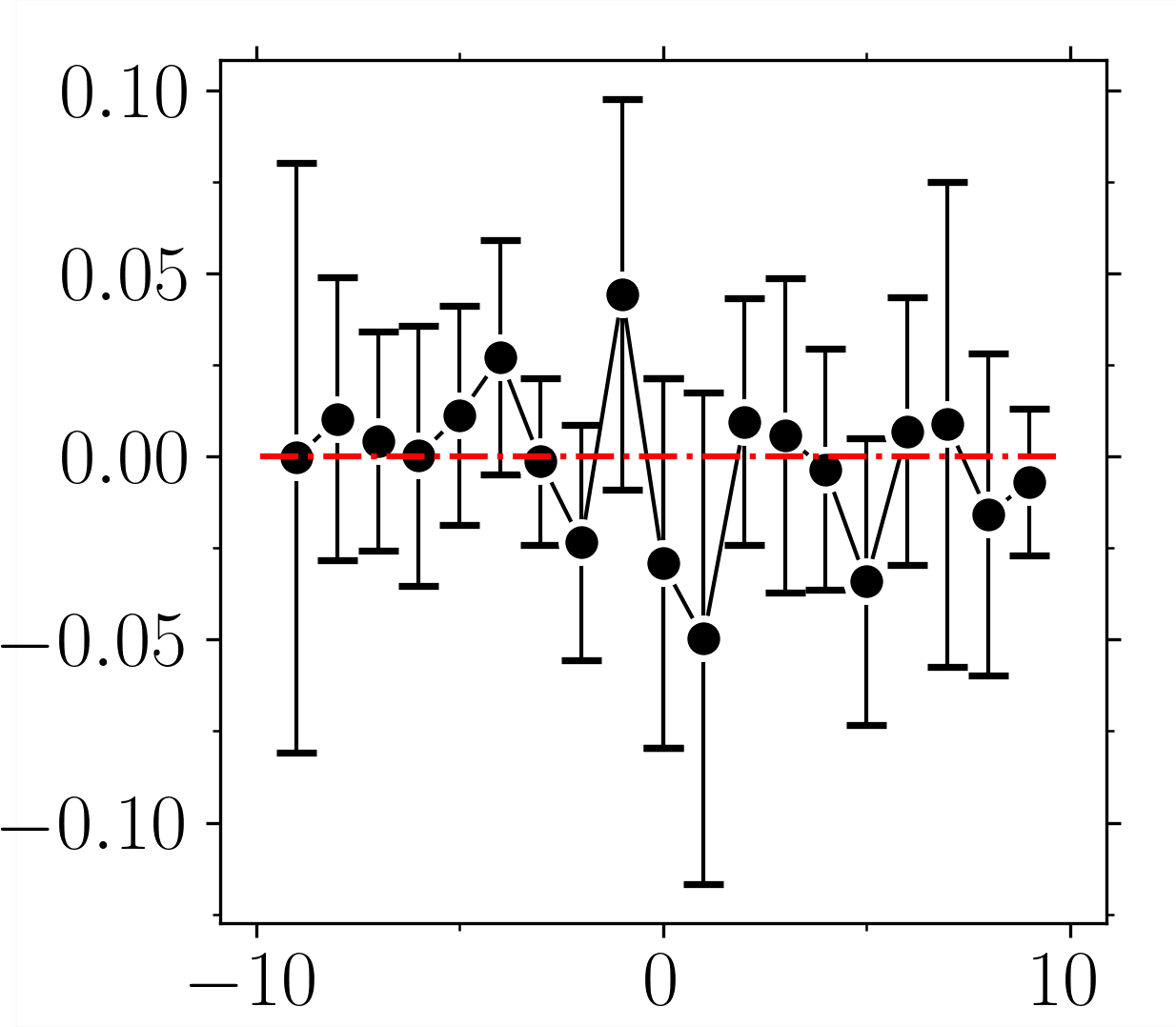}
        \LabelFig{16}{87}{$g)$}
        \Labelxy{50}{-6}{0}{$n$}
        \Labelxy{-7}{37}{90}{$c_{xy}$}
	\end{overpic}
    \begin{overpic}[width=0.23\textwidth]{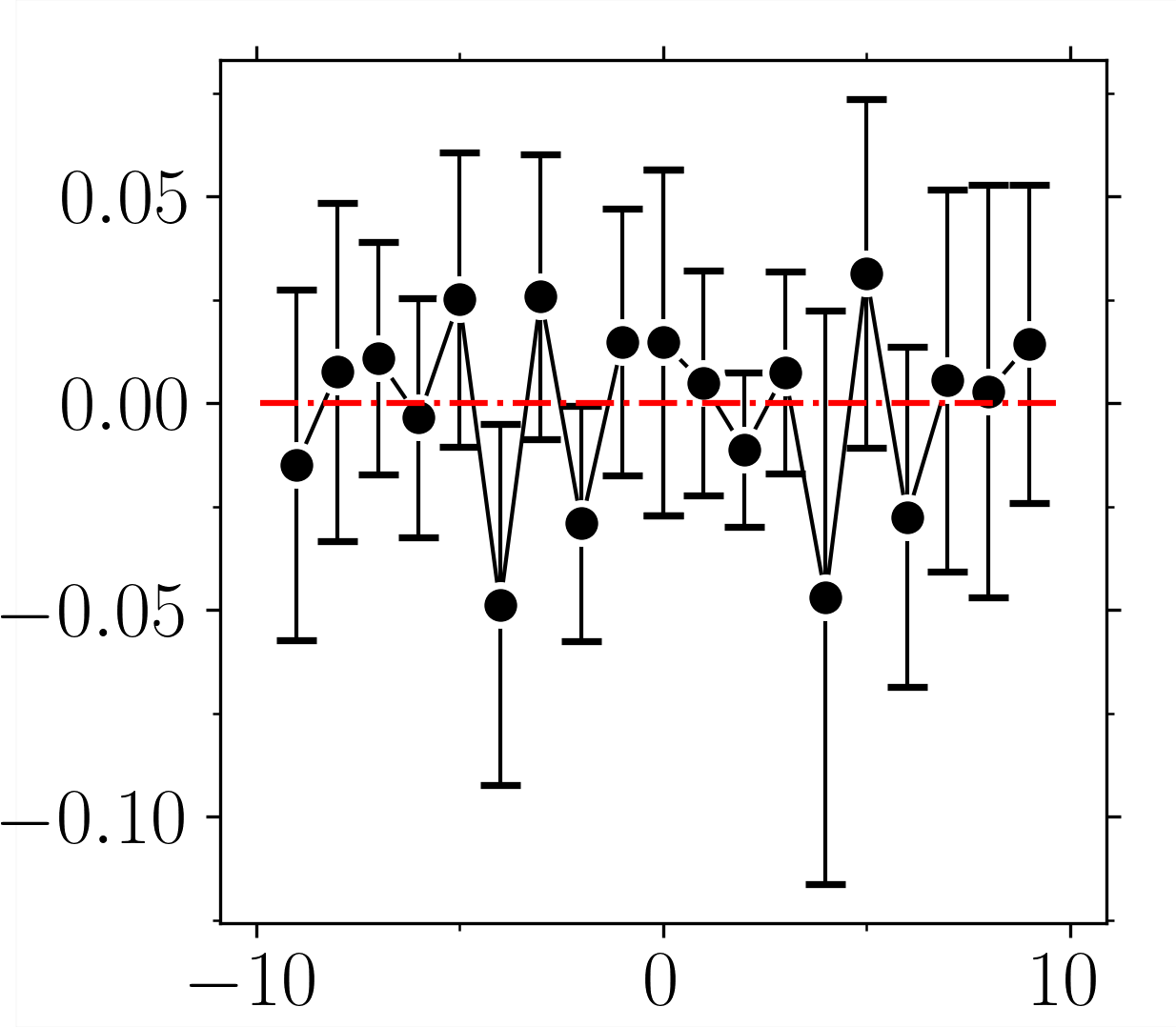}
        \LabelFig{16}{87}{$h)$}
        \Labelxy{50}{-6}{0}{$n$}
        \Labelxy{-4}{37}{90}{$c_{xz}$}
	\end{overpic}
	\vspace{8pt}
    \begin{overpic}[width=0.23\textwidth]{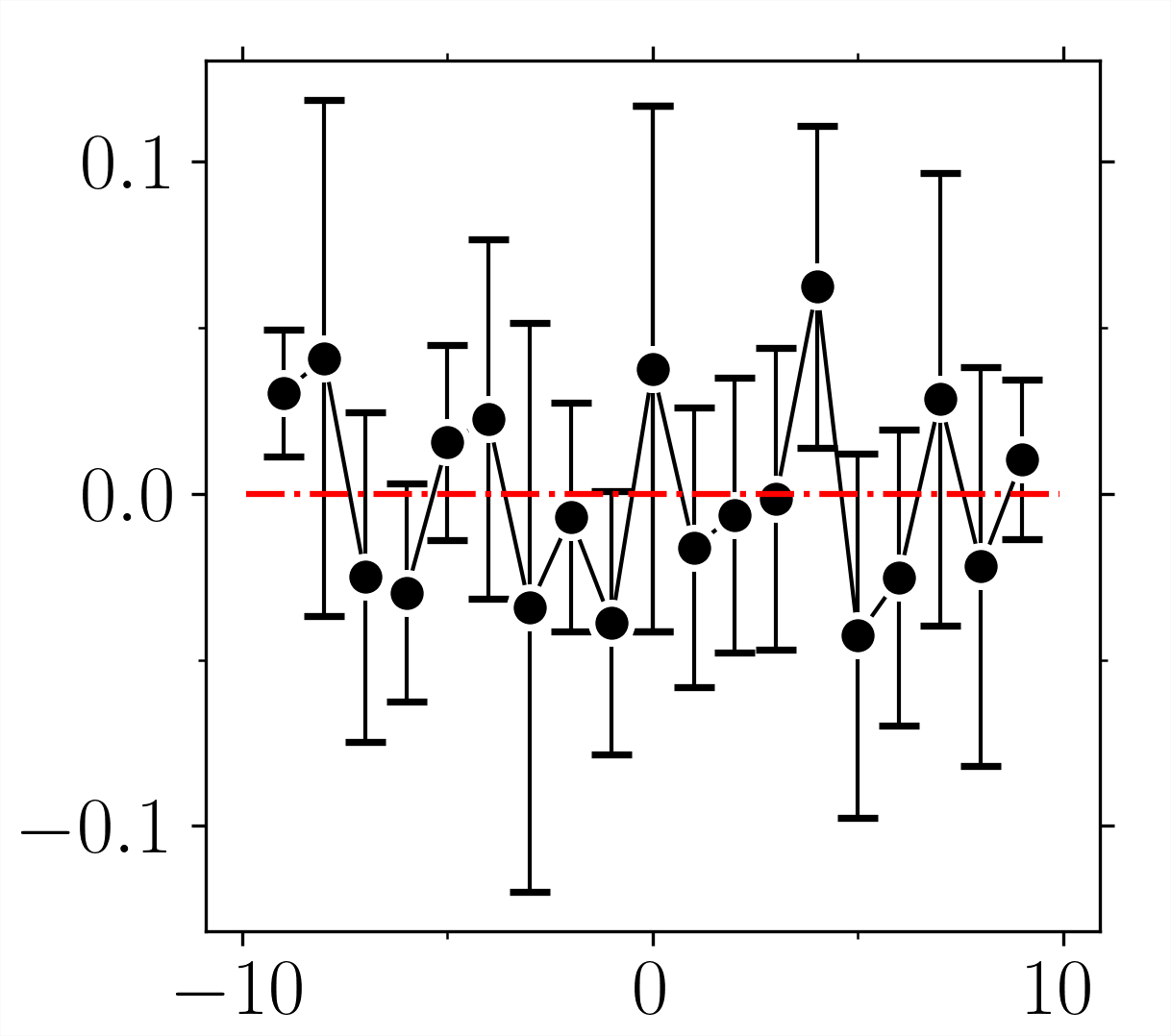}
        \LabelFig{16}{87}{$i)$}
        \Labelxy{50}{-6}{0}{$n$}
        \Labelxy{-4}{37}{90}{$c_{yz}$}
	\end{overpic}
    \caption{Dynamics of vacancy increments and associated temporal correlations in \glsix alloys at $T=1000$ K. \textbf{a}) $\Delta x(t)$ \textbf{b}) $\Delta y(t)$ \textbf{c}) $\Delta z(t)$  \textbf{d-i}) Noise correlations $c_{\alpha\beta}(n)$ with index shift $n$. The dashdotted lines indicate zero correlations. 
    }
    \label{fig:correlations}
\end{figure*}

To check the relevance of fBm, we performed a correlation analysis of the vacancy increments for multiple index shifts $n$ represented by the cross correlation function $c_{\alpha\beta}(\pm n)=\langle~\overline{\Delta r}_{k\alpha}~\overline{\Delta r}_{(k\pm n)\beta}~\rangle_k$ with $n\in(0,1,2,...)$.
Here $\langle.\rangle_k$ denotes averaging over the event index $k$ and $\overline{\Delta r}_{k\alpha}$ indicates the fluctuating part (with the mean value subtracted) normalized by the standard deviation associated with the vacancy jump ${\Delta r}_{k\alpha}$. The Greek letters $\alpha$ and $\beta$ denote Cartesian indices.
As shown in Fig.~\ref{fig:correlations}, the correlation functions associated with discrete increments are essentially indistinguishable from the noise floor suggesting no or very little ``memory" effects.

\section*{Jump Size Distributions}

\begin{figure*}[t]
    \centering
	
	%
    \begin{overpic}[width=0.23\textwidth]{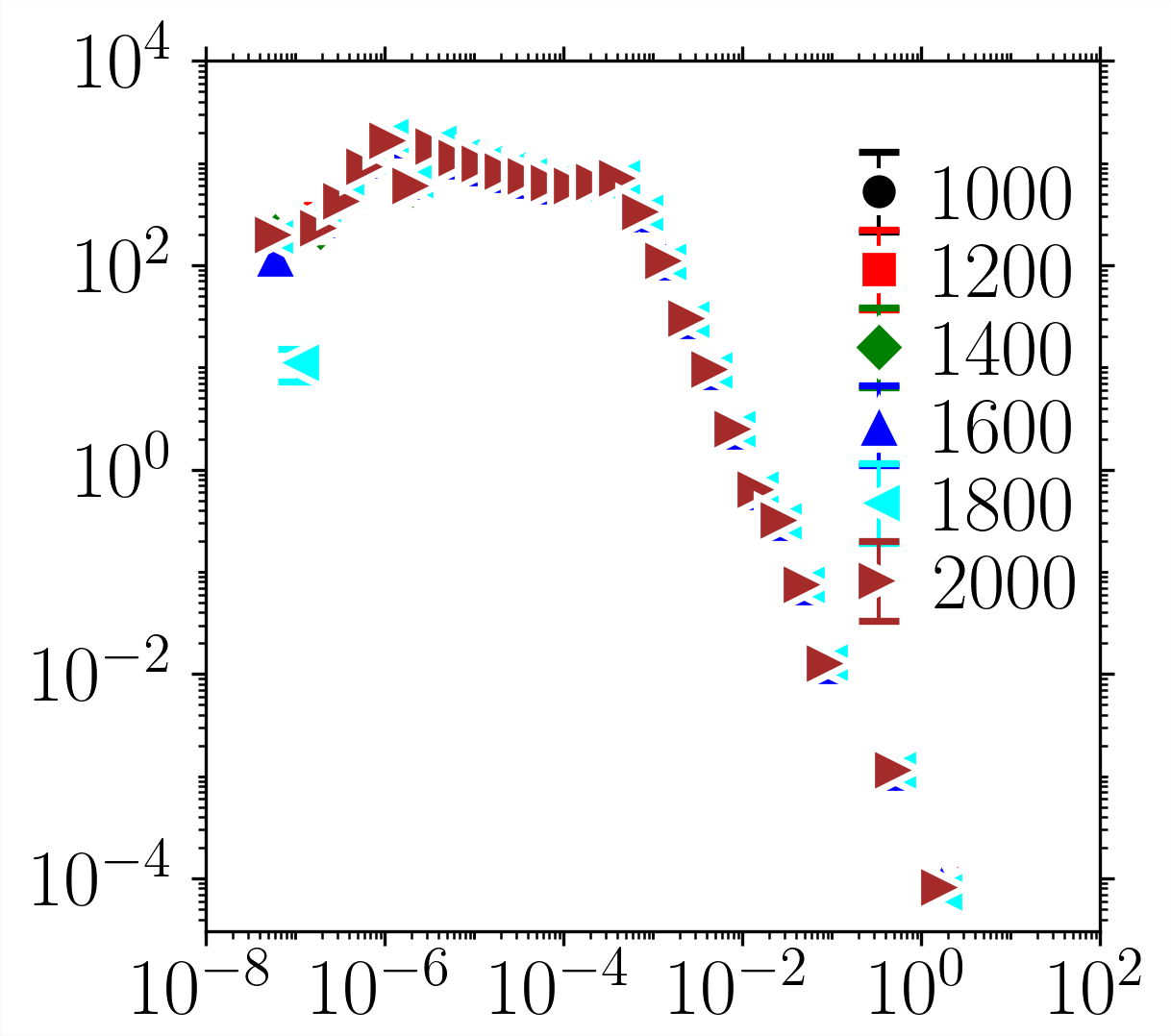}
        \put(19,10){\includegraphics[width=0.1\textwidth]{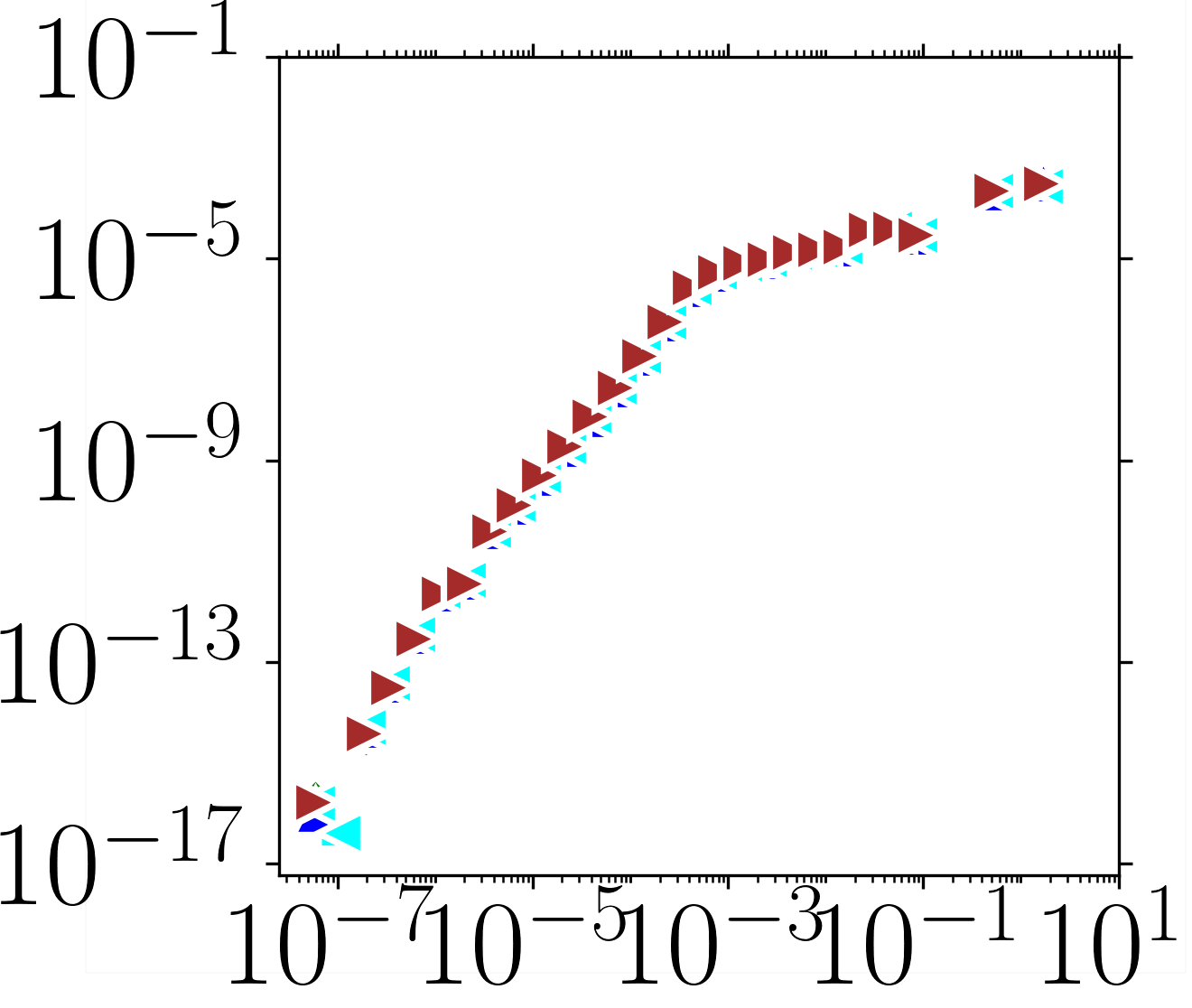}}
        \LabelFig{16}{87}{$a)$~Ni}
        \Labelxy{50}{-6}{0}{$|\Delta x|$(\r{A})}
        \Labelxy{-7}{36}{90}{$p(\Delta x)$}
        \Labelxy{75}{77}{0}{\scriptsize $T(\text{K})$}
        \Labelxy{40}{17}{0}{\tiny $|\Delta x|$}
	\end{overpic}
    \begin{overpic}[width=0.23\textwidth]{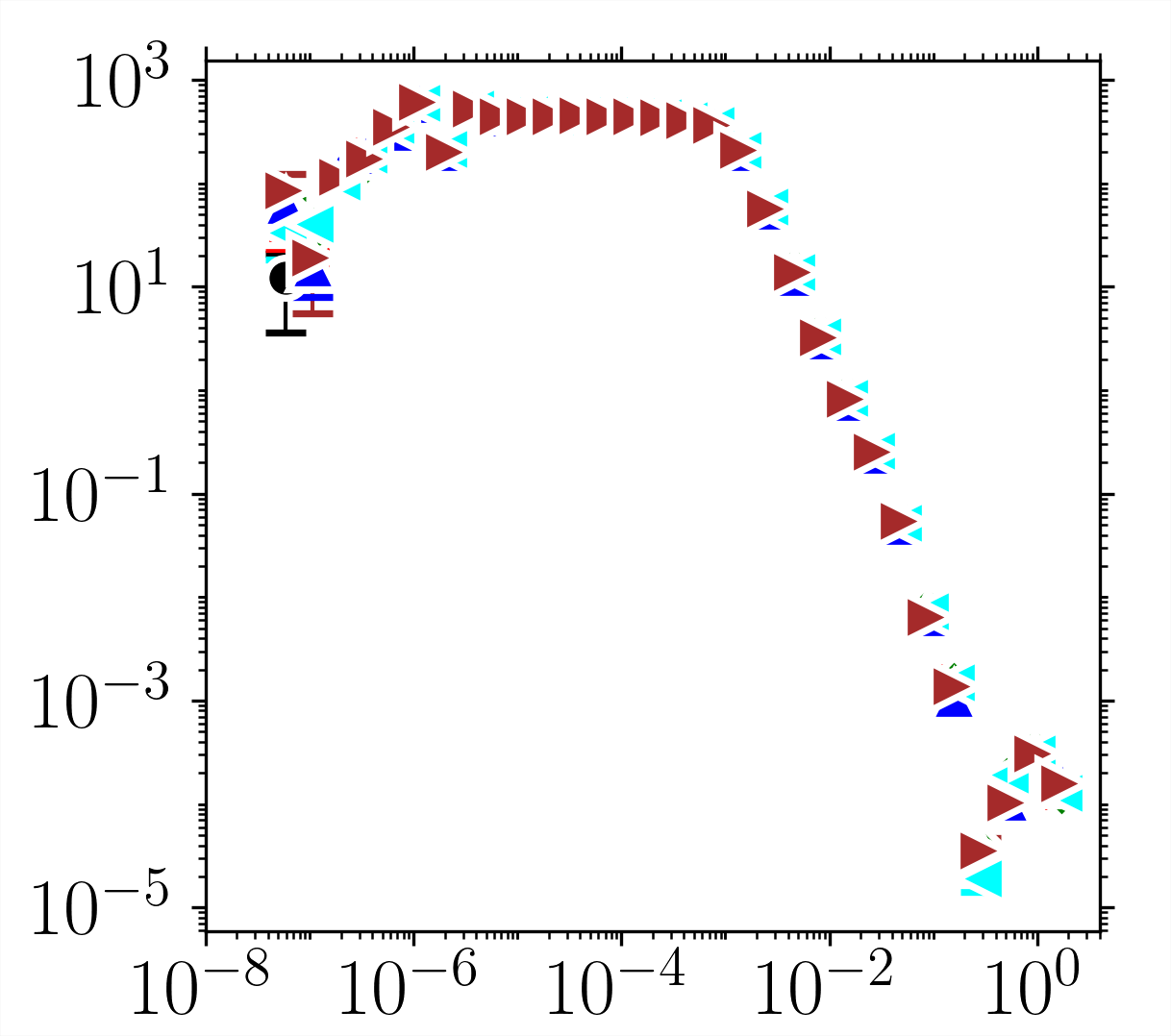}
        \put(19,10){\includegraphics[width=0.1\textwidth]{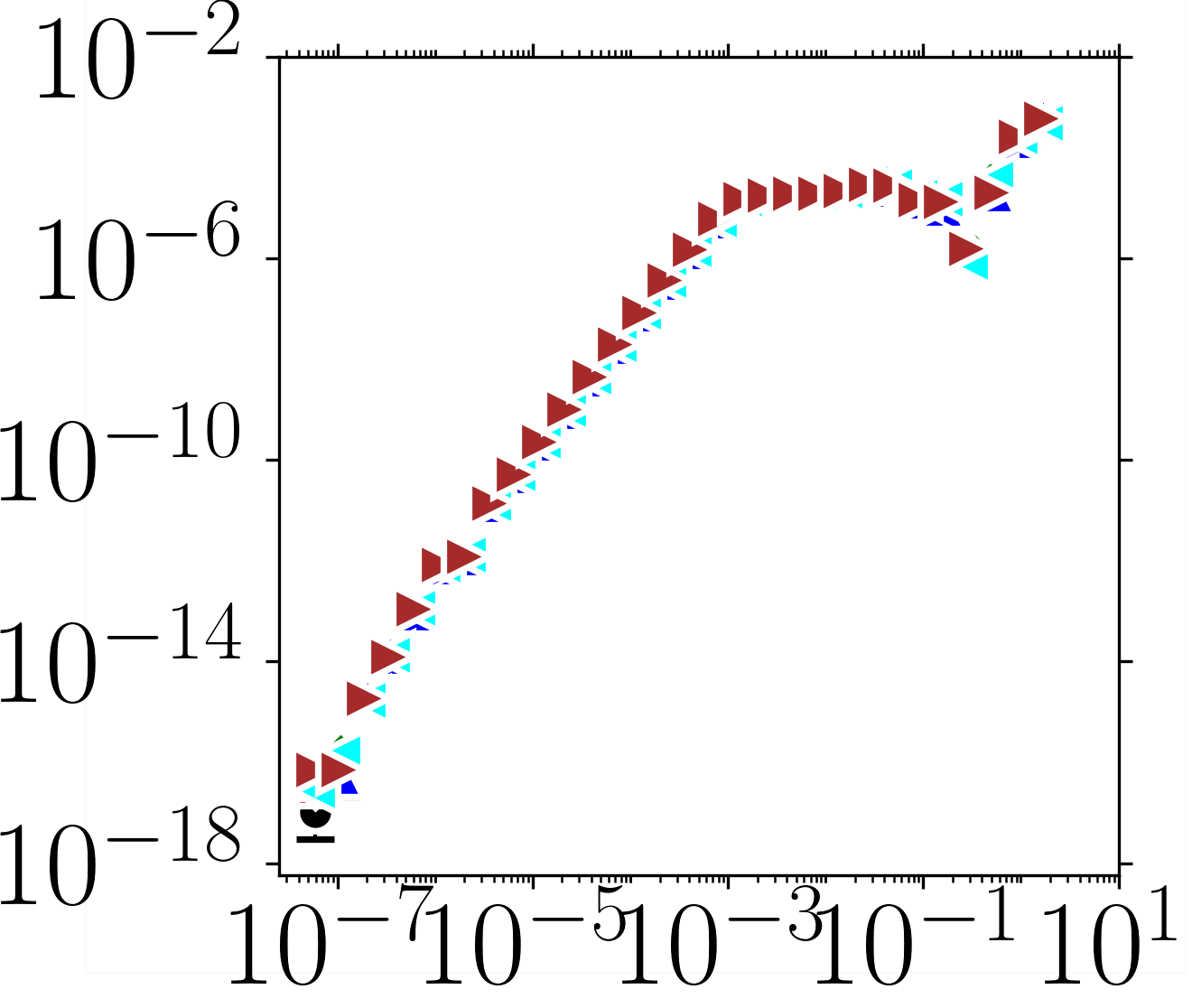}}
        %
        %
        \LabelFig{16}{87}{$b)$~\glsix}
        \Labelxy{50}{-6}{0}{$|\Delta x|$(\r{A})}
        \Labelxy{-7}{36}{90}{$p(\Delta x)$}
        \Labelxy{40}{17}{0}{\tiny $|\Delta x|$}
        \Labelxy{61}{13}{90}{ \tiny $p(\Delta x) |\Delta x|^{5/2}$}
	\end{overpic}
    \caption{Atom-scale statistics at various temperatures. Jump size distributions $p(\Delta x)$ corresponding to \textbf{a}) pure Ni \textbf{b}) \glsix alloy  as a function of time $t\text{(s)}$. 
    The insets in \textbf{a}) and \textbf{b}) are the same as the main graphs but with $p(\Delta x)$ rescaled by $|\Delta x|^{-1-\frac{d}{d-1}}$. 
    }
    \label{fig:jumpSize_nicocr}
\end{figure*}

Jump size distributions $p(\Delta x)$ are shown in Fig.~\ref{fig:jumpSize_nicocr}(a) and (b) considering a set of absolute increments $|\Delta x|=\{|\Delta x_i|,|\Delta y_i|,|\Delta z_i|\}_{i=1\cdots N}$ for pure Ni and \glsix\!.
The size distributions possess fairly long tails in both cases with a power-law decay that spans at least three orders of magnitude in $\Delta x$.
The strong tail and observed power-law behavior could be understood theoretically in the context of the long-range perturbation field that localized transformation zones embedded in an infinite elastic matrix generates in the far-field \cite{karimi2018correlation,karimi2019plastic}.
In the context of complex disordered alloys, local atomic misfits may be treated as random dilational/contractional point sources whose collective elastic-type effects govern the structure of residual strains/stresses (i.e. strength of fluctuations and associated disorder length) within the surrounding medium \cite{geslin2021microelasticity,geslin2021microelasticity_b}.
Such far-field effects are known to decay universally as inverse square distance $\Delta x \propto 1/r^{d-1}$ in three dimensions $d=3$ in a broad range of solids.
In the large $\Delta x$ limit, the theory further predicts $p(\Delta x)\propto |\Delta x|^{-(1+\mu)}$ with $\mu=d/(d-1)$ \cite{karimi2019self}.
This seems to be a relevant scaling behavior for the \glsix alloy (but not pure Ni) where the rescaled distributions $p(\Delta x) |\Delta x|^{5/2}$ show a nearly flat region over almost three orders of magnitude, as in the inset of Fig.~\ref{fig:jumpSize_nicocr}(b).
For the case of bounded mean times but diverging jump size variance with $0<\mu<2$, one essentially recovers a supper-diffusive dynamics with $\text{msd}(t) \propto t^{2/\mu}$\remove[KK]{, referred to as L\'evy flight}. 
We note a shallower-than-predicted decay of increments associated with pure Ni in the inset of Fig.~\ref{fig:jumpSize_nicocr}(a) with a characteristic peak around $2$ \r{A} that should correspond to the vacancy hopping distance (e.g. mean nearest-neighbor distance between atoms).

\bibliography{references}

%% file: macros.tex
\newcommand{\legSqTxt}[3]{
		\begin{tikzpicture}
			     \coordinate (center1) at (#1,#2); 
                \coordinate (b) at ($ (center1) - (.075,0) $);
                \coordinate (c) at ($ (b) - (.1,0) $); 
                \coordinate (d) at ($ (center1) + (.075,0) $);
                \coordinate (e) at ($ (d) + (.1,0) $); 
				 \draw[#3,fill=#3] ($(b)-(0.0,0.075)$) rectangle ($ (d) + (0.0,0.075) $); 
		\end{tikzpicture}		
}

\newcommand{\circTxtFill}[3]{
		\begin{tikzpicture}
                \coordinate (center1) at (#1,#2); 
				 \draw[#3,fill=#3] (center1) circle (2pt);
		\end{tikzpicture}		
}

\newcommand{\legDiamondTxt}[3]{
		\begin{tikzpicture}
			     \coordinate (center1) at (#1,#2); 
                \coordinate (b) at ($ (center1) + (0.075,0.0) $); 
                \coordinate (c) at ($ (center1) + (0,0.075) $); 
                \coordinate (d) at ($ (center1) + (-0.075,0) $);
                \coordinate (e) at ($ (center1) + (0.0,-0.075) $);
				 \draw[#3,fill=#3] (b) -- (c) -- (d) -- (e) -- (b);
                \coordinate (c) at ($ (b) + (.1,0) $); 
                \coordinate (e) at ($ (d) - (.1,0) $); 
		\end{tikzpicture}		
}
                
\newcommand{\legTriangleTxt}[3]{
		\begin{tikzpicture}
			     \coordinate (center1) at (#1,#2); 
                \coordinate (b) at ($ (center1) - 0.17*(0.5,0.289) $); 
                \coordinate (c) at ($(center1) + 0.17*(0.5,-0.289) $); 
                \coordinate (d) at ($ (center1) + 0.17*(0,0.366) $);
				\draw[black,fill=#3, line width=0.1mm] 
    (b) -- (c) -- (d) -- cycle;                 
    			\coordinate (b) at ($ (center1) - (.075,0) $);
                \coordinate (c) at ($ (b) - (.1,0) $); 
                \coordinate (d) at ($ (center1) + (.075,0) $);
                \coordinate (e) at ($ (d) + (.1,0) $); 
		\end{tikzpicture}		
}		
		
\definecolor{light-gray}{gray}{0.85}

\newcommand{\leftTriang}[3]{
		\begin{tikzpicture}
			     \coordinate (center1) at (#1,#2); 
                \coordinate (b) at ($ (center1) + 0.17*(0.289,0.5) $); 
                \coordinate (c) at ($ (center1) + 0.17*(0.289,-0.5) $); 
                \coordinate (d) at ($ (center1) + 0.17*(-0.366,0.0) $);
				 \draw[fill=#3,#3] (b) -- (c) -- (d) -- (b);
                \coordinate (b) at ($ (center1) - (.075,0) $);
                \coordinate (c) at ($ (b) - (.1,0) $); 
                \coordinate (d) at ($ (center1) + (.075,0) $);
                \coordinate (e) at ($ (d) + (.1,0) $); 
		\end{tikzpicture}		
}

\newcommand{\rightTriang}[3]{
		\begin{tikzpicture}
			    \coordinate (center1) at (#1,#2); 
                \coordinate (b) at ($ (center1) + 0.17*(-0.289,0.5) $); 
                \coordinate (c) at ($ (center1) + 0.17*(-0.289,-0.5) $); 
                \coordinate (d) at ($ (center1) + 0.17*(0.366,0.0) $);
				 \draw[black,fill=#3] (b) -- (c) -- (d) -- (b);
                \coordinate (b) at ($ (center1) - (.075,0) $);
                \coordinate (c) at ($ (b) - (.1,0) $); 
                \coordinate (d) at ($ (center1) + (.075,0) $);
                \coordinate (e) at ($ (d) + (.1,0) $); 
		\end{tikzpicture}		
 }
\newcommand{\drawSlope}[6]{ 
				\coordinate (center1) at (#1,#2); 
				 \FPeval{\nx}{cos(#4*pi/180)}%
				 \FPeval{\ny}{sin(#4*pi/180)}%
				 \FPeval{\absNx}{abs(\nx)}%
				 \FPeval{\absNy}{abs(\ny)}%
				\coordinate (b) at ($ (center1) + #3*(-\ny,+\nx) $);
				\coordinate (c) at ($ (center1) - #3*(-\ny,+\nx) $); 
				 \FPeval{\xx}{(-\nx*\ny)}%
				\ifdim\xx pt < 0pt 
				\coordinate (d) at ($ (c) -2*#3*\ny*(1,0) $);
				\node at ($(d) +(0,#3*\nx)-0.2*(\nx/\absNx,0)$) {{\color{#5}#6}}; 
				\else
				\coordinate (d) at ($ (c) +2*#3*\nx*(0,1) $);
				\node at ($(d)-#3*\nx*(0,1)-0.2*\nx/\absNx*(1,0)$) {{\color{#5}#6}}; 
				\fi
				\draw[#5,line width=0.1mm] (d) -- (b); 
				\draw[#5,line width=0.1mm] (d) -- (c); 
				\draw[#5,line width=0.1mm] (b) -- (c); 
} 

\newcommand{\coordSys}[9]{
			\coordinate (center2) at (#1,#2); 
            \coordinate (x1) at ($ (center2) + #3*(#4,#5) $); 
            \coordinate (x2) at ($ (center2) - .3*#3*(#4,#5) $); 
            \coordinate (y1) at ($ (center2) + #3*(#6,#7) $);
            \coordinate (y2) at ($ (center2) - .3*#3*(#6,#7) $);
            \coordinate (z1) at ($ (center2) + #3*(#8,#9) $); 
            \coordinate (z2) at ($ (center2) - .3*#3*(#8,#9) $); 
            \draw[->,>=stealth,black][line width=0.4mm] (x2) -- (x1);
			\node at ($(x1)+(0.2,0)$){{\color{black} $x$}};
            \draw[->,>=stealth,black][line width=0.4mm] (y2) -- (y1);
            \node at ($(y1)+(0,0.2)$){{\color{black}  $y$}};
            \draw[->,>=stealth,black][line width=0.4mm] (z2) -- (z1);
			\node at ($(z1)+(0,0.2)$){{\color{black}  $z$}};
    	}
    	
\newcommand{\arrow}[6]{
			\coordinate (center2) at (#1,#2); 
            \coordinate (x1) at ($ (center2) + #3*(#4,#5) $); 
            \coordinate (x2) at ($ (center2) - .3*#3*(#4,#5) $); 
            \draw[->,>=stealth,black][line width=0.4mm] (x2) -- (x1);
			\node at ($(x1)+(0.2,0)$){#6}; 
}

\newcommand{\lline}[6]{
			\coordinate (center2) at (#1,#2); 
            \coordinate (x1) at ($ (center2) + #3*(#4,#5) $); 
            \coordinate (x2) at ($ (center2) - .3*#3*(#4,#5) $); 
            \draw[black][line width=0.4mm] (x2) -- (x1);
			\node at ($(x1)+(0.2,0)$){#6}; 
}